\documentclass{aa}
\newcommand{\hMsun}{{\ifmmode{h^{-1}{\rm {M_{\odot}}}}\else{$h^{-1}{\rm{M_{\odot}}}$}\fi}}  
\newcommand{\hMpc}{{\ifmmode{h^{-1}{\rm Mpc}}\else{$h^{-1}$Mpc }\fi}}  
\newcommand{\hkpc}{{\ifmmode{h^{-1}{\rm kpc}}\else{$h^{-1}$kpc }\fi}}
\newcommand{\kms}{\,km~s$^{-1}$} 

\usepackage{graphicx, subfig}
\usepackage{rotating}
\usepackage{txfonts}
\usepackage{natbib}
\usepackage{lscape}
\usepackage{array}
\usepackage{color}

\begin{document}
%
\title{Characterizing SL2S galaxy groups using the Einstein radius
\thanks{SL2S: Strong Lensing Legacy Survey}$^{,}$  
\thanks{Based on observations obtained with MegaPrime/MegaCam, a joint project of CFHT and CEA/DAPNIA, at the Canada-France-Hawaii Telescope (CFHT) which is operated by the National Research Council (NRC) of Canada, the Institut National des Sciences de l'Univers of the center National de la Recherche Scientifique (CNRS) of France, and the University of Hawaii. This work is based in part on data products produced at TERAPIX and the Canadian Astronomy Data center as part of the Canada-France-Hawaii Telescope Legacy Survey, a collaborative project of NRC and CNRS. Also based on \emph{Hubble Space Telescope} (HST) data as well as Magellan (IMACS) and VLT (FORS\,2) data.} }
   	\subtitle{}
   \author{T. Verdugo\inst{1,2},
                V. Motta\inst{2},
                G. Fo{\"e}x\inst{2},
                J. E. Forero-Romero\inst{3},
                R. P. Mu\~{n}oz\inst{4},                
                R. Pello\inst{5},
                M. Limousin\inst{6,7},
                A. More\inst{8},
                R. Cabanac\inst{5},
                G. Soucail\inst{9},
                J. P. Blakeslee\inst{10},
                A. J. Mej\'ia-Narv\'aez\inst{1,11},
                G. Magris\inst{1},
                J. G. Fern\'andez-Trincado\inst{1,11,12}
                }
   \offprints{verdugo$@$cida.ve}

   \institute{Centro de Investigaciones de Astronom\'ia, Apartado Postal 264, M\'erida 5101-A, Venezuela
   \and
   Universidad de Valpara\'{\i}so, Departamento de F\'{\i}sica y Astronom\'{\i}a, Avenida Gran Breta\~{n}a 1111, Valpara\'{\i}so, Chile
       \and
         Departamento de F\'{i}sica, Universidad de los Andes, Cra. 1 No. 18A-10, Edificio Ip, Bogot\'a, Colombia
        \and
        Instituto de Astrof\'isica, Facultad de F\'isica, Pontificia Universidad Cat\'olica de Chile, Av.~Vicu\~na Mackenna 4860, 7820436 Macul, Santiago, Chile
         \and
        Laboratoire d'Astrophysique de Toulouse-Tarbes, Universit\'e de Toulouse, CNRS,
     57 Avenue d'Azereix, 65 000 Tarbes, France
     \and
                 Laboratoire d'Astrophysique de Marseille, Universit\'e de Provence,\\ CNRS, 38 rue Fr\'ed\'eric Joliot-Curie, F-13388 Marseille Cedex
13, France
                \and
        Dark Cosmology Center, Niels Bohr Institute, University of Copenhagen,
       Juliane Marie Vej 30, 2100 Copenhagen, Denmark 
     \and
      Kavli Institute for the Physics and Mathematics of the Universe (Kavli IPMU), The University of Tokyo, 5-1-5 Kashiwanoha, Kashiwa-shi, Chiba, 277-8583, Japan
      \and
       Universit\'e de Toulouse, UPS-Observatoire Midi-Pyr\'en\'ees, IRAP, Toulouse, France
             \and
       Herzberg Institute of Astrophysics, National Research Council of Canada, Victoria, BC V9E 2E7, Canada 
     \and
     Universidad de Los Andes, Posgrado de F\'isica Fundamental, La Hechicera, M\'erida, Venezuela. 
     \and
     Institute Utinam, CNRS UMR6213, Universit\'e de Franche-Comt\'e, OSU THETA de Franche-Comt\'e-Bourgogne, Besan\c{c}on, France  
   }

 \date{Accepted for publication }

 \abstract
  {}
  {We aim to study the reliability of $R_A$ (the distance from the arcs to the center of the lens) as a measure of the Einstein radius in galaxy groups. In addition, we want to analyze the possibility of using $R_A$ as a proxy to characterize some properties of galaxy groups, such as luminosity (L) and richness (N).}
  {We analyzed the Einstein radius, $\theta_E$, in our sample of  Strong Lensing Legacy Survey (SL2S) galaxy groups, and compared it with $R_A$, using three different approaches: 1) the velocity dispersion obtained from weak lensing assuming a singular isothermal sphere profile ($\theta_{E,I}$), 2) a strong lensing analytical method ($\theta_{E,II}$)  combined with a velocity dispersion-concentration relation derived from numerical simulations designed to mimic our group sample, and 3) strong lensing modeling ($\theta_{E,III}$)  of eleven groups (with four new models presented in this work) using Hubble Space Telescope (HST) and Canada-France-Hawaii Telescope (CFHT) images. Finally,  $R_A$ was analyzed as a function of redshift $z$  to investigate possible correlations with L, N, and the richness-to-luminosity ratio (N/L).}
 {We found a correlation between $\theta_{E}$ and $R_A$, but with large scatter.
We estimate $\theta_{E,I}$ = (2.2 $\pm$ 0.9) + (0.7 $\pm$ 0.2)$R_A$, $\theta_{E,II}$ = (0.4 $\pm$ 1.5) + (1.1 $\pm$ 0.4)$R_A$, and $\theta_{E,III}$ = (0.4 $\pm$ 1.5) + (0.9 $\pm$ 0.3)$R_A$ for each method respectively. We found weak evidence of anti-correlation between $R_A$ and $z$, with Log$R_A$ = (0.58$\pm$0.06) - (0.04$\pm$0.1)$z$, suggesting a possible evolution of the Einstein radius with $z$, as reported previously by other authors.  Our results also show that $R_A$ is correlated with L and N (more luminous and richer groups have greater $R_A$),  and a possible correlation between $R_A$ and the N/L ratio.}
 {Our analysis indicates that $R_A$ is correlated with $\theta_E$ in our sample, making $R_A$ useful for characterizing properties like L and N (and possibly N/L) in galaxy groups. Additionally, we present evidence suggesting that the Einstein radius evolves with $z$.}


   \titlerunning{Einstein radii in SL2S}
   \authorrunning{Verdugo et~al.}

  \maketitle
%

 
 \section{Introduction}\label{Intro}

Since most of the galaxies in the Universe belong to galaxy groups \citep[][]{Eke2004}, the systematic examination of this intermediate regime of the mass-spectrum (between large elliptical galaxies and clusters) will shed light on the formation and evolution of structures in the hierarchical framework. Although galaxy groups have been the subject of study from different approaches, such as optical \citep[e.g.,][]{Wilman2005a,Wilman2005b,Yang2008,Knobel2009,Cucciati2010,Balogh2011,Li2012},  X-ray \citep[e.g.,][]{Helsdon2000a,Helsdon2000b,Osmond2004,Willis2005,Finoguenov2007,Rasmussen2007,Sun2012}, and numerical simulations \citep[e.g.,][]{Sommer-Larsen2006,Romeo2008,Cui2011};  the systematic investigation of such a mass regime from a lensing perspective has recently started  \citep[e.g.,][]{Mandelbaum2006,paperI,More2012}.

The Strong Lensing Legacy Survey \citep[SL2S\footnote{\tt http://www-sl2s.iap.fr/},][]{Cabanac2007}  selects  its sample from
the Canada-France-Hawaii Telescope Legacy Survey (CFHTLS)\footnote{\tt http://www.cfht.hawaii.edu/Science/CFHLS/}.  The SL2S 
has allowed us to find and study a large sample of group-scale lenses  \citep{More2012}, as well as galaxy-scale gravitational lenses \citep{Gavazzi2012}. Some galaxy groups discovered in the SL2S have been studied in detail using different  techniques \citep[e.g.,][]{Tu09,paperI,Limousin2010,Thanjavur2010,Verdugo2011}, further highlighting the importance of  SL2S. \citet{More2012} showed the first compilation of lens candidates, the SL2S-ARCS (SARCS) sample, consisting of 127 objects, with 54 systems labeled as promising lenses.  The authors also present the first constraints on the average mass density profile of groups using strong lensing. One of the main goals of the SL2S is to accurately determine  the characteristics of the lensing groups through various methods, for example with dynamics using spectroscopy \citep{Roberto2013} as well as weak lensing analysis and luminosity density maps \citep{Gael2013}. In particular, the latter work combines lensing and optical analysis to further constrain the sample, and presents a list of the 80 most secure lens candidates. Even though most of the objects in the sample must be confirmed, these objects present a weak-lensing signal (detection at the 1$\sigma$ level), and show an over-density in their luminosity density maps \citep[see][for a detailed discussion]{Gael2013}. Thus, these candidates give us the opportunity to test a wide range of astrophysical problems and to probe diverse phenomena.  
 
For  instance, \citet{Zitrin2012} analyzed the universal distribution of the Einstein radius on 10000 clusters in the SDSS, discussing the possibility of an Einstein radius evolution with redshift. These authors reported that the mean effective Einstein radius decreases between $z$ = 0.1 to $z$ = 0.45, and argue that such a decrease is possibly  related to cluster evolution, since clusters at lower $z$ are expected to have more concentrated mass distributions, thus they are stronger lenses. Considering only geometrical effects, they demonstrated that a  profile steeper than  the singular isothermal sphere (SIS) is necessary to explain the decline of $\sim$40$\%$. \citet{Zitrin2012} explain the tentative increase in the Einstein radius towards $z$ = 0.5 invoking an increase in the size of the critical curves, as a result of merging subclumps in clusters \citep[e.g.,][]{Torri2004,Dalal2004,Redlich2012}.

Our sample of secure lens candidates \citep{Gael2013}  can be useful to test such an assertion, namely the $\theta_E$ evolution with redshift.  Although it is clearly in a distinct mass range, our sample has a larger range in redshift and the benefit of being selected by their strong lensing features.  In our analysis we assume that $R_A$ (the distance between the more extended lensed image and the brightest lens galaxy in the group) is roughly the Einstein radius \citep[obtained by][]{More2012}. This is justified since the Einstein radius provides a natural angular scale to describe the lensing geometry \citep{Narayan1996}; the typical angular separation of images is on the order of  2$\theta_E$. However, we need to be cautious because there are some factors that could bias the comparison. For example, the sample selected by \citet[][]{More2012} is made up of groups that display small arcs and giant arcs  (see Section ~\ref{WeakLensing}). For giant arcs, comparing $R_A$ and  $\theta_E$ is a rough estimation \citep[e.g.,][]{MiraldaEscude1995}, since this kind of arc tends to appear close to the critical curve in a spherically symmetric mass distribution model (although in general lenses are elongated). On the other hand, comparing $R_A$ and  $\theta_E$ could be inaccurate for those images (which are not giant arcs)  that appear, for example, along the major-axis critical curve. 
In this sense, it is important to note that arc radial positions  could extend beyond the Einstein radius \citep[depending on the Einstein radius definition, e.g.,][]{Puchwein2009,Richard2009}. Furthermore,  comparison between the expected $\theta_E$ in the Lambda cold dark matter ($\Lambda$CDM) paradigm 
 and observations may lead to different conclusions  depending on the assumption of spherical or triaxial dark matter halos \citep[e.g.,][]{Broadhurst08,Oguri2009}, and on which $\theta_E$ geometrical definition is used \citep[see the discussion in the review of][]{Meneghetti2013}.  Although from a lensing perspective galaxy groups are not as complex as galaxy clusters, 
some natural questions arise: Is $R_A$ a reliable estimation for  $\theta_E$? What effect does asphericity or substructure have on such an assumption? The aim of the present work is to answer  these questions and  test the viability  of using $R_A$ as a proxy to characterize or even quantify some properties such as luminosity or richness in galaxy groups. As scaling relations are naturally expected (and have been observed at different redshifts) between the mass and optical properties in groups and clusters \citep[e.g.,][]{Lin2003,Popesso2005,Becker2007,Reyes2008,Rozo2009,Andreon2010,Gael2012,Gael2013}, a correlation between $R_A$ and these properties follows clearly because $R_A$ scales with the mass of the halos.

\begin{table*}
\caption{Photometric data for galaxies and arcs used in the four new gravitational lensing models.}
\label{tbl-1} 
\centering 
\begin{tabular}{lccccccc}
\hline\hline 
\\
ID     &  Galaxy/Arc   &
$u^{*}$ &
$g'$  &
$r'$  &
$i'$  & $z'$  &  $k_s$  \\
\hline 
\\
SL2S J08591$-$0345 (SA72) &  G  &  25.4 $\pm $ 0.1 & 23.51 $\pm $ 0.02 & 21.96 $\pm $ 0.01 & 20.681 $\pm $ 0.004 & 20.245  $\pm $ 0.005 & --    \\[3pt]
  &    A & 25.0 $\pm $ 0.1 & 24.1 $\pm $ 0.1  & 23.4 $\pm $ 0.1 & 22.7 $\pm $ 0.1 & 22.4 $\pm $ 0.1 & 20.7 $\pm $ 0.1    \\[3pt]
SL2S J08520$-$0343  (SA63) &  G1  &  23.62 $\pm $ 0.06 & 21.437 $\pm $ 0.009 & 19.808 $\pm $ 0.004 & 18.971 $\pm $ 0.003 & 18.577  $\pm $ 0.005 & --    \\[3pt]
  &    A & 27.2 $\pm $ 0.9 & 24.42 $\pm $ 0.07  & 23.39 $\pm $ 0.04 & 22.56 $\pm $ 0.03 & 22.52 $\pm $ 0.08 & --   \\[3pt]
SL2S J09595$+$0218 (SA80) &  G  &  25.59 $\pm $ 0.05 & 24.69 $\pm $ 0.02 & 23.098 $\pm $ 0.007 & 21.753 $\pm $ 0.003 & 20.987  $\pm $ 0.004 & --    \\[3pt] 
     &  A  &  25.23 $\pm $ 0.04 & 24.97 $\pm $ 0.03 & 25.20 $\pm $ 0.05 & 24.94 $\pm $ 0.05 & 24.14  $\pm $ 0.07 & --    \\[3pt]
SL2S J10021$+$0211 (SA83) &  G  &  25.59 $\pm $ 0.05 & 24.69 $\pm $ 0.02 & 23.098 $\pm $ 0.007 & 21.753 $\pm $ 0.003 & 20.987  $\pm $ 0.004 & --    \\[3pt]
       &  B  &  25.83 $\pm $ 0.07 & 24.04 $\pm $ 0.01 & 23.42 $\pm $ 0.01 & 23.04 $\pm $ 0.01 & 22.68  $\pm $ 0.01 & --    \\[3pt] 
\\
\hline 
\end{tabular}
\tablefoot{Column (1) is the identification for each object (see text). Column (2) is the object type. Columns (3-7) are the CFHTLS magnitudes. Column (8) is the magnitude in the $k_s$ band from  WIRCam.
}
\end{table*}

Nowadays we have an extraordinary amount of data available in the search for and study of lensing galaxy groups from the Sloan Digital Sky Survey \citep{Abazajian2003} or the CFHTLS, for example. There will be even more data with the upcoming long-term big survey projects such as LSST \citep{LSST2012}, the Dark Energy Survey (DES\footnote{http://www.darkenergysurvey.org/}), and EUCLID \citep{Boldrin2012}. Even though automated software is used to look for strong lensing candidate detection \citep[e.g.,][]{Alard2006,Seidel2007,Marshall2009,Sygnet2010,Maturi2013,Joseph2014,Gavazzi2014}, we still lack  crucial information for accurate lens modeling. For instance, the impossibility (in most cases) of spectroscopically confirming the lensing nature of arcs in groups, or even dynamically confirming that the members of the  group-lensing candidates are gravitationally-bound structures \citep[e.g.,][]{Thanjavur2010, Roberto2013}. The study presented by \citet{Gael2013}, and the present work, try to tackle this lack of spectroscopic information, by analyzing the strong-lensing group candidates using complementary approaches.

\begin{figure}[h!]
\begin{center}
\includegraphics[scale=0.5]{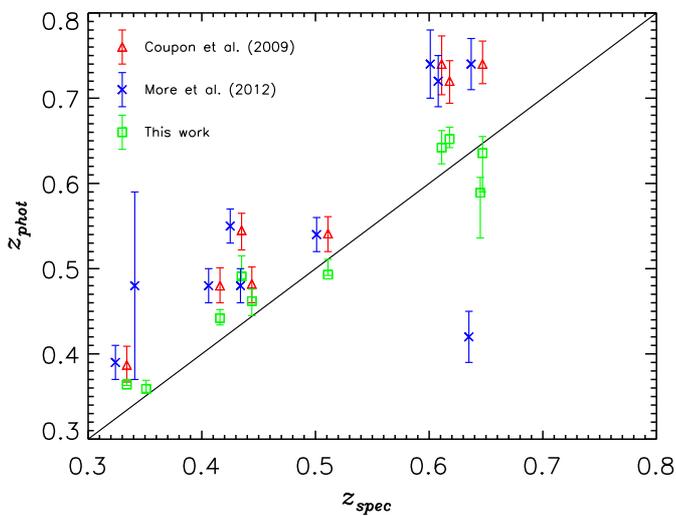}
\caption{Comparison between spectroscopic \citep[from][]{paperI,Roberto2013} and photometric redshifts. Red triangles depict the $z_{phot}$ from \citet{Coupon2009}, blue crosses the values reported in \citet{More2012}, and green squares those from this work. Blue crosses were shifted in $z_{spec}$ for clarity.\label{ZphotZspc} }
\end{center}\end{figure}

To this end, we present the Einstein radius analysis of the secure sample of galaxy groups in the SARCS sample. 
We consider three methods that use 1)  the velocity dispersion from  a weak lensing analysis, following \citet{Gael2013}; 2) strong lensing models following \citet{Broadhurst08},   together with numerical simulations that mimic the properties of our group sample; and 3) strong lensing modeling using a ray-tracing code (coupled with new spectroscopic data for one of the groups). Finally, we analyze the correlations between $R_A$ and the optical properties of the groups. Our paper is arranged as follows: In Sect.\,\ref{Data} we present the observational data images and spectroscopy. We describe the numerical simulations in Sect.\,\ref{Simulations}. In Sect.\,\ref{EinsteinRadius} we explain the methodology used to calculate the Einstein radius with the three different methods. We summarize and discuss our results in Sect.\,\ref{Discussion}. Finally in Sect.\,\ref{Conclusions}, we present the conclusions. All our results are scaled to a flat, $\Lambda$CDM cosmology with $\Omega_{\rm{M}} = 0.3, \Omega_\Lambda = 0.7$ and a Hubble constant \textsc{H}$_0 = 70$
km\,s$^{-1}$ Mpc$^{-1}$. All images are aligned with WCS coordinates, i.e., north is up, east is left. Magnitudes are given in the AB system.

 \section{Data}\label{Data}

The objects presented in this work have been imaged by ground-based telescopes and in some cases by the Hubble Space Telescope (HST). From the ground, the groups were observed in five filters ($u^*,g',r',i',z'$) as part of the CFHTLS \citep[see][]{Gwyn2011} using the wide-field imager \textsc{MegaPrime}, which covers $\sim$\,1 square degree on the sky, with a pixel size of 0.186$\arcsec$.  The galaxy group SL2S J08591--0345 (SA72) was observed by WIRCam (near infrared mosaic imager at CFHT) as part of the proposal 08BF08 (P.I. G. Soucail). From space, the lens was followedup with the  HST in snapshot mode (C\,15, P.I. Kneib) in three Advanced Camera For Surveys (ACS) filters (F814, F606, and F475). In addition, spectroscopic follow-up of the arcs in SL2S\,J08591--0345 (SA72) and SL2S\,J08520-0343 (SA63) were carried out with IMACS at Las Campanas Observatory.  Throughout the present paper we will keep both names for some of the discussed lensing groups, the long name, e.g. SL2S  \citep[see][]{Cabanac2007} because it gives us the object's coordinates, and the compact SARCS name, e.g. SA, to be consistent with \citet[][]{More2012}.

\begin{figure*}\begin{center}
  \centering 
  \subfloat{\includegraphics[width=0.45\textwidth]{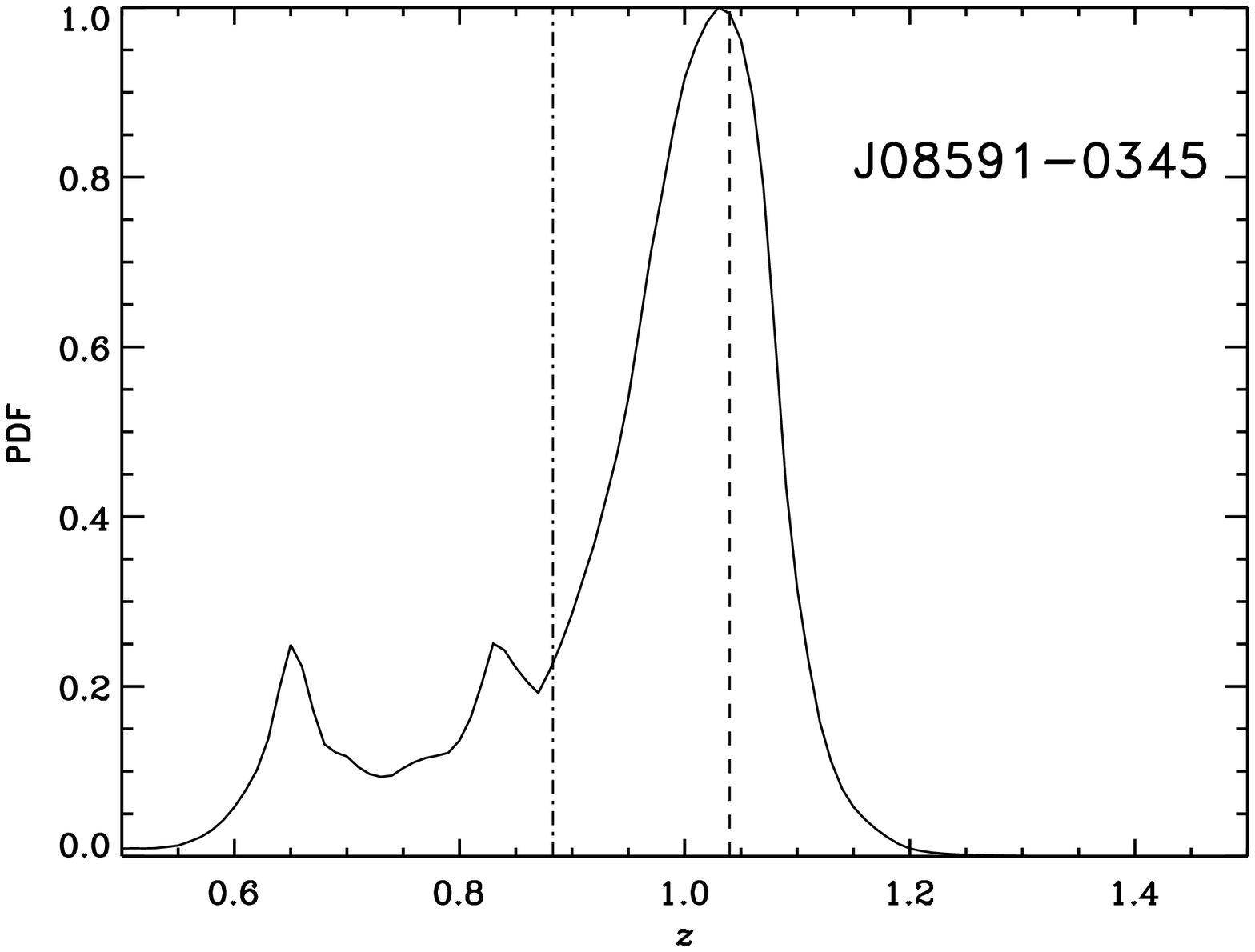}}%
  \subfloat{\includegraphics[width=0.45\textwidth]{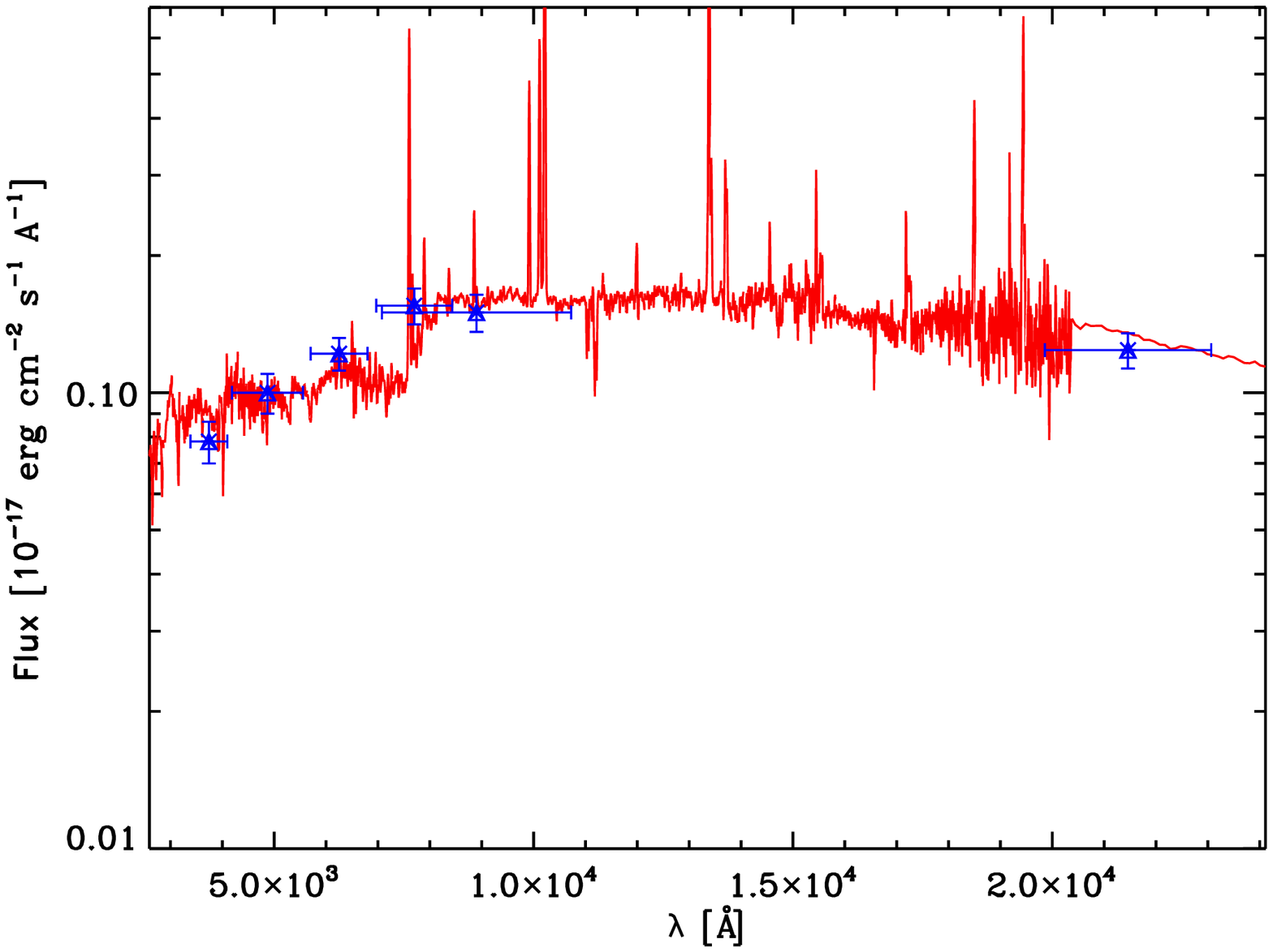}}\\ 
  \subfloat{\includegraphics[width=0.45\textwidth]{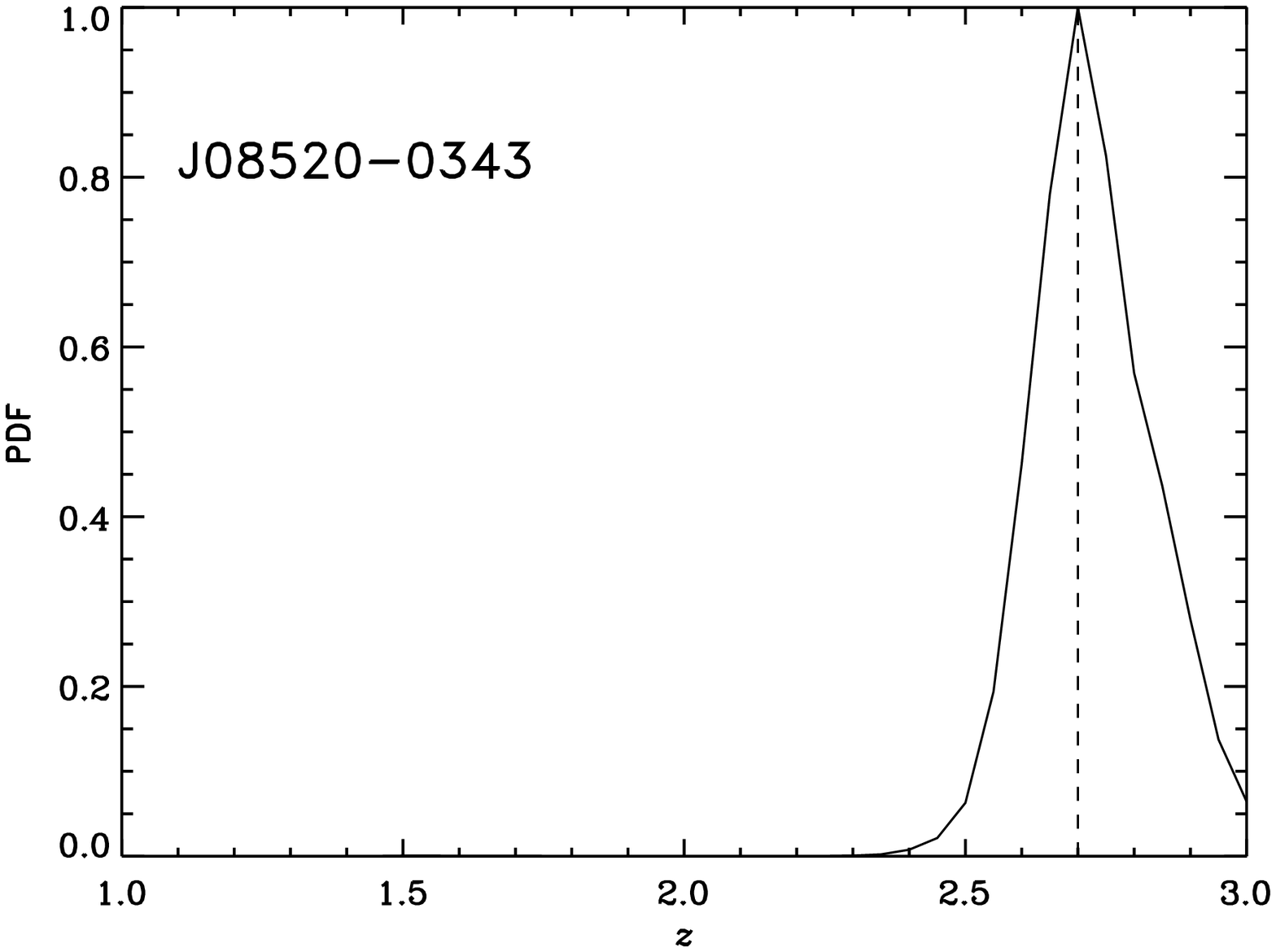}}%
  \subfloat{\includegraphics[width=0.45\textwidth]{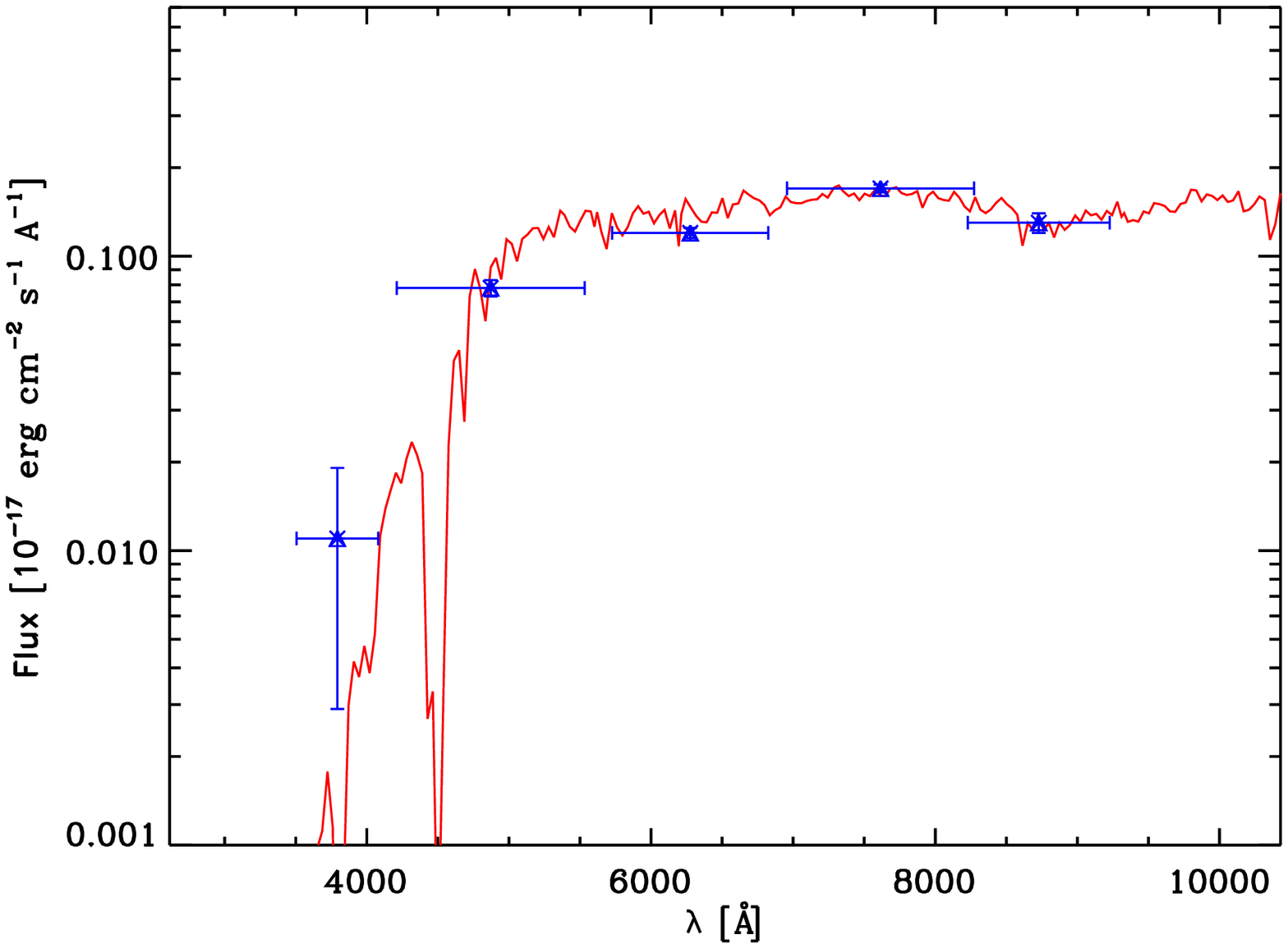}}\\  
  \qquad 
  \caption{\textit{Left column.} Output photometric redshift PDF for the selected arcs (see text). The dashed vertical lines corresponds to the best redshift estimation. In SL2S J08591$-$0345 the dash-dotted line corresponds to the spectroscopic value. \textit{Right column.} Best fit spectral energy distribution. Points with error bars are the observed CFHTLS broadband magnitudes and $k_s$ from  WIRCam  (vertical error bars correspond to photometric errors, horizontal error bars represent the range covered by the filter).}
  \label{SED_PDF}
\end{center}\end{figure*}

\begin{figure*}\begin{center}
  \ContinuedFloat 
  \centering 
  \subfloat{\includegraphics[width=0.45\textwidth]{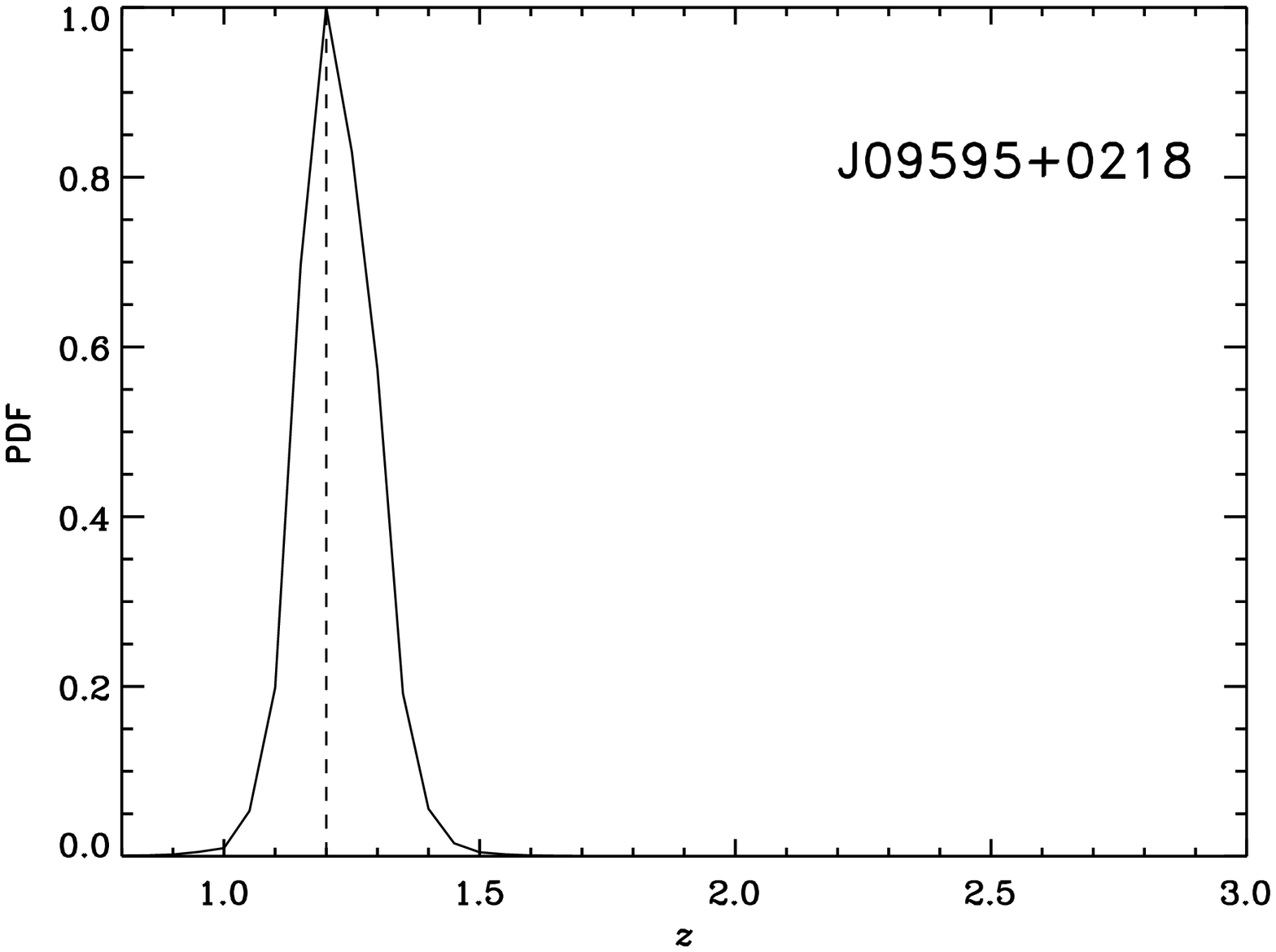}}%
  \subfloat{\includegraphics[width=0.45\textwidth]{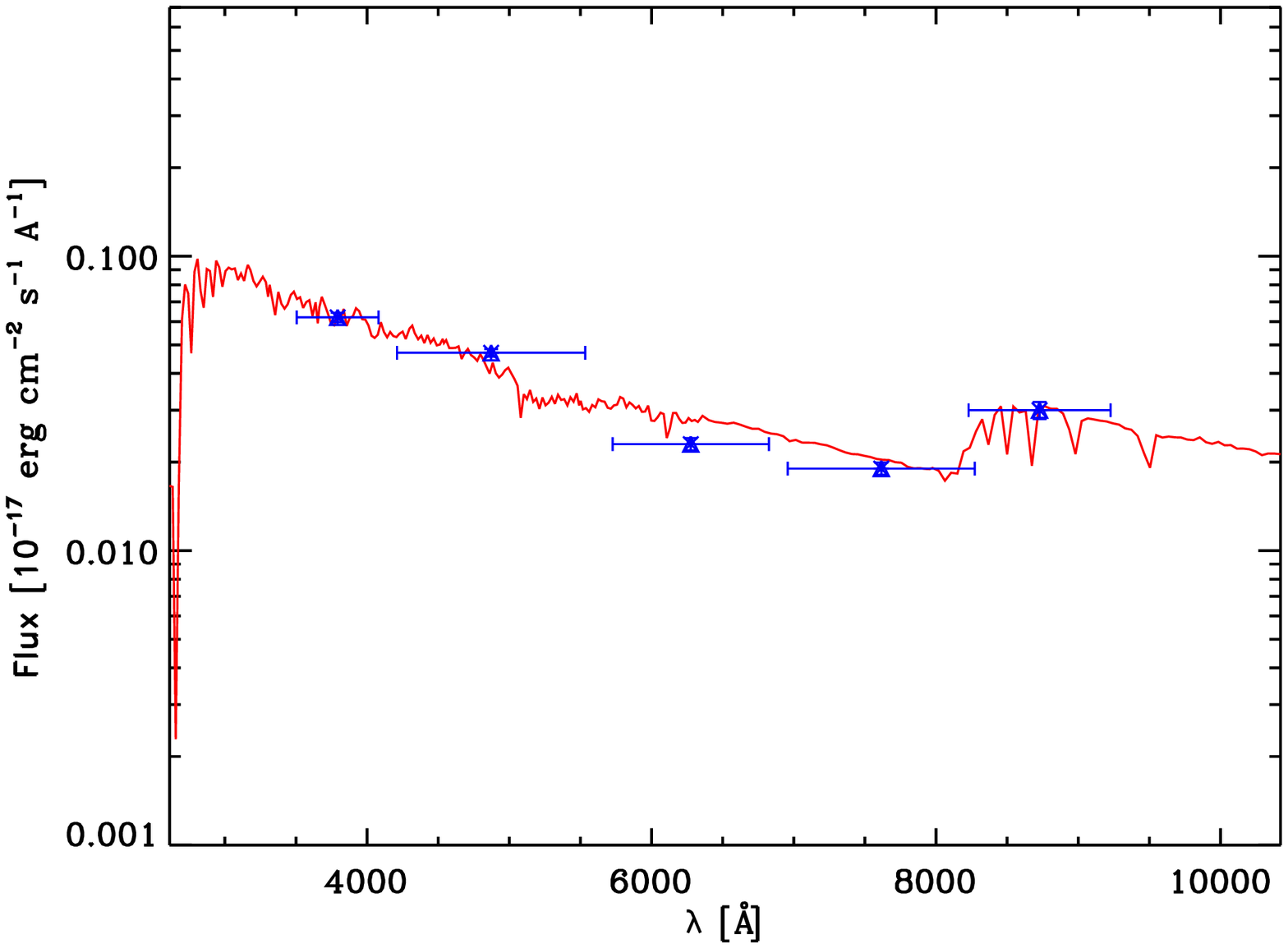}}\\ 
  \subfloat{\includegraphics[width=0.45\textwidth]{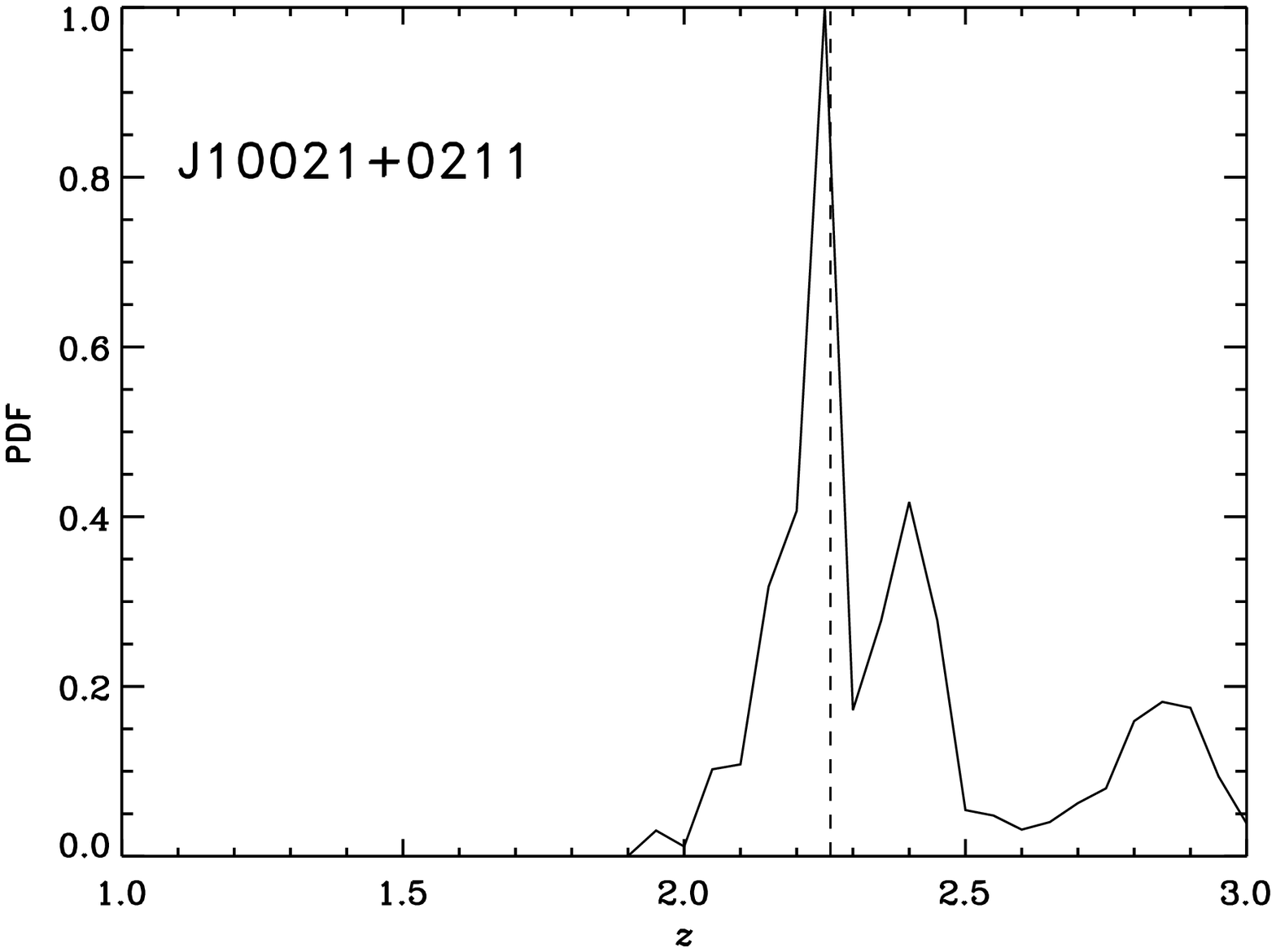}}%
  \subfloat{\includegraphics[width=0.45\textwidth]{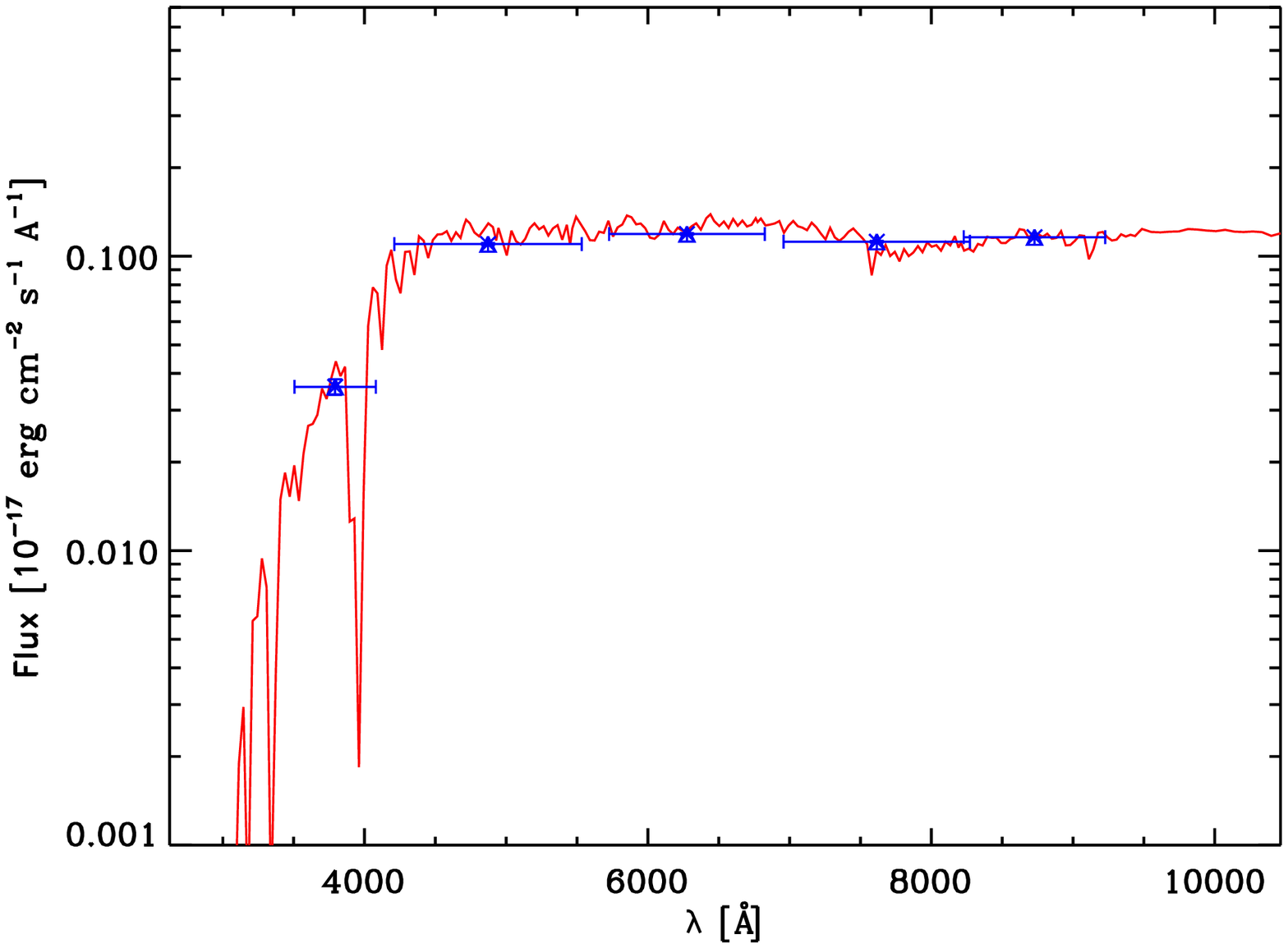}}\\   
  \qquad 
  \caption[]{Continuation. \textit{Left column.} Output photometric redshift PDF for the selected arcs (see text). The dashed vertical lines correspond to the best redshift estimation. \textit{Right column.} Best fit spectral energy distribution. Points with error bars are the observed CFHTLS broadband magnitudes (vertical error bars correspond to photometric errors, horizontal error bars represent the range covered by the filter).}
  \label{SED_PDF}
\end{center}
\end{figure*}

 \subsection{Imaging} \label{Imaging}
 
Photometric redshifts for the group sample were reported in \citet{More2012}. However, for the groups discussed in Sect.\,\ref{RT_modeling}  the photometric redshifts ($z_{phot}$) were estimated for both the lens and the lensed galaxy. For these four lens groups we estimated their $z_{phot}$ using the magnitudes of the brightest galaxy populating the strong lensing deflector. The photometry for these galaxies was performed in all CFHT bands with the IRAF\footnote{IRAF is distributed by the National Optical Astronomy Observatory, which is operated by the Association of Universities for Research in Astronomy (AURA) under cooperative agreement with the National Science Foundation.} package \textit{apphot}. Considering that the magnitudes of the lens galaxy could be contaminated by the close and bright arcs (biasing the photometric redshift), we carefully measured the magnitudes using different apertures (5,8,11,13,15, and 18 pixels). Then each aperture measurement was used to estimate redshifts using  the HyperZ  software \citep{hyperz}.  The best redshifts were selected, i.e. those with the highest probability; using selected apertures with no contamination due to arcs or other galaxies.  The  photometric data and redshifts estimations  are presented in Table~\ref{tbl-1} and Table~\ref{tbl-4}. 

The method is tested estimating photometric redshifts for some groups with reported spectroscopy  \citep[][]{paperI, Roberto2013}.  These groups are SL2S J02215-0647 (SA39), SL2S J0854-0121  (SA66), SL2S J02140-0532 (SA22), SL2S J02141-0405 (SA23), SL2S J02180-0515 (SA33), SL2S J08591-0345 (SA72), SL2S J22214-0053 (SA127), SL2S J14081+5429 (SA97), SL2S J14300+5546 (SA112), and SL2S J02254-0737 (SA50). In Figure~\ref{ZphotZspc} we compare our $z_{phot}$ (as well as previously reported values) with the spectroscopic redshift. We note that the $z_{phot}$ values from \citet{More2012} and \citet{Coupon2009} are slightly overestimated, probably because of the automatic extraction of the magnitudes used in those works.

 Another effect to take into account  in $z_{phot}$ calculations is that the point-spread functions (PSF)  are different for each filter band, which makes it difficult to match physical apertures.  This is especially important when estimating the magnitudes of the arcs (Table~\ref{tbl-1}) because the different degree of blurring  in each filter band could produce an important bias in the redshift estimations \citep{Hildebrandt2012}. Thus, for the arcs, we proceed in another way.

\begin{figure}[h!]\begin{center}
\includegraphics[scale=0.5]{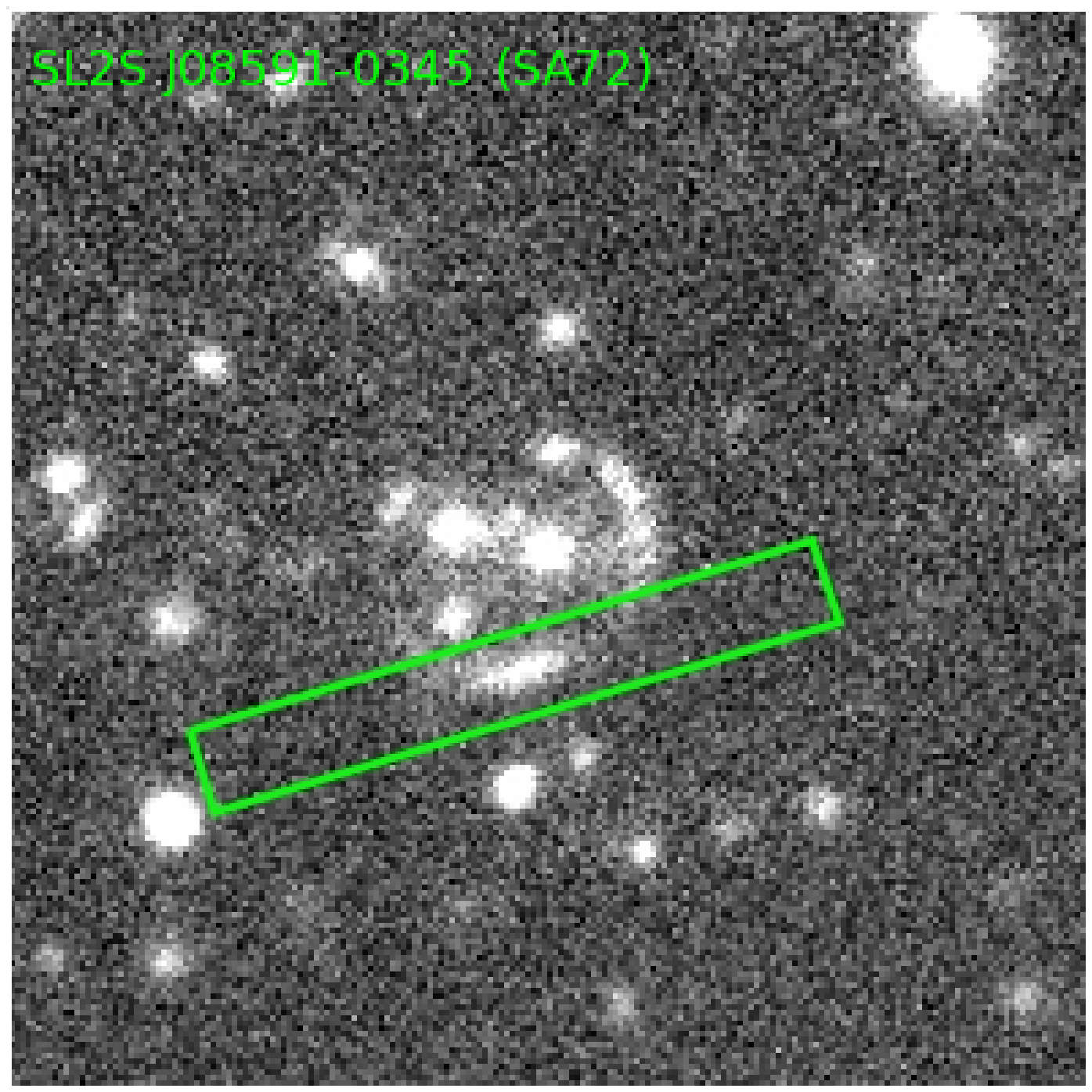}\\  
\includegraphics[scale=0.45]{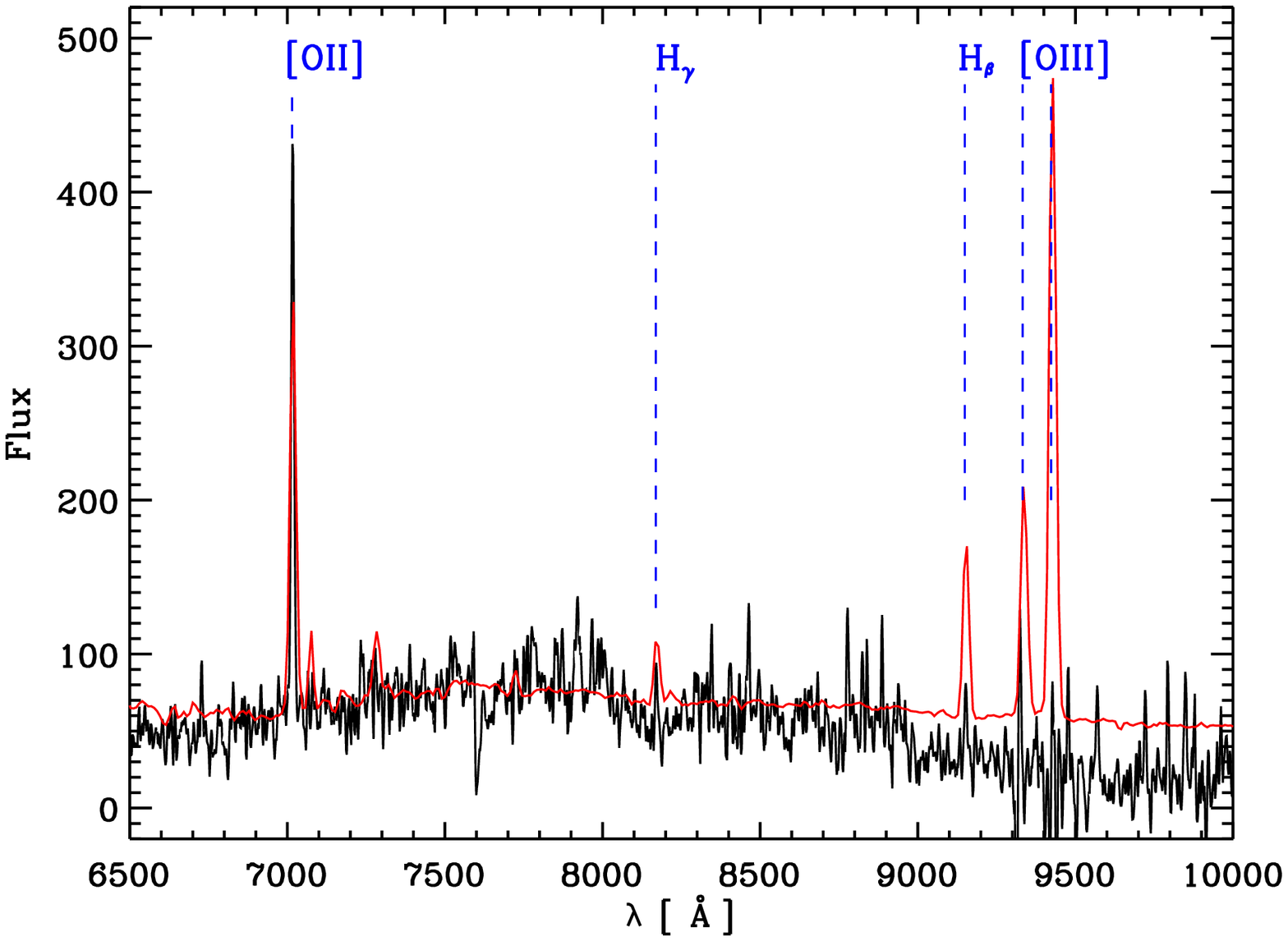}\\ 
\includegraphics[scale=0.27]{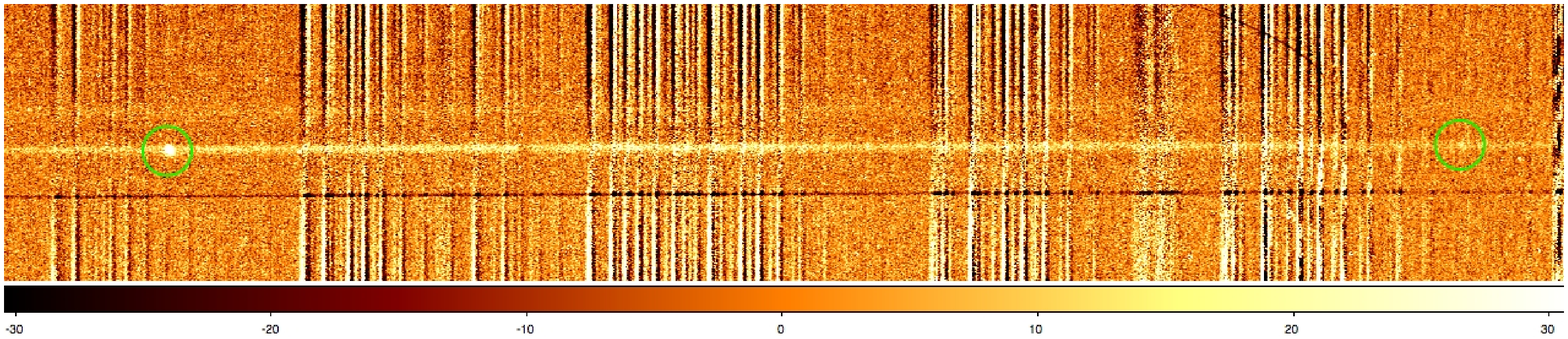}  
\caption{\textit{Top panel}. CFHTLS $g$-band image of SL2S\,J08591-0345 showing the slit position over arc A (see also second row, left panel in Figure~\ref{model}). \textit{Middle panel}. Observed spectrum of arc A (black continuous line). In red we depict a starburst template from \citet{Kinney1996} shifted at $z$ = 0.883. Some emission lines are identified, e.g., [OII]$\lambda$3727, H$_{\gamma}$, and H$_{\beta}$. \textit{Bottom panel}.  Two-dimensional spectra of the same arc. We note the [OII]$\lambda$3727 and H$_{\beta}$ emission lines.}
\label{spectrumJ08591}
\end{center}\end{figure}

Although photometric redshifts have been used extensively in clusters of galaxies to compute arc redshifts \citep[e.g.][]{Sand05}, in galaxy groups the process can be very challenging \citep[see][]{Verdugo2011} since the arcs are usually near the central galaxy (or galaxies) and skew the results by light contamination. In order to minimize this effect, we subtract the central galaxies in each group. Following \citet{mcl98}, we analyzed the $u$, $g$, $r$, $i$, $z$, and $k_s$ images and fitted a model convolved with a PSF (de Vaucouleurs profiles were fitted to the galaxies with synthetic PSFs). After the galaxy subtraction, we employed polygonal apertures to obtain more accurate flux measurements of the selected arcs. The vertices of the polygons for each arc were determined using the IRAF task \textit{polymark}, and the magnitudes inside these apertures were calculated using the IRAF task \textit{polyphot}. 
It is important to stress that the computed redshifts (see Table ~\ref{tbl-4}) for the arcs have a substantial uncertainty;  the worst case is SL2S\,J08520$-$0343 (SA63) at 2-$\sigma$, $\delta$$z_s$ $\sim$ 0.16. In spite of this deterrent, these values are used in our strong lensing models, since these errors do not have a strong influence on the Einstein radius estimations. As we will see in  Sect.\,\ref{RT_modeling}, the  $\delta$$z_s$ has more weight when the lensing source is located  close (in redshift) to the lens. We present the output probability distribution function (PDF) for the selected arcs in the left column of  Fig.~\ref{SED_PDF}. In the right column we depict the best fit spectral energy distribution obtained from HyperZ where we superimposed the observed data  as points with error bars. For SL2S\,J08591--0345 (SA72), the PDF is rather wide, which likely reflects the complexity of  accurately removing the light contamination from the four galaxies that lie in the center of the lensing group (see top panel of Figure~\ref{spectrumJ08591}). Nonetheless, the spectroscopic value lies within 2-$\sigma$ of the photometric redshift value.

 \subsection{Spectroscopy}
 
Low resolution spectra for SL2S J08520-0343 (SA63) and SL2S J08591--0345 (SA72) were obtained with the Inamori-Magellan Areal Camera and Spectrograph (IMACS Short-Camera) at the Magellan telescope. Observations were carried out on February 19, 2012, and consisted of long-slit spectroscopy of the two systems (two exposures of 23 minutes for each target). The \textit{grism} 200GR was used (2.037 \AA /pix, 5000--9000\AA \, wavelength range) since we were interested in the lensed source redshifts. Spectra were reduced using standard IRAF procedures consisting of dark correction, flat-fielding, and wavelength calibration ($RMS=0.23$). Advanced steps in the data reduction consisted of a two-exposure combination to remove cosmic ray, two-dimensional sky-subtraction, and one-dimensional spectra extraction also using IRAF tasks.

The spectrum obtained for SL2S J08520-0343 (SA63) did not show any clear features (emission or absorption lines) making it impossible to obtain a redshift estimation. On the other hand, the analysis of the spectrum in one of the arcs on SL2S J08591--0345 (SA72) shows some features. 
We show the long slit configuration for this object, as well as the obtained spectrum in Figure~\ref{spectrumJ08591}. Some emission lines are clearly visible in the spectrum, like OII$\lambda$3727, OIII, and H$_{\beta}$. Considering a starburst galaxy \citep{Kinney1996} as comparison, 
we found that the spectrum is consistent with an object at $z=0.883\pm0.001$. Two-dimensional spectra clearly shows OII$\lambda$3727 and H$_{\beta}$ (see bottom panel of Figure~\ref{spectrumJ08591}). After applying a Gaussian fitting  to those emission lines we obtain a redshift estimation of $z=0.883\pm0.001$.

\begin{table}
\caption{Results from simulations}
\label{tbl-2} 
\centering 
\begin{tabular}{ccc}
\hline\hline 
\\

\multicolumn{1}{c}{Center}  & \multicolumn{1}{c}{Simulation}  & \multicolumn{1}{c}{Total Number}    
\\

\multicolumn{1}{c}{redshift bin} & \multicolumn{1}{c}{Snapshot} & \multicolumn{1}{c}{of Halos} 
       
\\
\hline 
\\
0.05 & 82 & 781764\\
0.16 & 76 & 816422\\
0.28 & 70 & 854495\\
0.40 & 66 & 869768\\
0.52 & 62 & 901588\\
0.64 & 60 & 909205\\
0.75 & 56 & 920882\\
0.87 & 54 & 923512\\
0.99 & 52 & 904938\\
1.10 & 50 & 898774\\
\\
\hline 
\end{tabular}
\tablefoot{Column (1) lists the center redshift bins used to query the MultiDark data
  base, Col. (2) the corresponding snapshot number in the
  simulation, Col. (3) the number of halos in the whole volume box
  with rms velocities in the range $300$\kms-$1000$\kms.}
\end{table}

\begin{figure}[h!]\begin{center}
\includegraphics[scale=0.44]{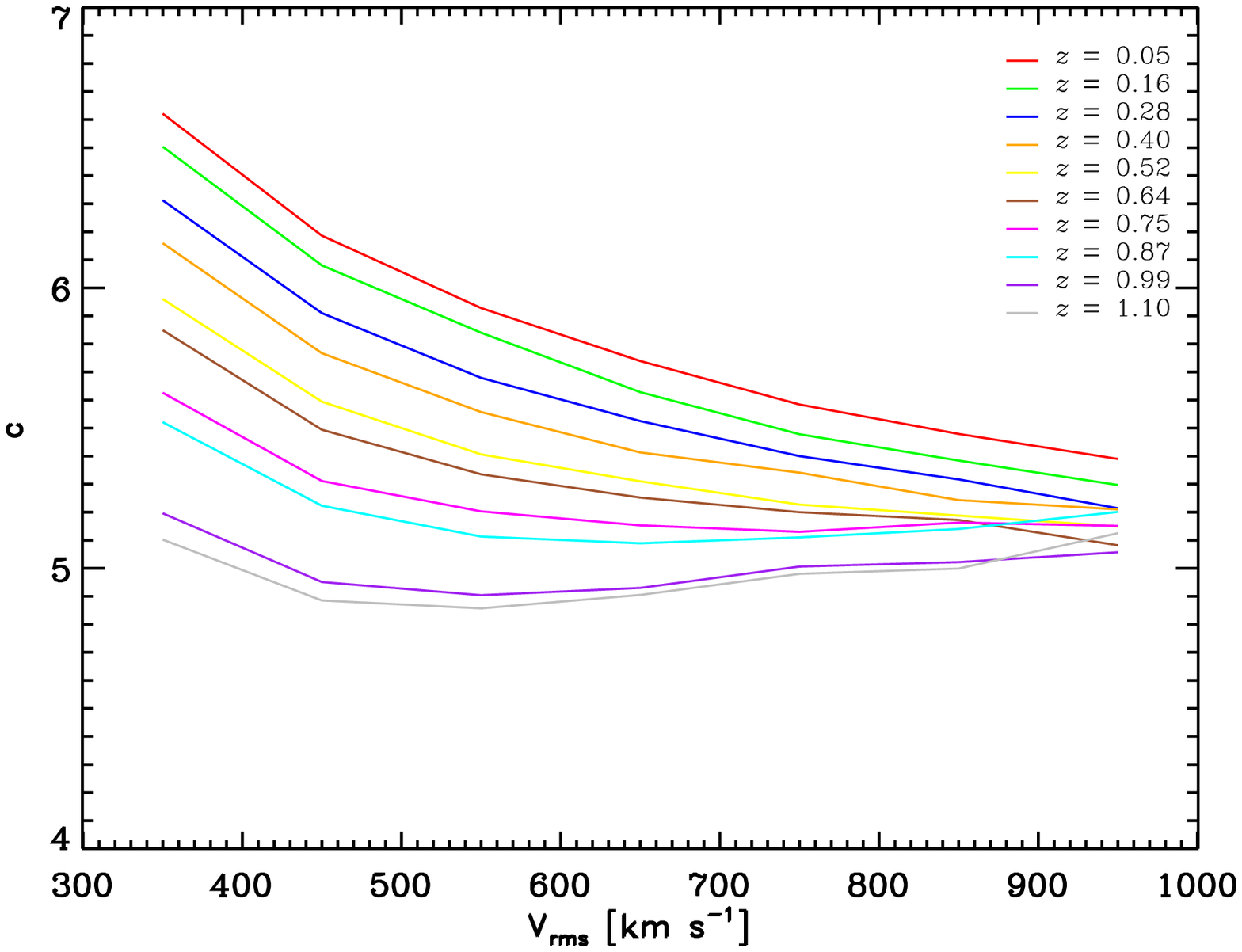}\\ 
\includegraphics[scale=0.44]{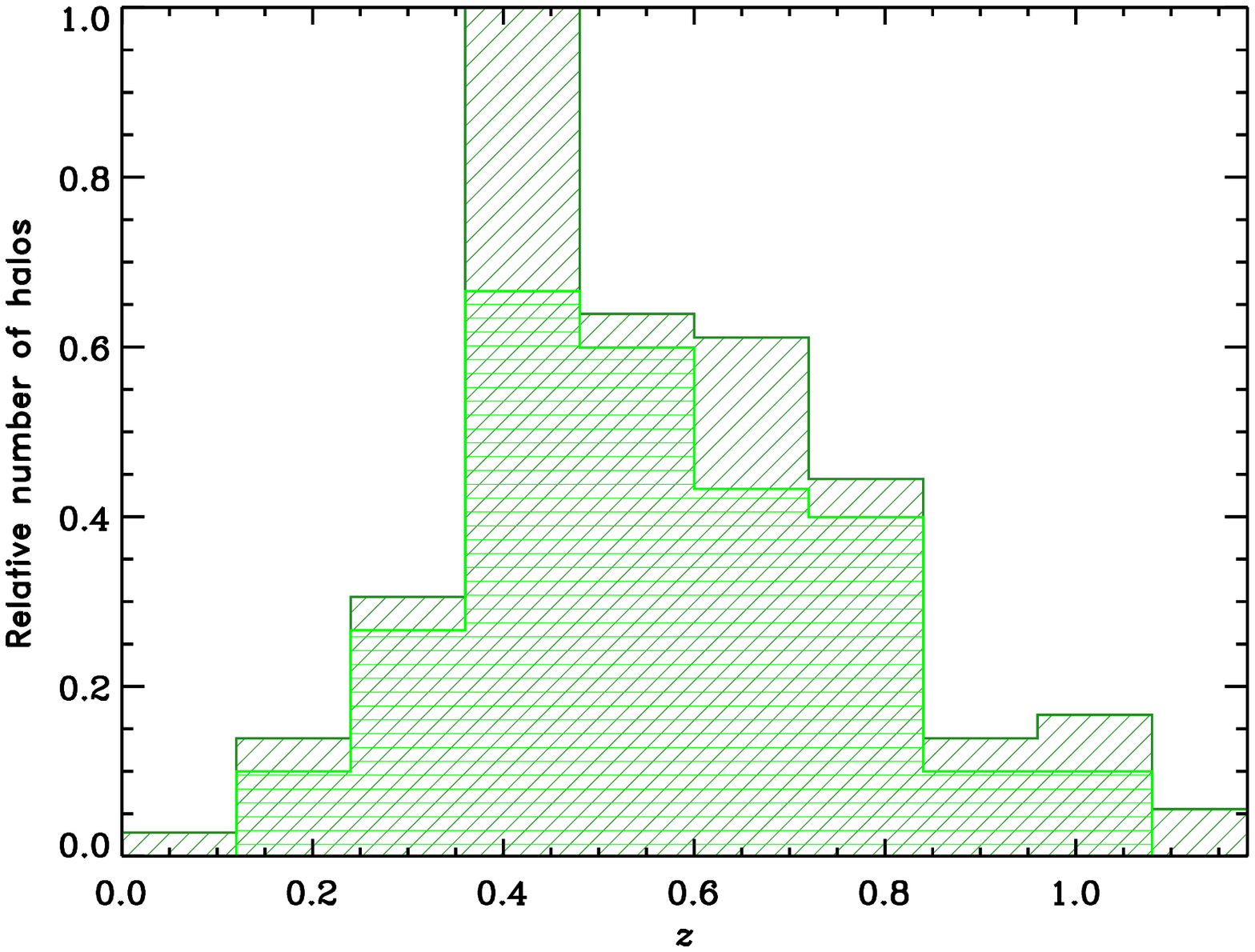}\\ 
\includegraphics[scale=0.42]{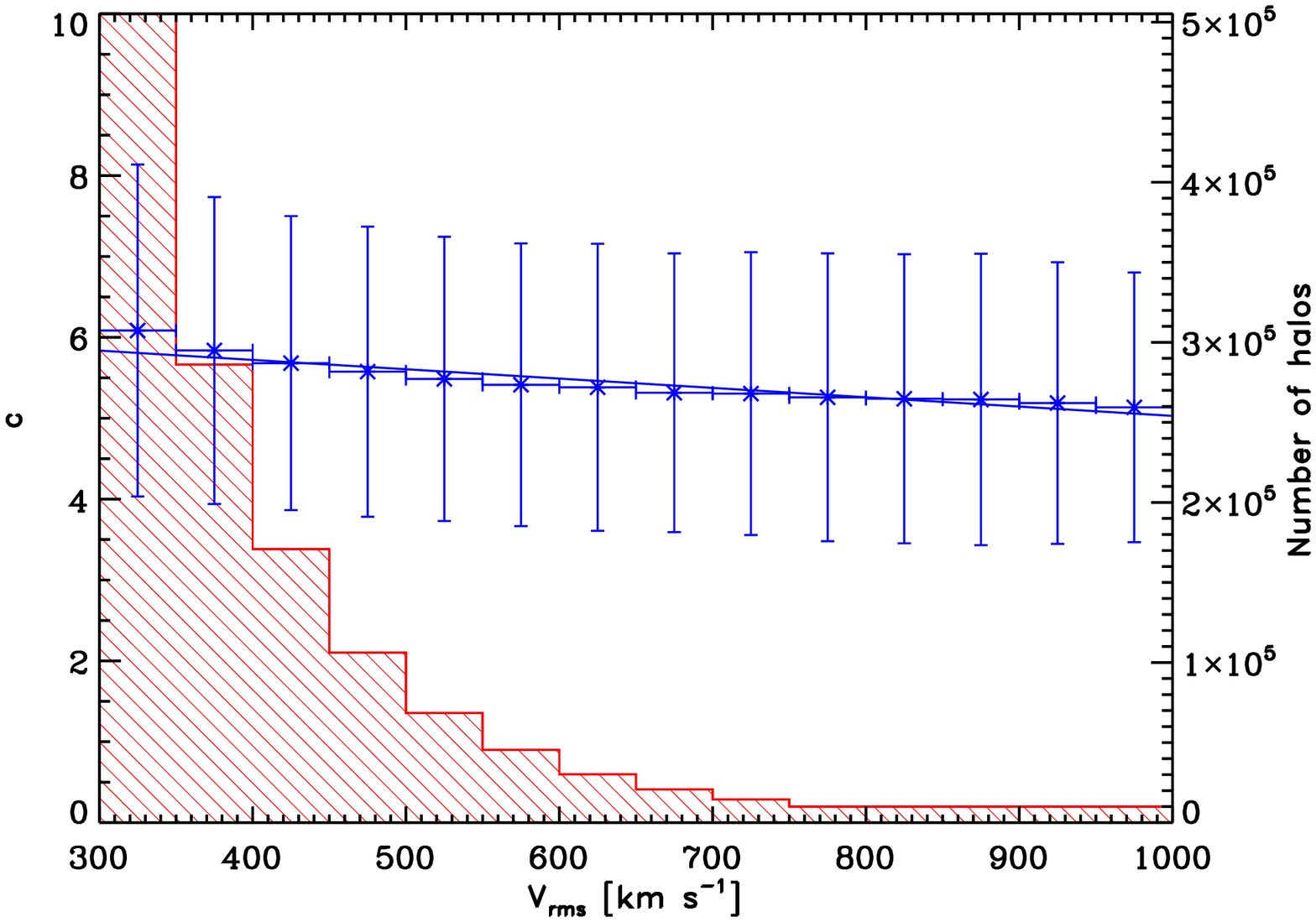}
\caption{ \textit{Top panel}.- $c-V_{\rm rms}$ relationship for different redshifts calculated from our simulations (see discussion in the text). \textit{Middle panel} .- Relative number of halos calculated from the number of strong lensing groups observed in each redshift bin \citep[according to][]{Gael2013}. Dark green histogram with tilted lines shows the results from the whole sample, light green histogram with horizontal lines those for the secure candidates. \textit{Bottom panel} .-  $c-V_{\rm rms}$ obtained mixing, in each velocity bin, halos at different redshifts, where the relative number of halos at each redshift are equal to the observed ones, i.e., given by the fraction depicted in the middle panel. The histogram in red show the total number of halos in each velocity bin.}
\label{CvsVel}
\end{center}\end{figure}

\section{Simulations}\label{Simulations}

In this section we present the characteristics of the simulation used in this work.  We explain how the observational properties of our groups \citep[][]{More2012,Gael2013} are used to select the dark matter halo sample that will be used in the next sections to infer the expected Einstein radii in our galaxy groups.

\subsection{Simulation characteristics}

We used a large N-body simulation called \verb"Multidark"
to extract statistics of halo parameters on cosmological scales. The
data we use for this paper are publicly available through a database
interface first presented by \cite{Riebe11}. Here we summarize the
main characteristics of the \verb"Multidark" volume. More details can
be found in \cite{Prada12}.  

The simulation was run using  an adaptive-mesh refinement code
called ART. Details about the technical aspects and comparisons with
other N-body codes are given in \cite{Klypin09}.   The simulations
follow the non-linear evolution of a dark matter density field
sampled with $2048^3$ particles in a volume of $1000$\hMpc. The
physical resolution of the simulation is almost constant in time $\sim
7$\hkpc between redshifts $z=0-8$. The cosmological parameters in the
simulation are $\Omega_m=0.27$, $\Omega_{\Lambda}=0.73$, $n_{s}=0.95$,
$h=0.70$, and $\sigma_8=0.82$ for the matter density, dark energy
density, slope of the matter fluctuations, the Hubble constant at
$z=0$ in units of 100km s$^{-1}$ Mpc$^{-1}$, and the normalization of
the power spectrum, respectively.  These cosmological parameters are
consistent with the results from WMAP5 and WMAP7
\citep{Komatsu2009,Jarosik2011}. With these characteristics the mass
per simulation particle is $m_p=8.63\times  10^{9}$\hMsun, which means
that group-like halos of masses $\sim 10^{13}$\hMsun~  are sampled with
at least 1100 particles.   

Dark matter halos are identified using a bound-density-maxima
algorithm (BDM). The code starts by finding the density maxima at the
particles' positions in the simulation volume. For each maxima
it finds the radius $R_{200}$ of a sphere containing a mass
over-density given by 

\begin{equation}\label{eq:M200}
M_{200} = \frac{4\pi}{3}\Delta \rho_{\rm cr}(z)R_{200}^{3},
\end{equation}

\noindent where $\rho_{\rm cr}$ is the critical density of the Universe and
$\Delta=200$ is the desired over-density. This procedure allows for the
detection of both halos and subhalos. In our analysis we kept only
the halos.

\subsection{Concentration estimates}

The estimation for the concentration values is done using an
analytical property of the NFW profile (see Sect.\,\ref{SL_Ana}) that relates the circular
velocity at the virial radius,
\begin{equation}
V_{200} = \left(\frac{GM_{200}}{R_{200}}\right)^{1/2},
\end{equation}
with the maximum circular velocity, 
\begin{equation}
V_{\rm max}^{2} = {\rm max}\left[\frac{GM(<r)}{r}\right].
\end{equation}

The $V_{\rm max}/V_{200}$ velocity ratio is used to determine the halo
concentration $c$ (the ratio between $R_{200}$ and the scale radii of
the NFW profile), using the following relation \citep{Klypin2001,Bolshoi},

\begin{equation}
\frac{V_{\rm max}}{V_{200}} = \left(\frac{0.216 c}{f(c)}\right)^{1/2},
\end{equation}

\noindent where $f(c)$ is
\begin{equation}
f(c) = \ln(1+c) - \frac{c}{(1+c)}.
\end{equation}

For each BDM overdensity the $V_{\rm max}/V_{200}$ ratio in calculated  in order to
find the concentration $c$ by solving numerically the previous two
equations.  This method provides a robust estimate of the concentration
compared to a radial fitting to the NFW profile, which is strongly
dependent on the radial range used for the fit
\citep{Bolshoi,Meneghetti13}. In particular, the NFW functional fit yields a small systematic offset of $(5-15)\%$, and a lower concentration value when compared with the velocity ratio method \citep{Prada12}.

\subsection{Halo sample selection}

We used the observational data to define ten redshift bins
as given in Table ~\ref{tbl-2} in order to construct a first catalog. We query the MultiDark database to obtain all the information for
halos with  \textit{root mean square} velocities (V$_{\rm rms}$) in the range $300$\kms $<$ V$_{\rm rms}$ $<$ 1000\kms.
 We use the values of $V_{\rm rms}$ as a proxy for
the velocity dispersion inferred in the observed lenses \citep[velocities were reported in][]{Gael2013}. For each redshift we construct a relationship $c-V_{\rm rms}$ by binning the halos in
the catalog in bins of 50\kms width. For each velocity bin the
average and standard deviation of the concentration is calculated (see top panel of Figure~\ref{CvsVel} ). We note that the concentration falls approximately in the range of 5 $\lesssim$   $c$ $\lesssim$ 6, for such intervals of redshift and velocity.

Since this range in concentration is not considerably large, we decided to test the effect of assuming a fixed value of concentration for a given  velocity bin.  Thus, we constructed a second halo catalog using all the halos, matching the shape of the observational redshift
distribution of the lenses  \citep[][see middle panel of Figure~\ref{CvsVel}]{More2012,Gael2013}. One of the main reasons for constructing this catalog is to have robust statistics from a single $\Lambda$CDM simulation volume, covering the mass range from groups up to clusters. Another motivation is that we will stack these objects using  their velocity dispersion (Foex et al. in prep.),  thus we need to know if there is any systemic difference in the estimates of $\theta_E$ when using different catalogs. We proceed as follows: for each redshift
bin we count the number of observed lenses and multiply it by
$10^5$, then we randomly select the same number of halos from the simulation. This
allows us to have a simulated sample $10^5$ times larger than the
 observed one, with the same redshift distribution. From this new
catalog we construct a new $c-V_{\rm rms}$ relationship in the same
way as described above. In the bottom panel of Figure~\ref{CvsVel} we show the $c-V_{\rm rms}$ relation, using halos at different redshifts for each velocity bin,   and using the same relative number of halos  for each redshift  as the ones that were observed. Taking into account the errors in concentration, our mimic sample has  5 $\lesssim$   $c$ $\lesssim$ 6 in the range $300$\kms $<$ V$_{\rm rms}$ $<$ 1000\kms. This is consistent with the results depicted in the top panel of Figure~\ref{CvsVel}. Our values are in agreement with those reported by \citet{Faltenbacher2007}, who investigate the concentration-velocity dispersion relation in galaxy groups using  cold dark matter N-body simulations.

\begin{figure*}\begin{center}
\includegraphics[scale=0.49]{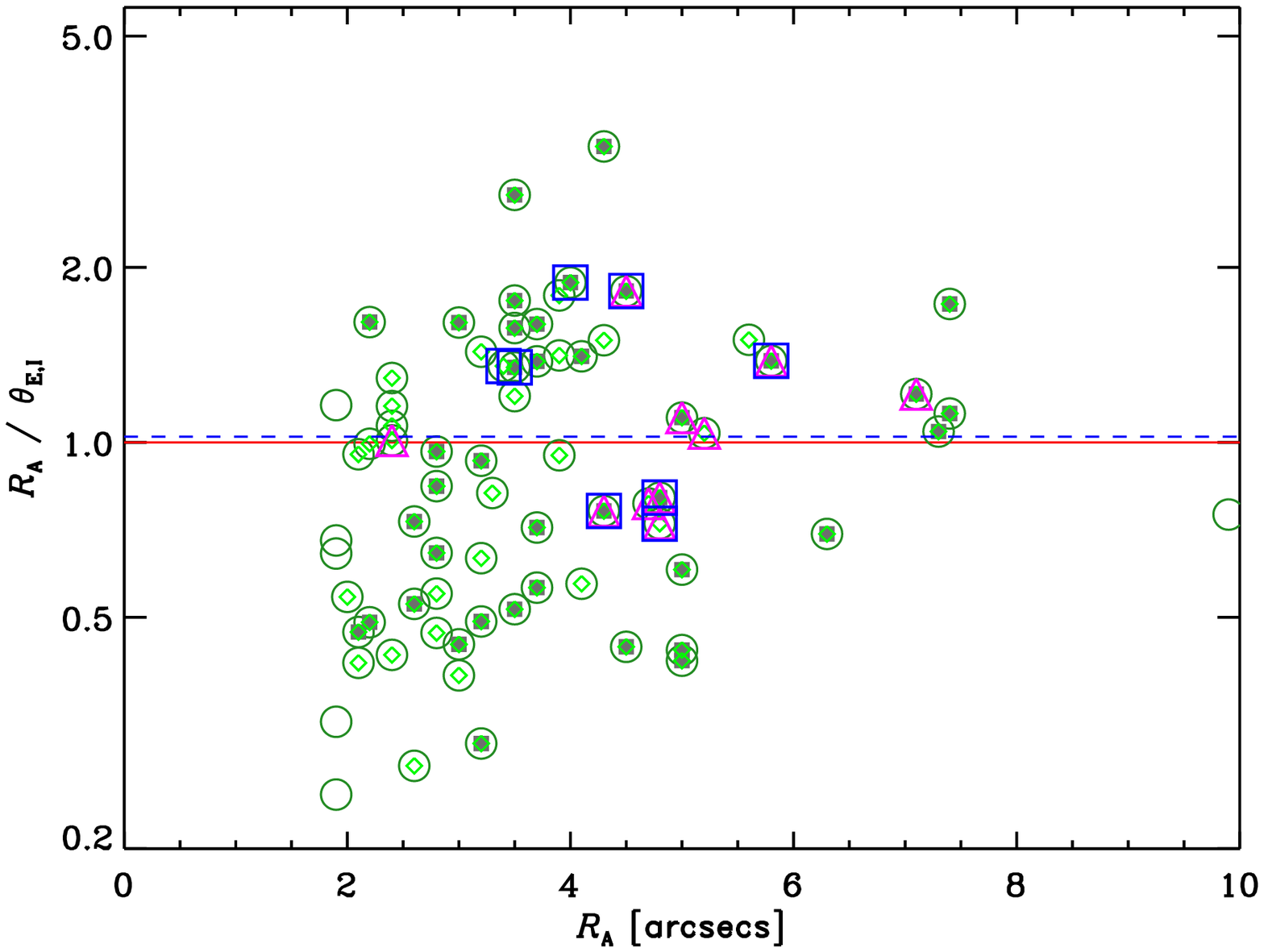}
\includegraphics[scale=0.49]{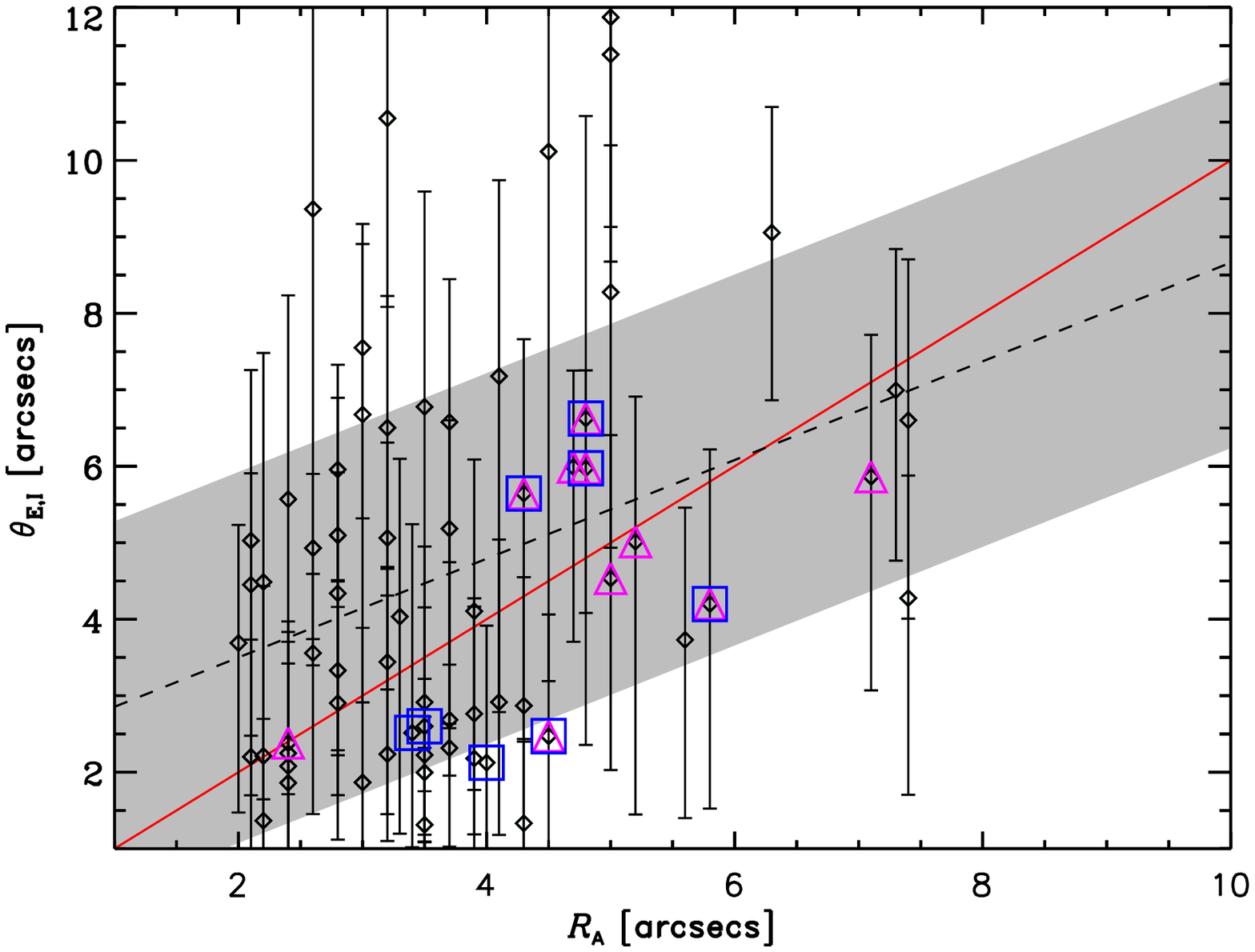}\\ 
\includegraphics[scale=0.49]{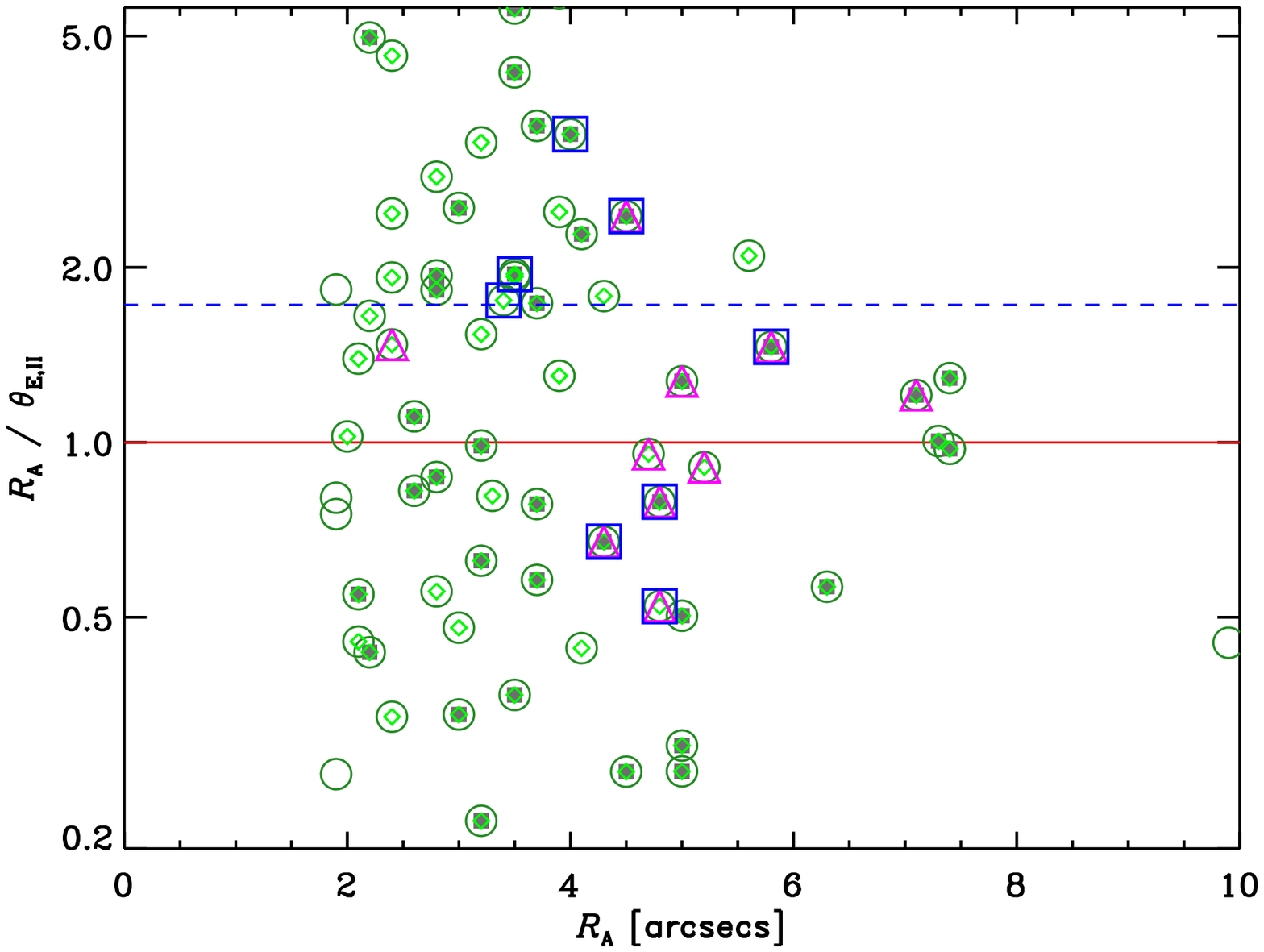}                   
\includegraphics[scale=0.49]{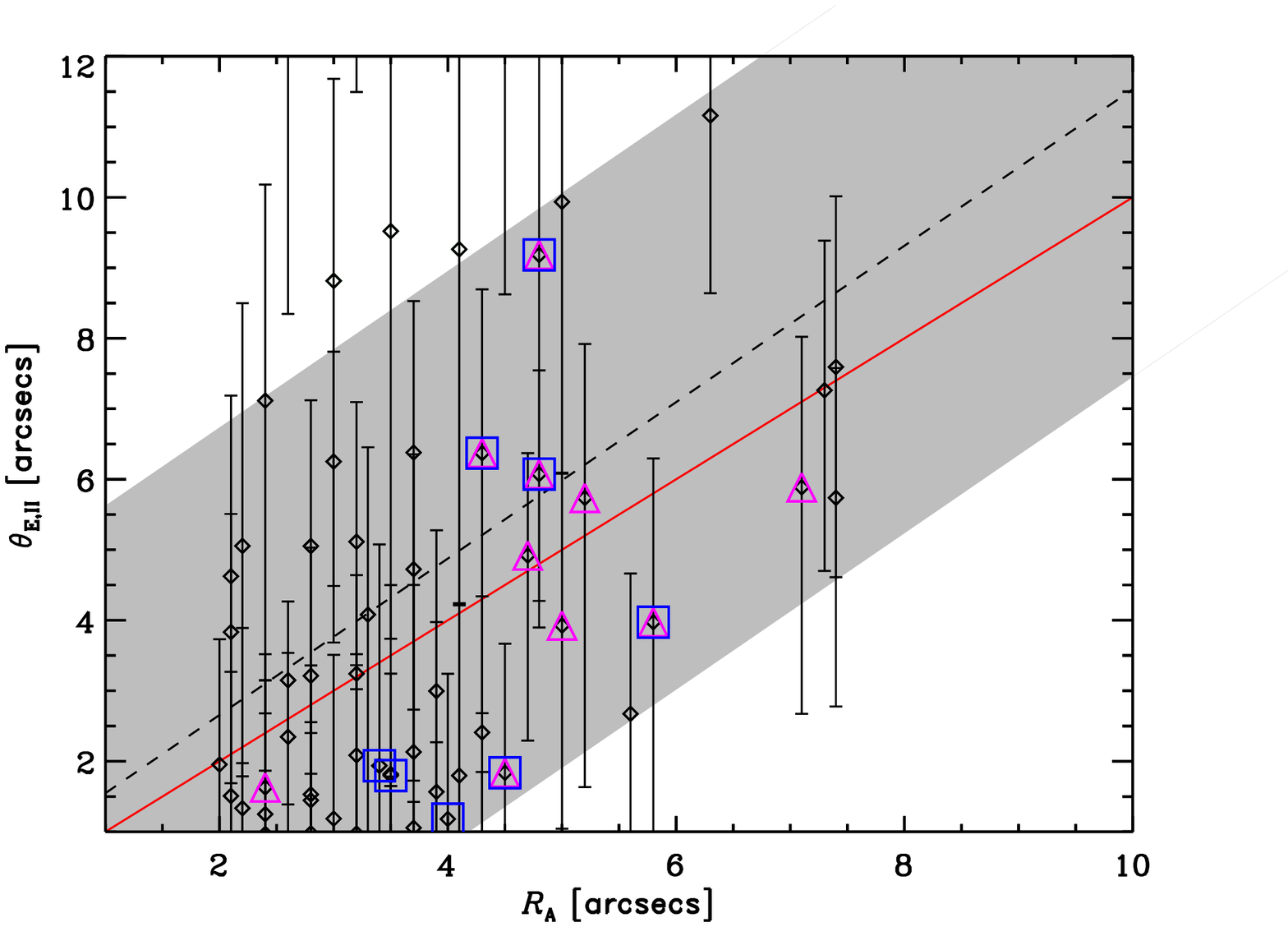}\\  
\includegraphics[scale=0.49]{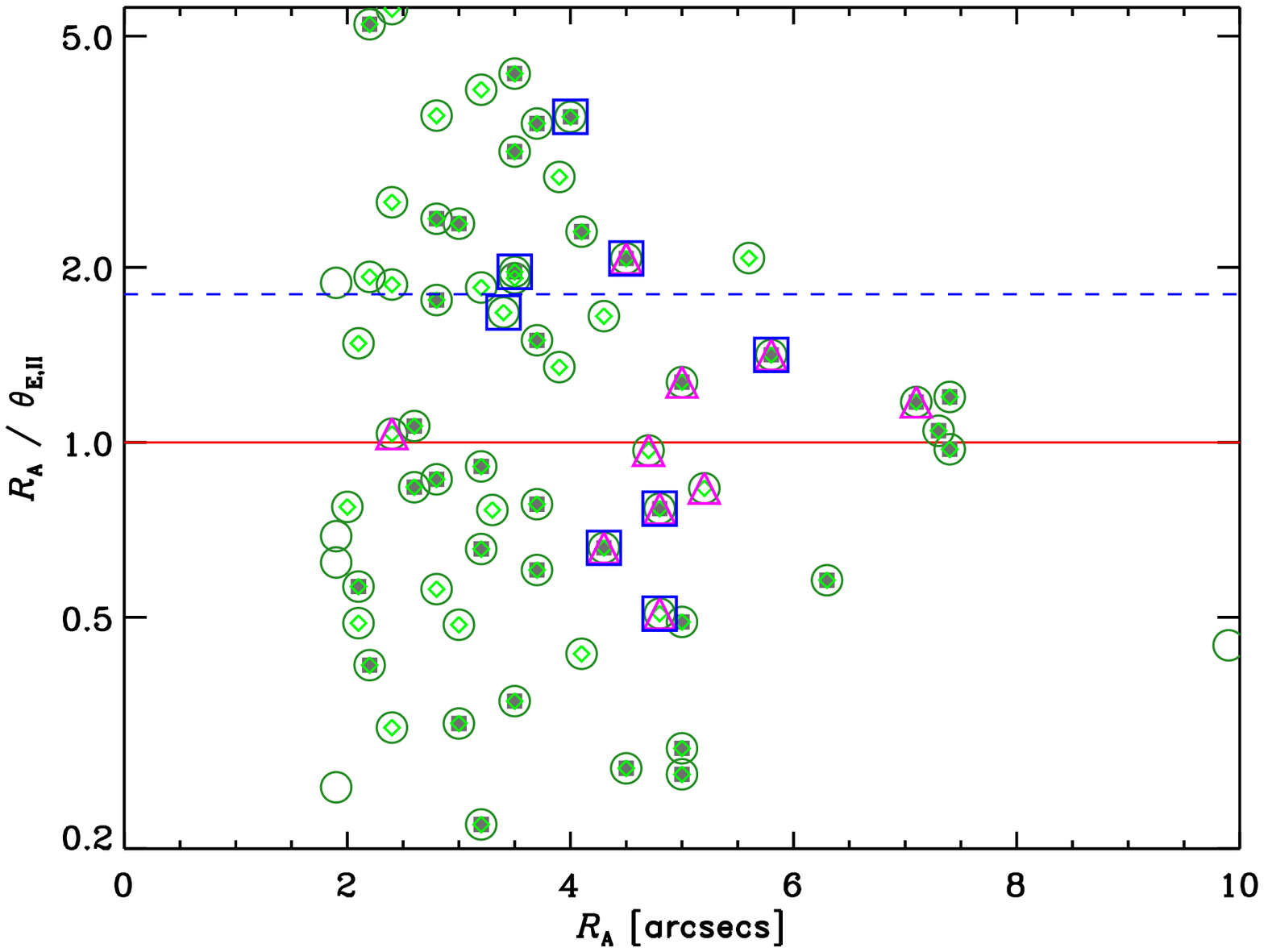}                     
\includegraphics[scale=0.49]{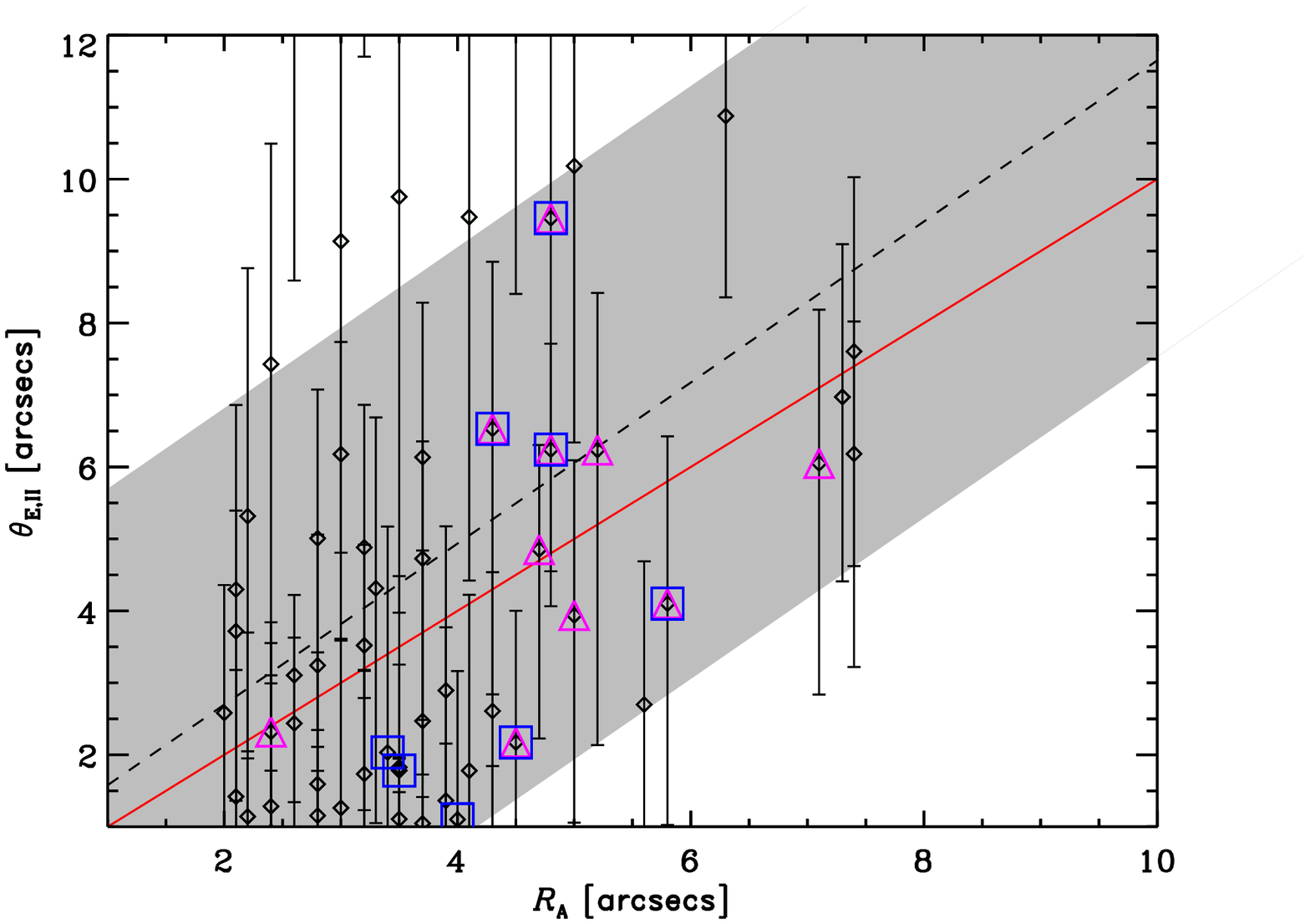}     
\caption{\textit{Left column.} $R_A$/$\theta_{E}$ $vs$ $R_A$ for the two first methods discussed in this work. Top panel:   $R_A$/$\theta_{E,I}$ $vs$ $R_A$. Green circles depict the secure group candidates \citep{Gael2013}, green diamonds those with  2.0" $\leq$ $R_A$ $\leq$ 8.0". Green filled squares are groups with multi-component features or very elongated light contours (see Section ~\ref{Discussion}). The blue dashed line depicts the mean $R_A$/$\theta_{E,I}$, the red continuous line shows the one-to-one relation, blue squares are groups with giant arcs (see Section ~\ref{WeakLensing}), and magenta triangles represent those groups with strong lensing models (see Section ~\ref{RT_modeling}). The error bars  were omitted for clarity.   Middle panel: $R_A$/$\theta_{E,II}$ $vs$ $R_A$ using the catalog constructed using the $V_{\rm rms}$ as proxy for the velocity dispersion measured in the observed lenses. 
Bottom panel: $R_A$/$\theta_{E,II}$ $vs$ $R_A$ using the catalog built to match the shape of the observational redshift distribution of the lenses.  
\textit{Right column.} $\theta_{E}$ $vs$ $R_A$. Black triangles show all the groups (regular or irregular) that satisfy 2.0" $\leq$ $R_A$ $\leq$ 8.0".  The black dashed line shows the fit to the data, with the 1$\sigma$-error depicted as a gray shaded region.  As before, the red continuous line shows the one-to-one relation.}
\label{ThetaEWL}
\end{center}\end{figure*}

\begin{figure}[h!]
\begin{center}
\includegraphics[scale=0.5]{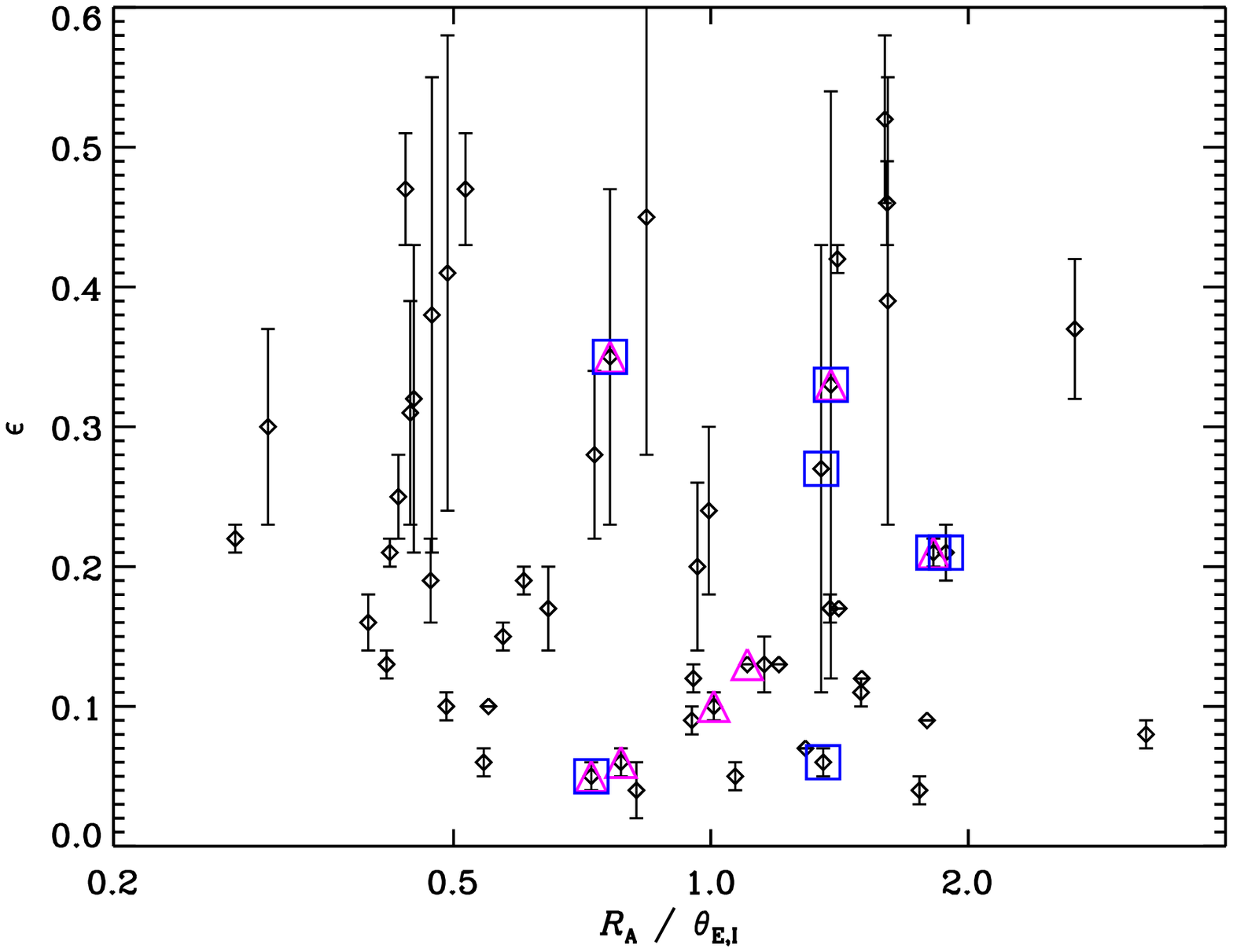}
\caption{The figure shows ellipticity as a function of the $R_A$/$\theta_{E,I}$ ratio. Blue squares are groups with giant arcs (see Section ~\ref{WeakLensing}), and magenta triangles are those groups with strong lensing models (see Section ~\ref{RT_modeling}).}
\label{Elip_Theta}
\end{center}\end{figure}

\section{Einstein radius} \label{EinsteinRadius}

We calculated the Einstein radius in three different ways using 1) weak lensing, following \citet{Gael2013} and assuming a SIS model; 2) strong lensing, considering a modified version of  the method in \citet{Broadhurst08}  together with the $c-V_{\rm rms}$ relation obtained in the previous section; and 3) strong lensing modeling of eleven SL2S groups with HST images. The fitting is done with the LENSTOOL\footnote{\tt This software is publicly available at: http://projets.lam.fr/projects/lenstool/wiki} ray-tracing code \citep{Kneib1993},  which uses a Bayesian  Markov chain Monte Carlo (MCMC) method to search for the most likely parameters in the lens modeling \citep{jullo07}.

\subsection{Weak lensing method}\label{WeakLensing}
 
The methodology used in this work follows \citet{Bardeau2007}, also applied by \citet{paperI} on the first SL2S group sample. A full description of the procedure can be found in \citet{Gael2012}, and of its applications in the last SL2S sample in \citet{Gael2013}. Here we outline the method briefly.

To detect and select the lensed background galaxies, we used {\tt SExtractor} \citep{Bertin1996} on the i-band images, which computes the best seeing among all the available filters.  Considering object size, its  magnitude, and central flux, we perform a first selection to build separate catalogs  for stars and galaxies. The shape parameters of  each object  are estimated using the {\tt Im2shape} software \citep{Bridle2002}  in i-band images. Stars are used to derive the PSF field at each galaxy position, simply by taking the average shape of the five nearest stars (the catalog of stars being first cleaned to keep only objects with similar shapes). This PSF field is convolved by {\tt Im2shape} to a given model of the galaxy shapes, in our case a simple elliptical Gaussian \citep[see][]{Gael2012, Gael2013}. Exploring the space of free parameters with a MCMC sampler, the code finds the best model that minimizes the residuals between the PSF-distorted model of the galaxy (including noise and background level, treated as well as free parameters in the modeling) to its observed shape. For each galaxy, {\tt Im2shape} returns an estimate of its shape parameters along with robust statistical uncertainties.\\
The next step of the analysis consists in selecting the lensed galaxies to estimate the shear signal. Here we used two selections. First, we kept galaxies with $21<m_{i'}<m_{comp}+0.5$, i.e., we removed the brightest objects, which are most likely foreground field galaxies. We also removed the faintest ones as they are too faint to derive reliable shape parameters. By staying close to the completeness magnitude, $m_{comp}$, of the observed galaxy distribution, we also keep a certain control over the redshift distribution, which is required to derive the lensing strength. To remove most of the group and cluster members, we also used the classical red sequence selection: starting from the color of the central galaxy, we defined a region within the color-magnitude space where the red elliptical galaxies of the groups and clusters are located. Only galaxies outside this region were included in the catalog of the lensed galaxies.\\
To estimate masses from the weak-lensing signal, we computed shear profiles  using the shape of the lensed galaxies. They were built with logarithmic bins centered around the lens galaxy. We fitted them with the SIS mass model within a fixed physical aperture, from 100 kpc to 2 Mpc with a classical $\chi^{2}$-minimization. To propagate the uncertainties on the shear profiles $\sigma_{\gamma}(r )$ (intrinsic ellipticity of lensed galaxies and measurement errors on their shape parameters), we generated 1000 Monte Carlo profiles, drawn assuming that each point of the observed shear profile $\gamma(r)$ follows a normal distribution $\mathcal{N}(\gamma(r ),\sigma_{\gamma}(r )$. The distribution of the best-fit parameters is chosen to characterize the model that  best describes our observations and the associated errors (68$\%$ confidence interval around the mode of the distribution). The shear signal was translated into physical units through the lensing strength $D_{LS}/D_{OS}$, which was derived from the photometric redshift distributions of the CFHTLS Deep fields provided by R. Pello. The same selection criteria (magnitude limits and color-magnitude)  were applied to these catalogs in order to match the redshift distribution of our lensed galaxies. In doing so, we also accounted for the dilution of the shear signal by the residual foreground galaxies \citep[see][for more details]{Gael2013}.

\begin{table}
\caption{Fitting results for $\theta_{E}$ vs $R_A$.}
\label{tbl-3} 
\centering 
\begin{tabular}{lcccc}
\hline\hline 
\\

\multicolumn{1}{c}{Correlation}  & \multicolumn{1}{c}{a $\pm$ $\delta$ a}  & \multicolumn{1}{c}{b $\pm$ $\delta$ b}    
& \multicolumn{1}{c}{$R$}      & \multicolumn{1}{c}{$P$}    \\

\multicolumn{1}{c}{} & \multicolumn{1}{c}{ } & \multicolumn{1}{c}{ } & \multicolumn{1}{c}{ } & \multicolumn{1}{c}{ }
       
\\
\hline 
\\
$\theta_{E,I \phantom{II}}$ - $R_A$  & 2.2 $\pm$ 0.9 & 0.7 $\pm$ 0.2  &  0.33  &  6$\times$10$^{-3}$ \\
$\theta_{E,II \phantom{I}}$ - $R_A$  & 0.4 $\pm$ 1.5 & 1.1 $\pm$ 0.4  &  0.40  &  1$\times$10$^{-3}$ \\
$\theta_{E,II \phantom{I}}$ - $R_A$$^{\dagger}$  & 0.4 $\pm$ 1.5 & 1.1 $\pm$ 0.4  &  0.40  &  1$\times$10$^{-3}$ \\
$\theta_{E,III}$ - $R_A$  & 0.4 $\pm$ 1.5 & 0.9 $\pm$ 0.3  &  0.60  &  6$\times$10$^{-2}$ \\
\\
\hline 
\end{tabular}
\tablefoot{
($\dagger$): Second catalog\\
Column (1) lists the correlation. Columns (2) and (3) list the coefficient values in the relation $Y$ = $a$ + $b$$X$. Columns (4) and (5) list the Spearman's rank correlation coefficient and the statistical significance, respectively.}
\end{table}

Given the SIS velocity dispersion obtained from weak lensing, we calculate the Einstein radius through the expression

\begin{equation}\label{eq:theta_eWL}
\theta_{E,I} = 4 \pi \frac{\sigma^2_{WL}}{c^2} \frac{D_{LS}}{D_{OS}},
\end{equation}

\noindent where $\sigma_{WL}$ is the  line-of-sight velocity dispersion calculated from weak lensing data, and $D_{ LS }$ and $D_{ OS }$ are the angular diameter distances between the lens and the source  and the observer and the source, respectively. These distances are estimated using the most likely redshift of the source \citep{Turner1984}, with an upper limit given by Eq. 2 in \citet[][]{More2011}. In the top-left panel of Fig.~\ref{ThetaEWL} we show the ratio of $\theta_{E,I}$ calculated through Eq.~\ref{eq:theta_eWL} and their respective $R_A$ values. The points are uniformly distributed on both sides of $R_A$/$\theta_{E,I}$ =   1, with a mean of 1.02 and a standard deviation of 0.56 (indicating a large scatter). We also note that groups with multi-components or with a high degree of elongation (see Section ~\ref{Discussion}) are also uniformly distributed in the plot. Groups with small values of $\theta_{E,I}$ with respect to $R_A$ are always irregular. And likewise, those with extremely high $\theta_{E,I}$ values are also not regular groups. In the figure we highlight with different symbols those groups with strong lensing models (see Section ~\ref{RT_modeling}) and those with giant arcs \citep[with a length-to-width ratio larger than 10, according  to][]{More2012}. We want to point out that the errors in $R_A$ (measured directly from the images) are small compared to the errors in $\theta_{E,I}$. The former are around two pixels, which is roughly 0.4$\arcsec$, thus, unless otherwise specified, we will omit the error bars of $R_A$ in the plots.

To further investigate quantitatively this effect, we use the task \textit{ellipse} in IRAF to fit ellipses to the luminosity maps of the groups \citep[see][]{Gael2013} and to obtain the ellipticity. For 14 groups it was not possible to obtain a fit because they present either multi-components or a high degree of elongation. In  Fig.~\ref{Elip_Theta} we plot the ellipticity as a function of the ratio $R_A$/$\theta_{E,I}$. This result is consistent with  Fig.~\ref{ThetaEWL} (top-left panel),  i.e. elongated groups  ($\epsilon$ $>$ 0.3) are evenly distributed in the $R_A$/$\theta_{E,I}$ axis. We note that two groups with giants arcs have $\epsilon$ $>$ 0.3, and three more do not appear in the plot (because they belong to the groups for which it was not possible to obtain a fit).

In the top-right panel of Fig.~\ref{ThetaEWL} we depicted $\theta_{E,I}$ vs $R_A$ for those groups with 2.0" $\leq$ $R_A$ $\leq$ 8.0". In Table ~\ref{tbl-3} we show the results  for our fit. We found  a low correlation between both variables. It is possible to show that, if we eliminate extreme values (probably outliers) in the ratio $R_A$/$\theta_{E,I}$, the correlation coefficient and the significance improve. 
For example, with a 0.5 $\leq$  $R_A$/$\theta_{E,I}$  $\leq$  2.0 cutoff \citep[ following][ who showed that around these values in the radial distribution of tangential arcs is where, approximately, the minimum and maximum in the cross-sections occurs]{Puchwein2009}, we found   $R$ = 0.6, and $P$ = 1$\times$10$^{-5}$. However the outliers were not eliminated,  keeping the sample as it is. In Section ~\ref{Discussion} we will explain the reason for such low correlation. We note that some groups with giant arcs and strong lensing models are far from the one-to-one correlation.

\begin{figure}[h!]
\begin{center}
\includegraphics[scale=0.5]{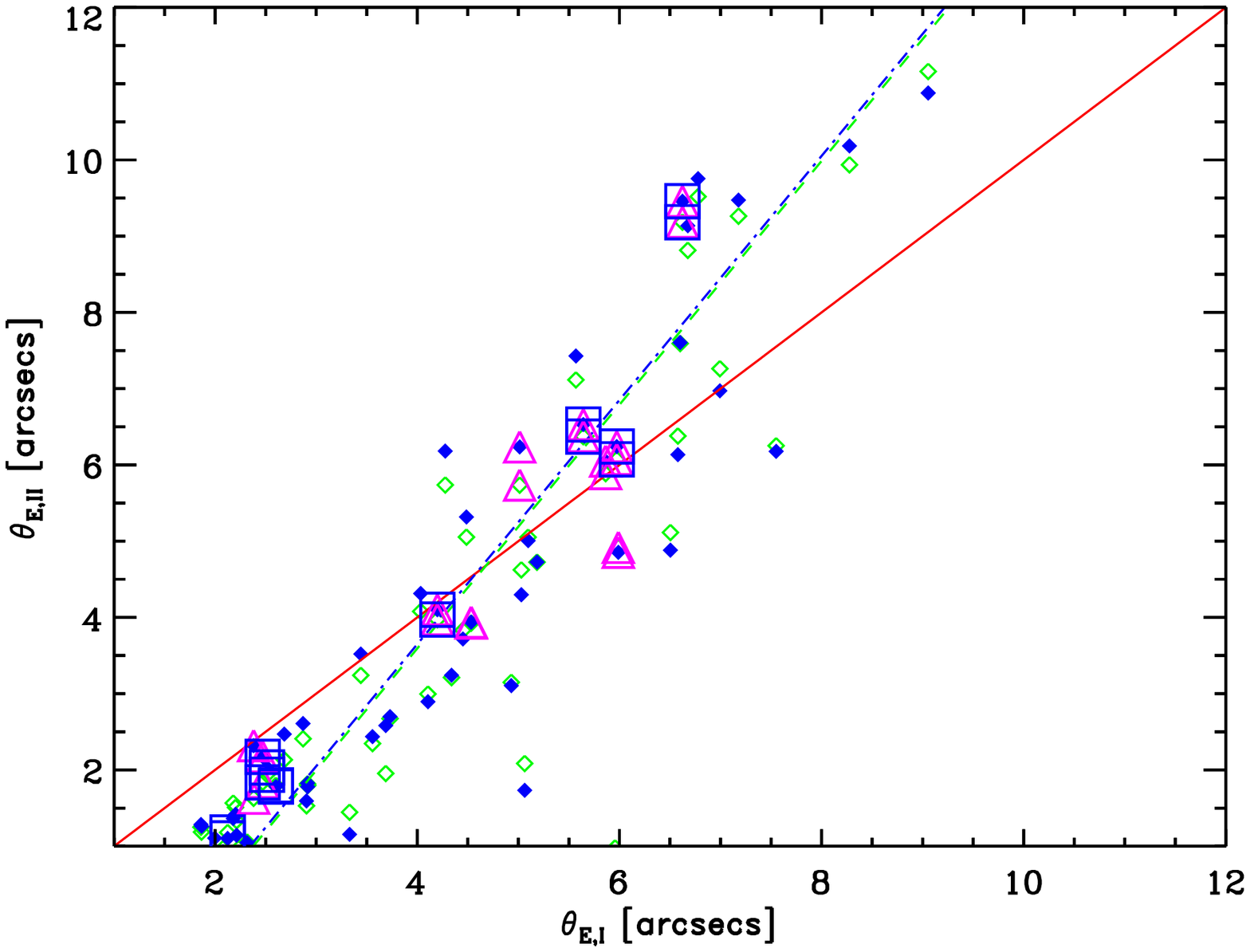}
\caption{$\theta_{E,II}$ $vs$ $\theta_{E,I}$. Green diamonds depict the objects used to construct the plots shown in the right column (top and middle panels) of Figure~\ref{ThetaEWL}. Green dashed line shows the fit to the data. Similarly, the blue filled diamonds are the objects associated with the correlations shown in the right column (top and bottom panels) of the same figure. Blue dash-dotted line shows the fit to the data. Blue squares are groups with giant arcs (see Section ~\ref{WeakLensing}), magenta triangles are those groups with strong lensing model (see Section ~\ref{RT_modeling}), and  the red continuous line shows the one-to-one relation. The error bars are omitted for clarity.}
\label{ThetaIIvsThetaI}
\end{center}\end{figure}

 \subsection{Analytically from a NFW profile}\label{SL_Ana}

The NFW universal density profile, predicted in
cosmological N-body simulations, has a two-parameter functional form \citep{nav97},

\begin{equation}\label{eq:rho}
\rho(r) = \frac{\rho_{\rm cr}(z)\delta_{c}(z)}{(r/r_s)(1+r/r_s)^{2}},
\end{equation}

\noindent where $\delta_c$ is a characteristic density
contrast, and $r_s$ is a
characteristic inner radius.

Integrating Eq.~(\ref{eq:rho}) and using Eq.~(\ref{eq:M200}), it is straightforward to show that the
concentration $c$ is related to $\delta_c$
by

\begin{equation}\label{eq:delta}
\delta_{c} = \frac{200}{3}
\frac{c^3}{ \left[ \ln(1+c) - c/(1+c) \right] }.
\end{equation}

\noindent Then, for a given halo redshift, mass $M_{200}$, and concentration $c$, we can specify the parameters of the NFW model.

We now consider a spherical NFW density profile acting like a lens. The analytical solutions for this lens were given by \citet{Bartelmann1996}, and have been studied by different authors
\citep{Wright2000,Golse2002,Meneghetti2003}. The positions of the source and the image are related through the equation

\begin{equation}\label{eq:beta}
\vec \beta = \vec \theta - \nabla\varphi( \vec \theta) =  \vec \theta -  \vec \alpha( \vec \theta),
\end{equation}

\noindent where $\vec\theta$ and $\vec\beta$ are the angular position in the image and in the source planes, respectively; $\vec \alpha$ is the reduced deflection angle between the image and the source; and $\varphi$ is the two-dimensional lens potential. We introduce the dimensionless radial coordinate $\vec x=\vec\theta/\theta_s$, where $\theta_s=r_s/D_{OL}$, and $D_{OL}$ is the angular diameter distance between the observer and the lens. In the case of an axially symmetric lens, the relation becomes simpler, as the position vector can be replaced by its norm.



\begin{sidewaystable}
\begin{minipage}[t][0.3cm]{1.0\linewidth}
\caption{Summary of the strong lensing modeling.}
\label{tbl-4} 
\centering 
\begin{tabular}{lccccccccccc}
\hline\hline 
\\

ID & Nc   &  $z_l$ & $z_s$  & \multicolumn{1}{c}{X} &
\multicolumn{1}{c}{Y}  &   \multicolumn{1}{c}{$\epsilon$}   &
\multicolumn{1}{c}{$\theta$} & \multicolumn{1}{c}{$\theta_{E,III}$} & \multicolumn{1}{c}{$R_{A}$$^{a}$}    &  \multicolumn{1}{c}{rms}  & $\chi^{2}_{DOF}$  
\\

&  &   & & \multicolumn{1}{c}{[\arcsec]} &
\multicolumn{1}{c}{[\arcsec]}  &        &
\multicolumn{1}{c}{[$^{\circ}$]} & \multicolumn{1}{c}{[$\arcsec$]} & \multicolumn{1}{c}{[$\arcsec$]}  &   &

\\
\hline 
\\
SL2S J08591$-$0345 (SA72)   &  12  &   0.642$\pm0.001$$^{*}$ & 0.883 $\pm$ 0.001 & 1.342 $\pm$ 0.002    & -0.233 $\pm$ 0.001   &   0.0158 $\pm$ 0.0004   & 138.6 $\pm$ 0.4    & 4.9 $\pm$ 0.2 & 4.5  &  0.57    &  19  \\[3pt]
SL2S J08520$-$0343  (SA63) &   6  &  0.457$_{-0.016}^{+0.016}$ &  2.70 $\pm$ 0.08 & 2.0 $\pm$ 0.1    & -0.98 $\pm$ 0.04   &   0.30 $\pm$ 0.03   & 157.0 $\pm$ 0.6    & 5.2 $\pm$ 0.1  & 5.0   &  0.03    &  0.3\phantom{0}  \\[3pt]
SL2S J09595$+$0218 (SA80) &   4   &   0.816$_{-0.013}^{+0.019}$ & 1.2$_{-0.04}^{+0.1}$  &  -0.48 $\pm$ 0.04   &  -0.77 $\pm$ 0.05   &  [0.30]     &  [30]    & 2.1 $\pm$ 0.1 & 2.4   & 0.05    & 0.89         \\[3pt]
SL2S J10021$+$0211 (SA83) &    2  &   0.801$_{-0.015}^{+0.022}$ & 2.26$_{-0.03}^{+0.01}$ &  [0]   & [0]   & [0.25]    & [105]  & 3.19 $\pm$ 0.04  & 2.6 & 0.00  & 0.00            \\
\\
\hline 
\end{tabular}
\tablefoot{(*): Spectroscopic redshifts from \citet{Roberto2013}.\\
(a):  \citet{More2012}\\
Columns: number of constraints Nc and optimized SIE parameters. Error bars represents 1$\sigma$ confidence level on the parameters inferred from the MCMC optimization. Values in brackets are not optimized. These values  correspond to models where the number of observational constraints is smaller than the number of free parameters characterizing the SIE profile. The goodness of the fit is quantified by the RMS in the image plane and the reduced $\chi^{2}_{DOF}$.}
\end{minipage}
\end{sidewaystable}

The reduced deflection angle then becomes \citep{Golse2002}

\begin{equation}\label{eq:alpha}
\vec \alpha(x) =  \left(\frac{4\rho_{\rm cr}\delta_c r_s}{\Sigma_{cr}}\right)\frac{\theta}{x^{2}}g(x)\hat{e_x},
\end{equation}

\noindent where $g(x)$ is a function related to the surface
density inside the dimensionless radius $x$, and is given by
\citep{Bartelmann1996}:

\begin{equation}\label{eq:g}
g(x) = \left\{ \begin{array}{ll}
\ln\frac{x}{2} + \frac{2}{\sqrt{1-x^{2}}} \, \rm{arctanh}\sqrt{\frac{1-x}{1+x}} & \textrm{if $x<1$,}\\

1+\ln\frac{1}{2} & \textrm{if $x = 1$,}\\

\ln\frac{x}{2} + \frac{2}{\sqrt{x^{2}-1}}\arctan\sqrt{\frac{x-1}{x+1}} & \textrm{if $x>1$.}\\
\end{array} \right.
\end{equation}

\noindent The quantity
$\Sigma_{crit}=(c^{2}/4\pi G)(D_{OS}/D_{OL}D_{LS})$ is the
critical surface mass density for lensing.

If we express the deflection angle in terms of $\bar{\kappa}$ = $(4\Sigma_{cr})\rho_{\rm cr}\delta_c r_sg(x)/x^{2}$, that is, the projected surface density $\Sigma$ measured in units of the critical surface mass density, or, in other words, the mean enclosed surface density, then Eq.~(\ref{eq:alpha}) can be expressed as:

\begin{equation}
\vec \alpha(x) =    \bar{\kappa} \theta \hat{e_x}.
\end{equation}

Hence, the lens Eq.~(\ref{eq:beta}) can be written
as:

\begin{equation}\label{eq:beta2}
\vec \beta = \vec \theta \left(1- \bar{\kappa} \right).
\end{equation}

\noindent Following \citet{Broadhurst08}, we define the Einstein radius as the projected radius where the mean enclosed surface density is equal to 1. Then, for a given halo concentration parameter $c$, mass, halo redshift $z_l$, and source redshift $z_s$, the Einstein radius is calculated by solving the equation 

\begin{equation}\label{eq:ThetaESL}
\theta_{E,II} = \left[\frac{4\rho_{\rm cr}\delta_c r_s}{\Sigma_{cr}}\theta_s^{2}g(\theta_E/\theta_s)\right]^{1/2}.
\end{equation}

\noindent Thus,  $\theta_{E,II}$  can be derived numerically using the $c-V_{\rm rms}$  relationship obtained in Sect.\,\ref{Simulations}.  Since we can relate $V_{\rm rms}$ to the $\sigma_{WL}$ value obtained previously for each group (Sect.\,\ref{WeakLensing}). Inasmuch as $M_{200}$ is an unknown variable, we assume it is given by the mass of the isothermal profile at the projected radius $R_{200}$, then

\begin{equation}\label{eq:Mass_IsoT}
M_{200} = \frac{ \pi r_{200}\sigma_{WL}^{2} }{G}.
\end{equation}

\noindent This normalization is arbitrary, but $R_{200}$ is usually taken as a measure of the cluster's virial radius. Although there is evidence from numerical simulations that the hydrostatic assumption is valid probably within $r_{500}$, the kinetic pressure to thermal  gas pressure ratio changes less than $\approx$15$\%$ between both radii \citep[see Fig. 3 in][]{Evrard1996}. Thus, we will keep the former radius. Therefore, using Eq.~(\ref{eq:M200})  and Eq.~(\ref{eq:Mass_IsoT}), the scale radius in Eq.~(\ref{eq:ThetaESL}) is given by

\begin{equation}\label{eq:Mass_cM}
r_{s} = \left[ \frac{3}{4G\rho_{\rm cr}\Delta}\right]^{1/2}\frac{\sigma_{WL}}{c}.
\end{equation}

\noindent No doubt this is an oversimplification of the problem, since we are taking the same $M_{200}$ for both profiles, the NFW and the isothermal. This is not a good assumption, and as we will see (Section ~\ref{Discussion}) it could skew the results, but it is useful to shed some light as a first-order approximation. In Fig.~\ref{ThetaEWL} (left column, middle and bottom panels) we show the ratio  between $\theta_{E,II}$, calculated through Eq.~(\ref{eq:ThetaESL}), and the respective $R_A$ for both catalogs discussed in Section~\ref{Simulations} (the first one constructed from the $V_{\rm rms}$ as proxy for the velocity dispersion inferred in the observed lenses and the second one built to match the shape of the observational redshift distribution of the lenses). For the first case, we obtained a mean of 1.7 and a standard deviation equal to 1.6, and for the second  one we obtained a mean of 1.8 with a standard deviation of 1.8, indicating a slightly larger scatter compared with the first method.

 \begin{figure*}\begin{center}
  \centering 
  \subfloat{\includegraphics[height=4.9cm,width=4.7cm]{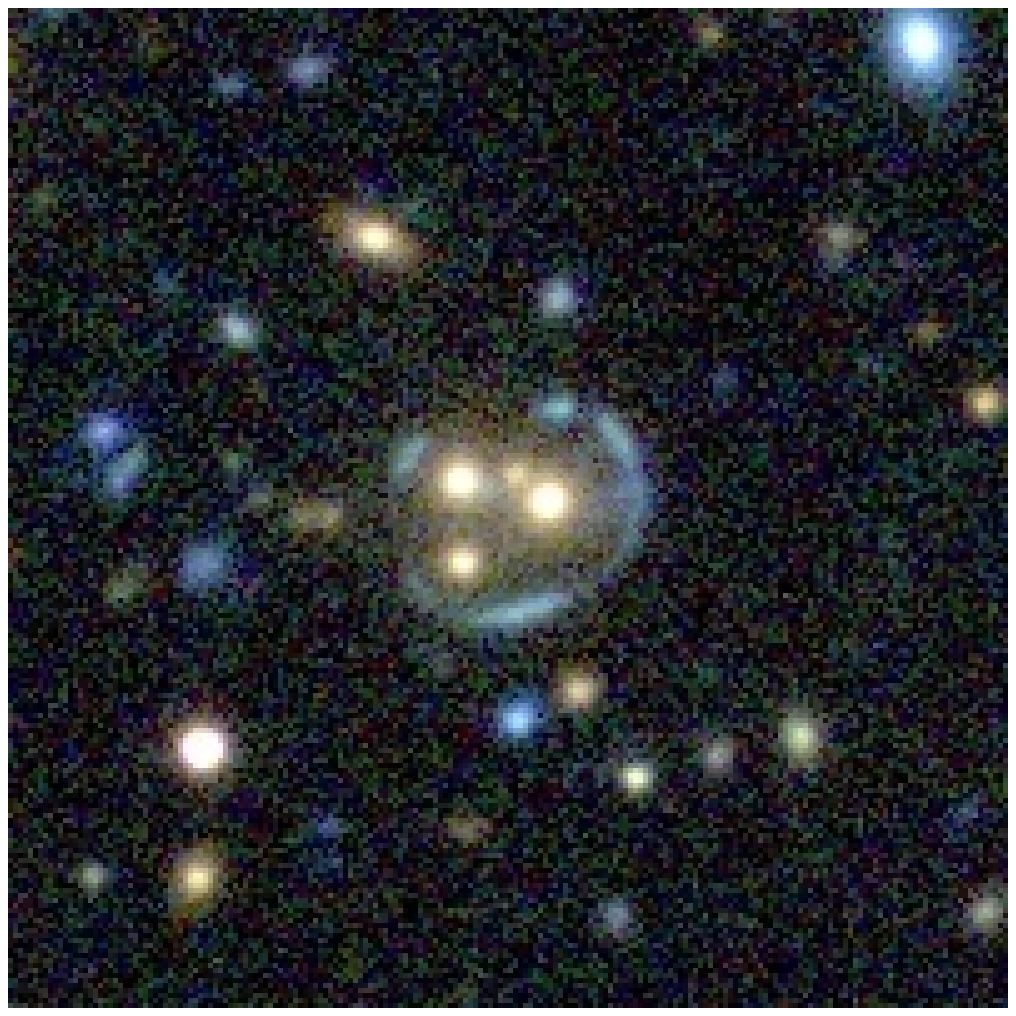}}%
  \subfloat{\includegraphics[height=4.9cm,width=4.7cm]{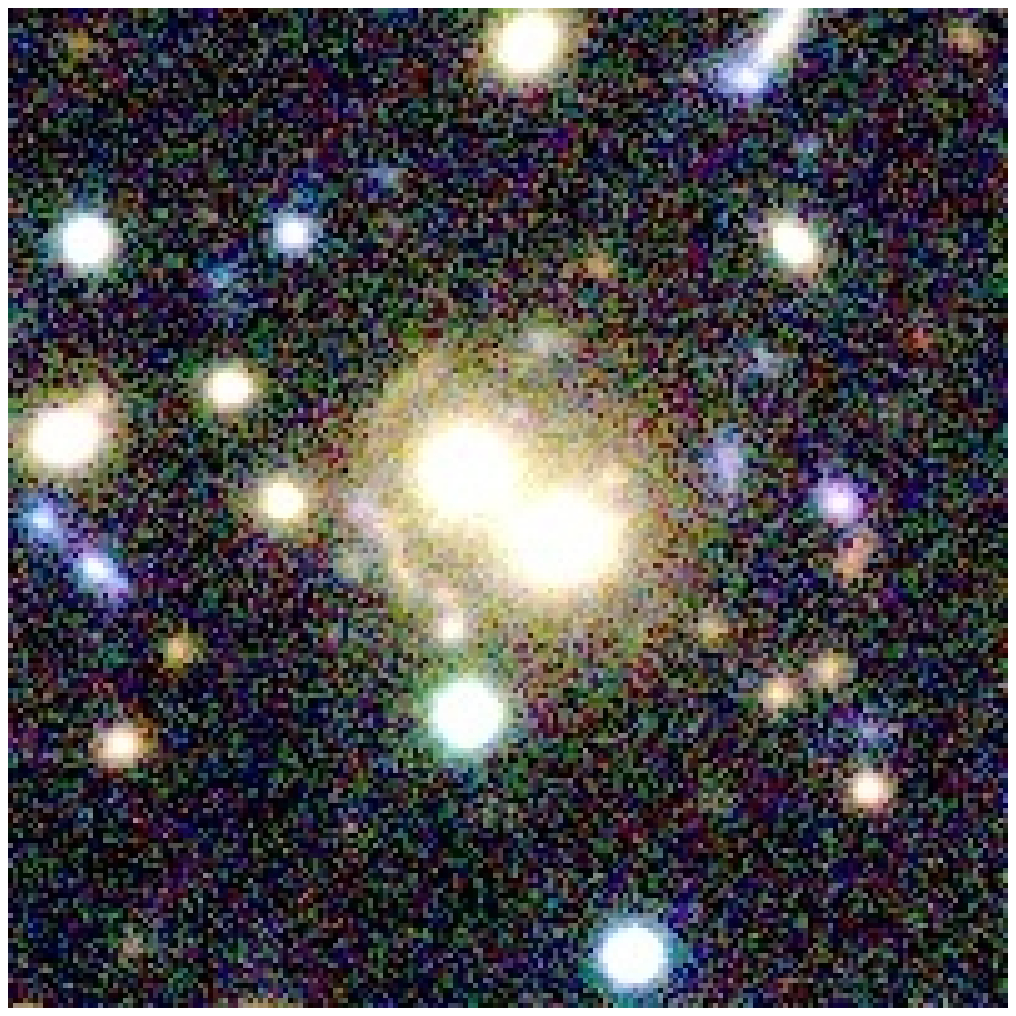}}\\ 
  \subfloat{\includegraphics[width=5.2cm,scale=0.35]{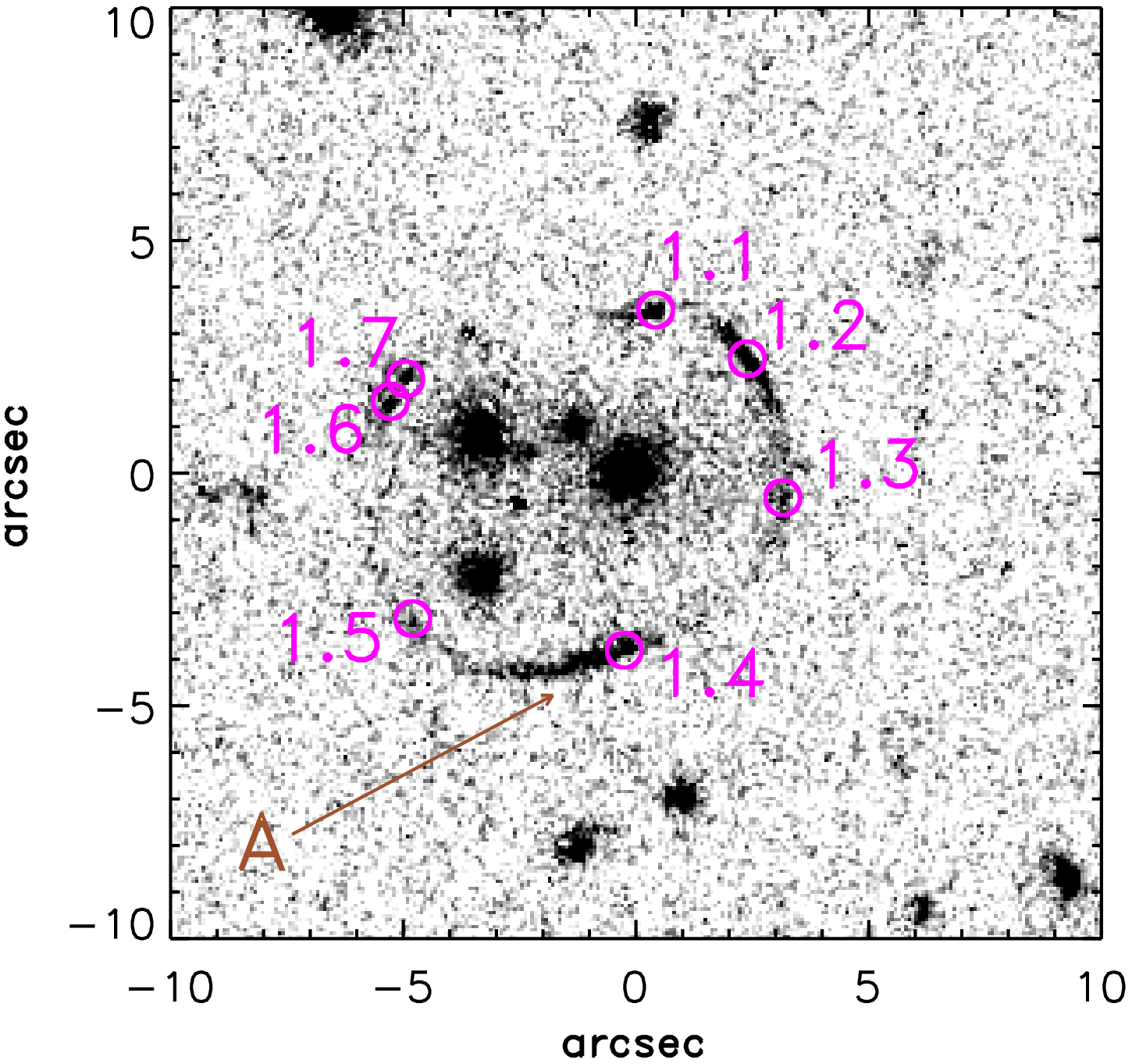}}%
  \subfloat{\includegraphics[width=5.2cm,scale=0.35]{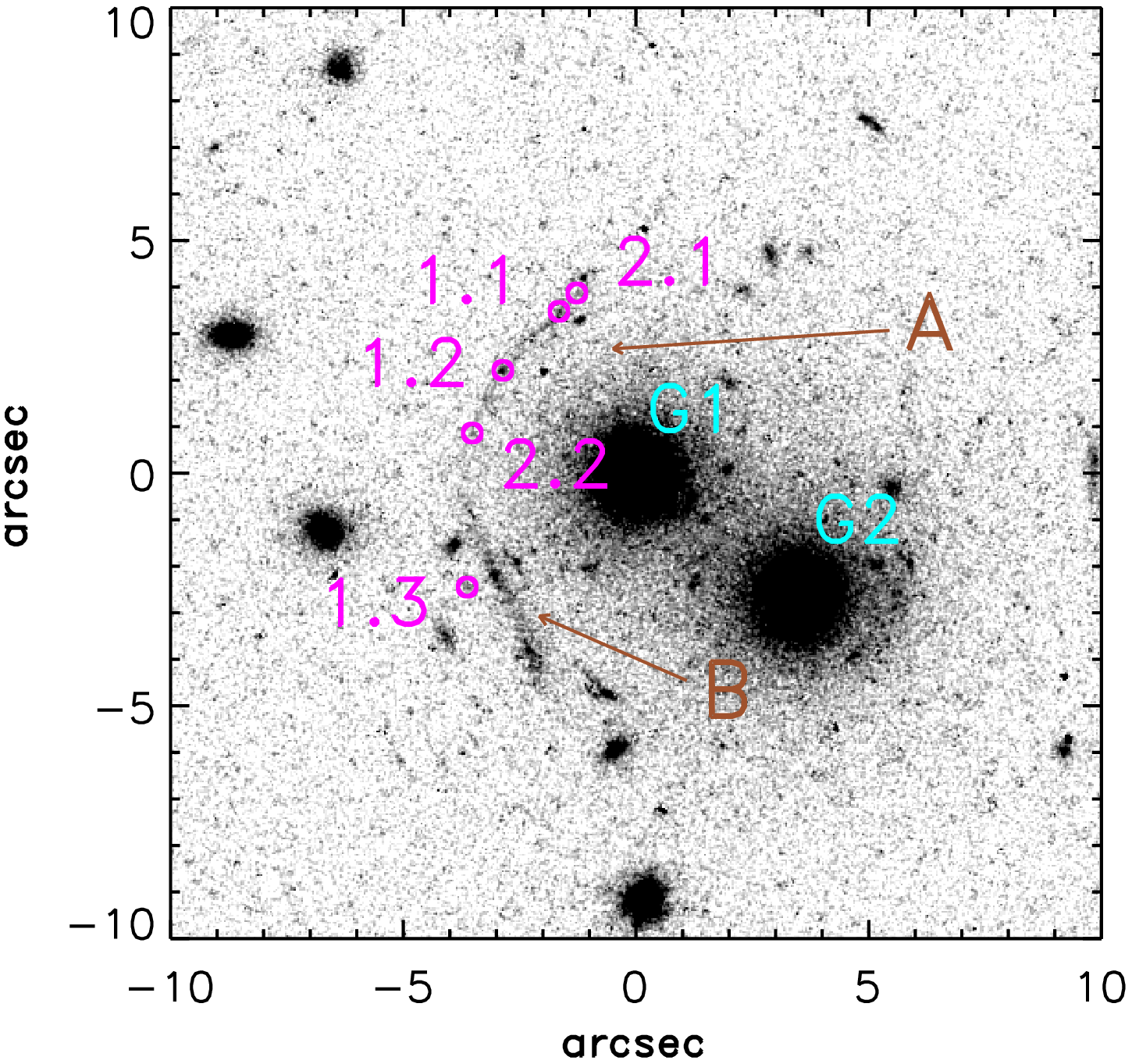}}\\ 
  \subfloat{\includegraphics[width=5.2cm,scale=0.35]{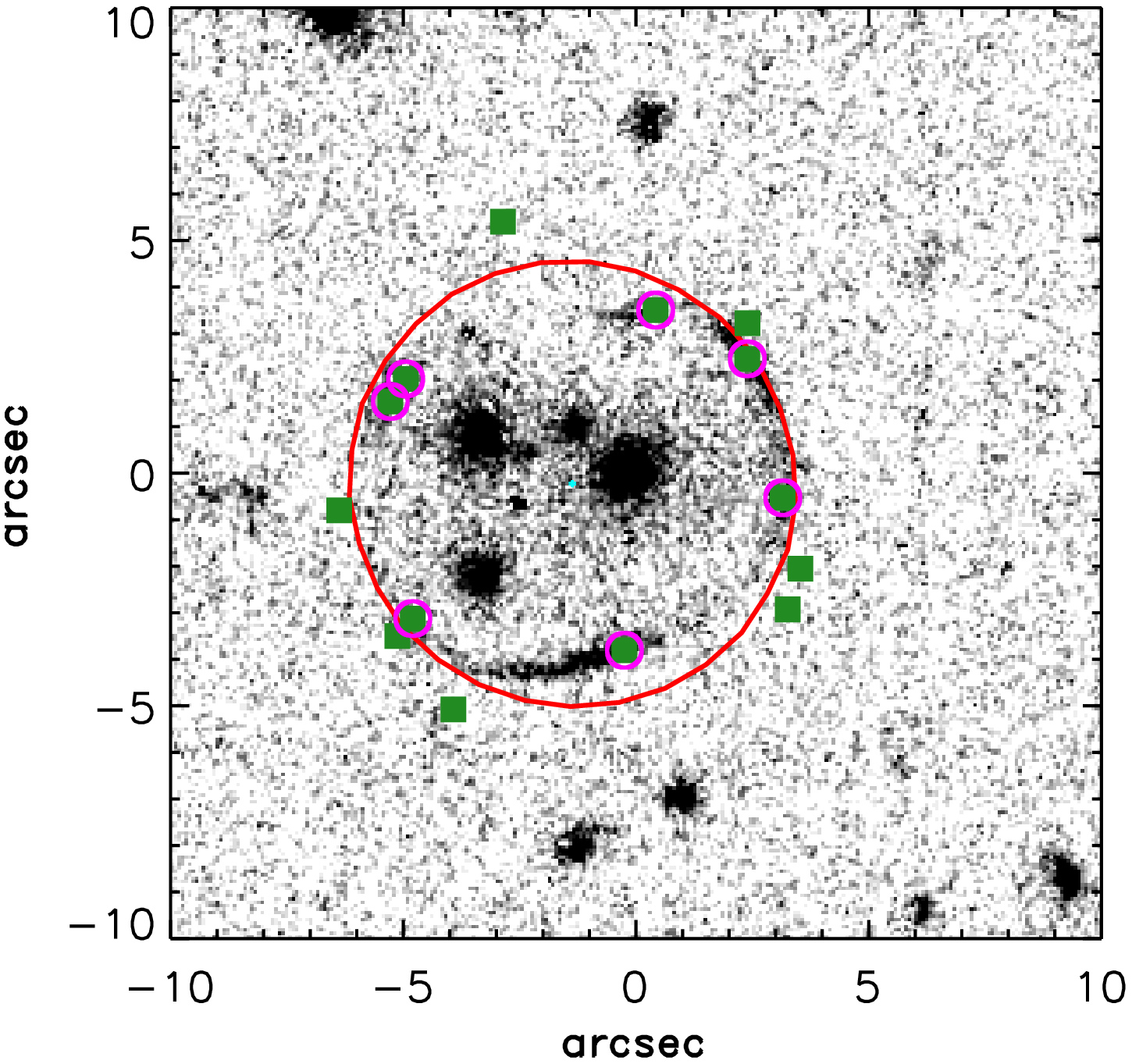}}%
  \subfloat{\includegraphics[width=5.2cm,scale=0.35]{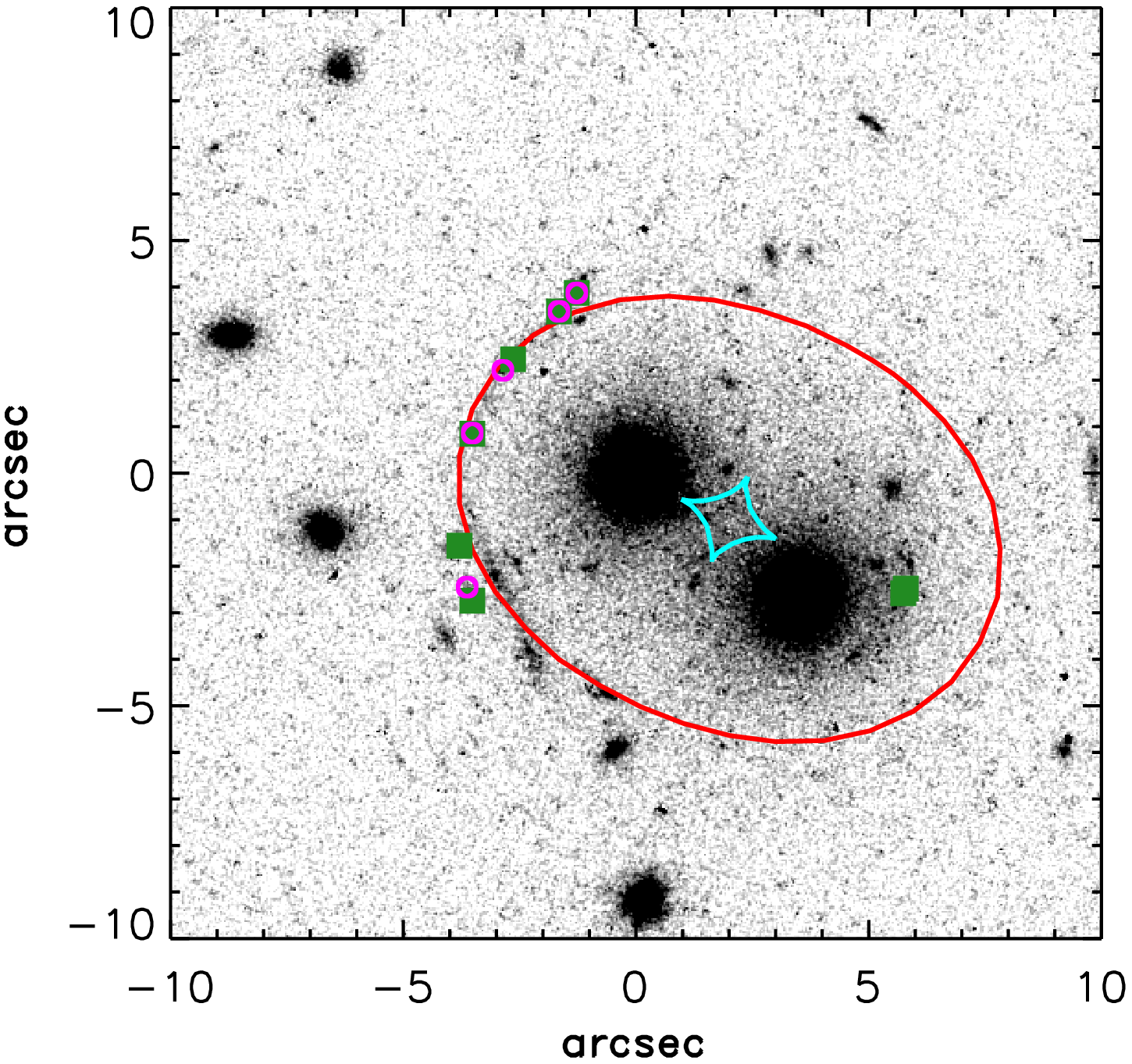}}\\ 
  \subfloat{\includegraphics[width=5.5cm,height=5.55cm]{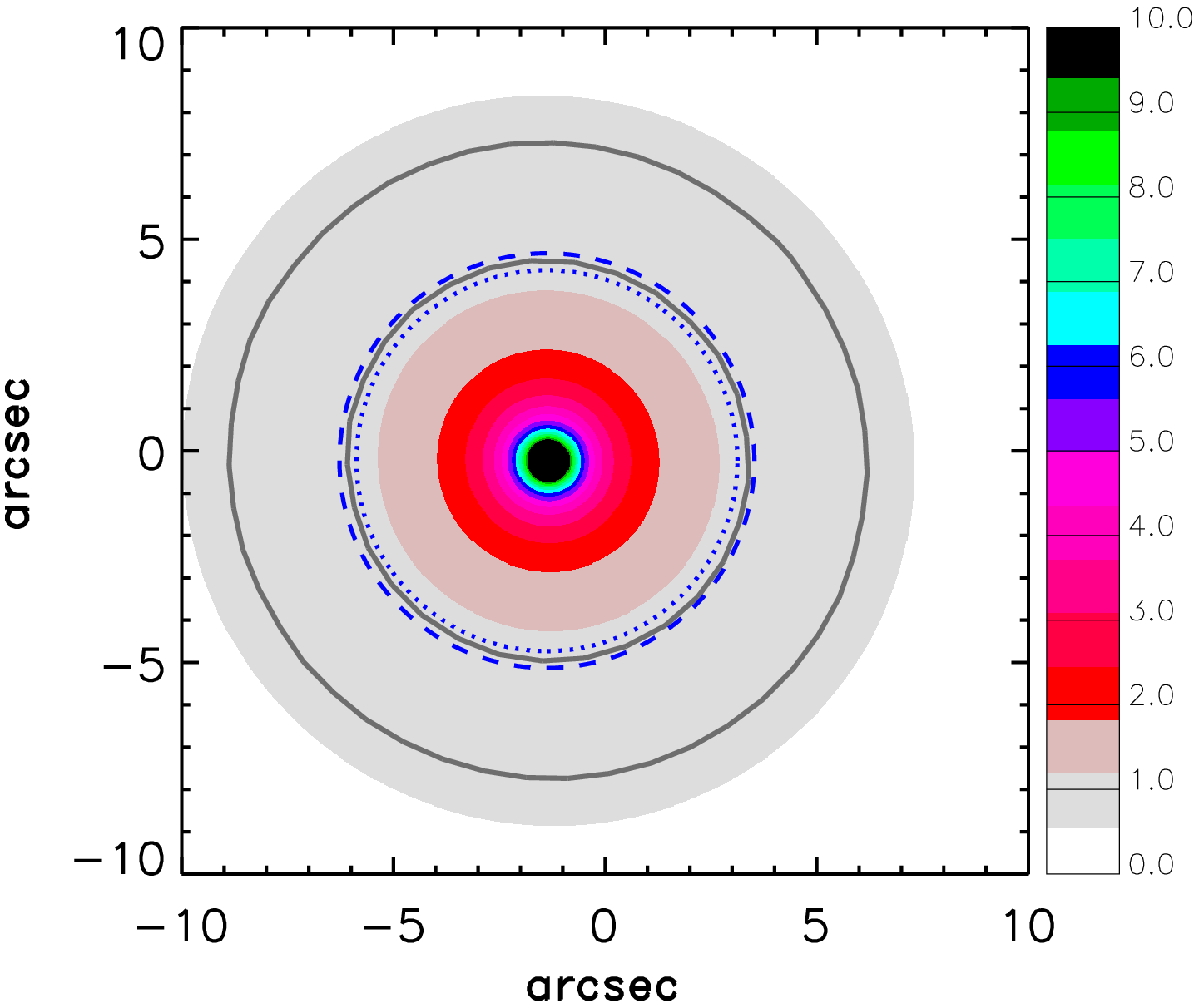}}%
   \subfloat{\includegraphics[width=5.5cm,height=5.55cm]{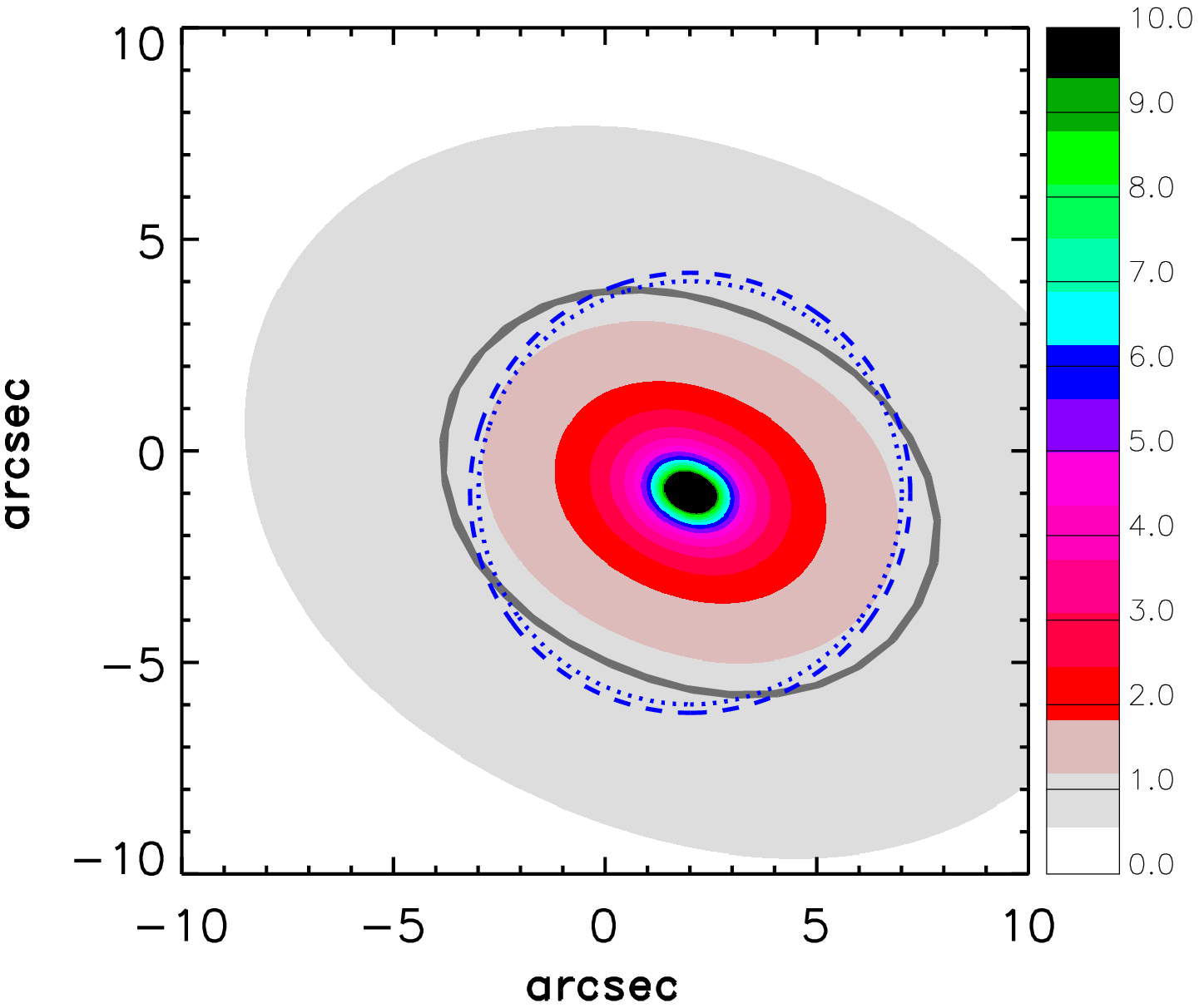}}\\  
  \qquad 
  \caption{SL2S J08591--0345 (SA72) and SL2S J08520--0343 (SA63), left and right columns, respectively. \textit{First row.-} Composite CFHTLS $g$, $r$,$i$ color images (30$\arcsec$$\times$30$\arcsec$). \textit{Second row.-} Identification of the arcs and their substructure in each lens (see Section ~\ref{RT_modeling}). \textit{Third row.-} Critical (red) and caustic lines (cyan) for the strong lensing models. The magenta circles show the measured positions of the image (input data for the model) and the green filled squares  are the model-predicted image positions. \textit{Fourth row.-} Convergence maps. For a source located at the respective redshifts, $z_s$, given in Table ~\ref{tbl-4}. Dark gray lines show $\kappa$ = 1 for sources located  in ($z_{phot}$ - $\delta z_{phot}$, $z_{phot}$ + $\delta z_{phot}$). Dotted blue lines and dashed blue lines depict the values for $R_A$, and $\theta_{E,III}$, respectively.}
  \label{model}
\end{center}\end{figure*}

\begin{figure*}\begin{center}
  \ContinuedFloat 
  \centering 
  \subfloat{\includegraphics[height=4.9cm,width=4.7cm]{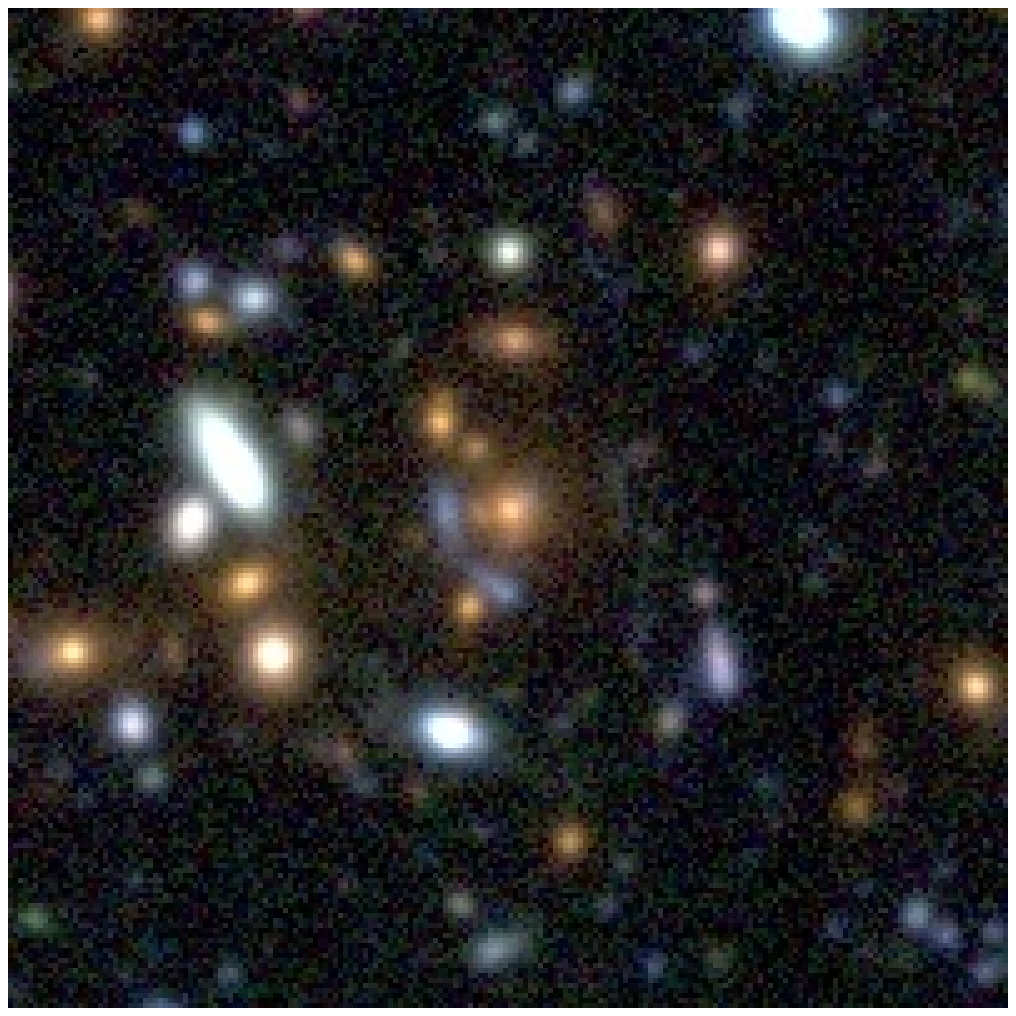}}%
  \subfloat{\includegraphics[height=4.9cm,width=4.7cm]{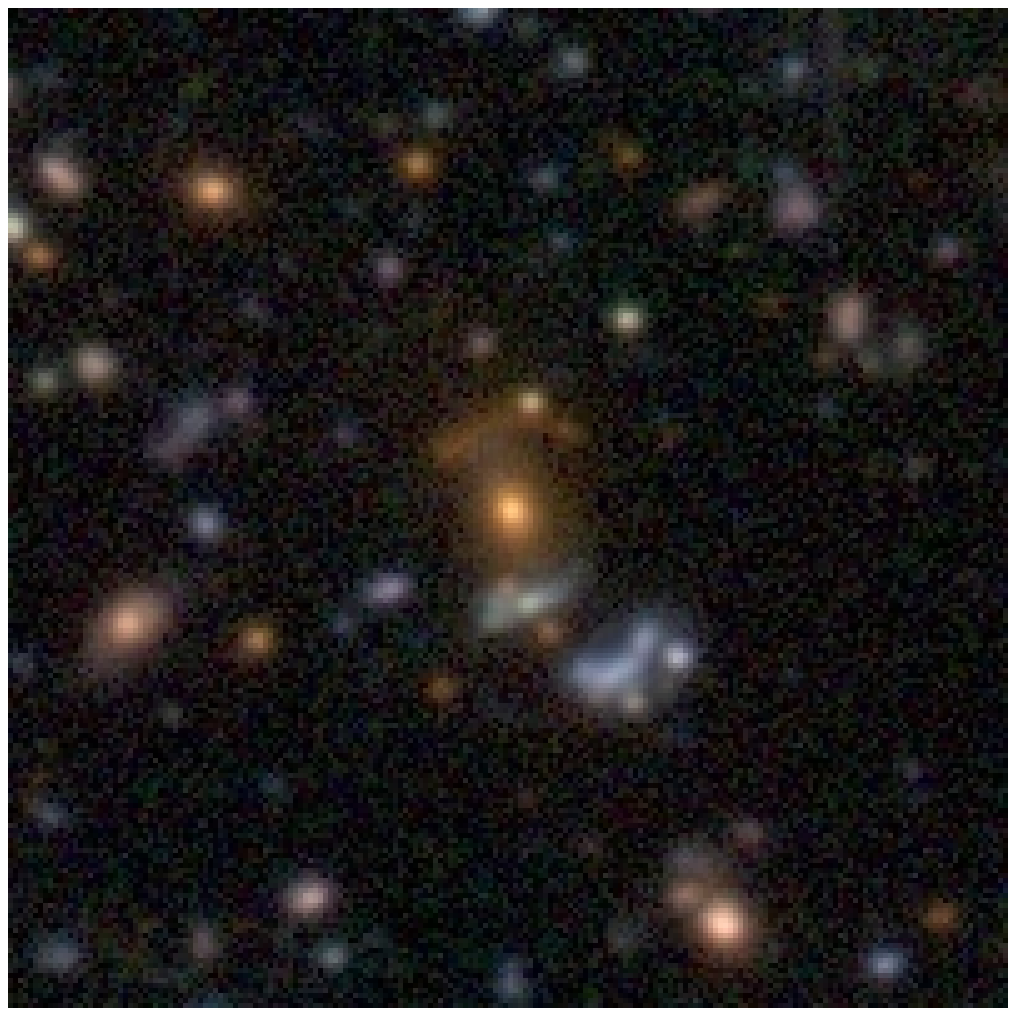}}\\ 
  \subfloat{\includegraphics[width=5.2cm,scale=0.35]{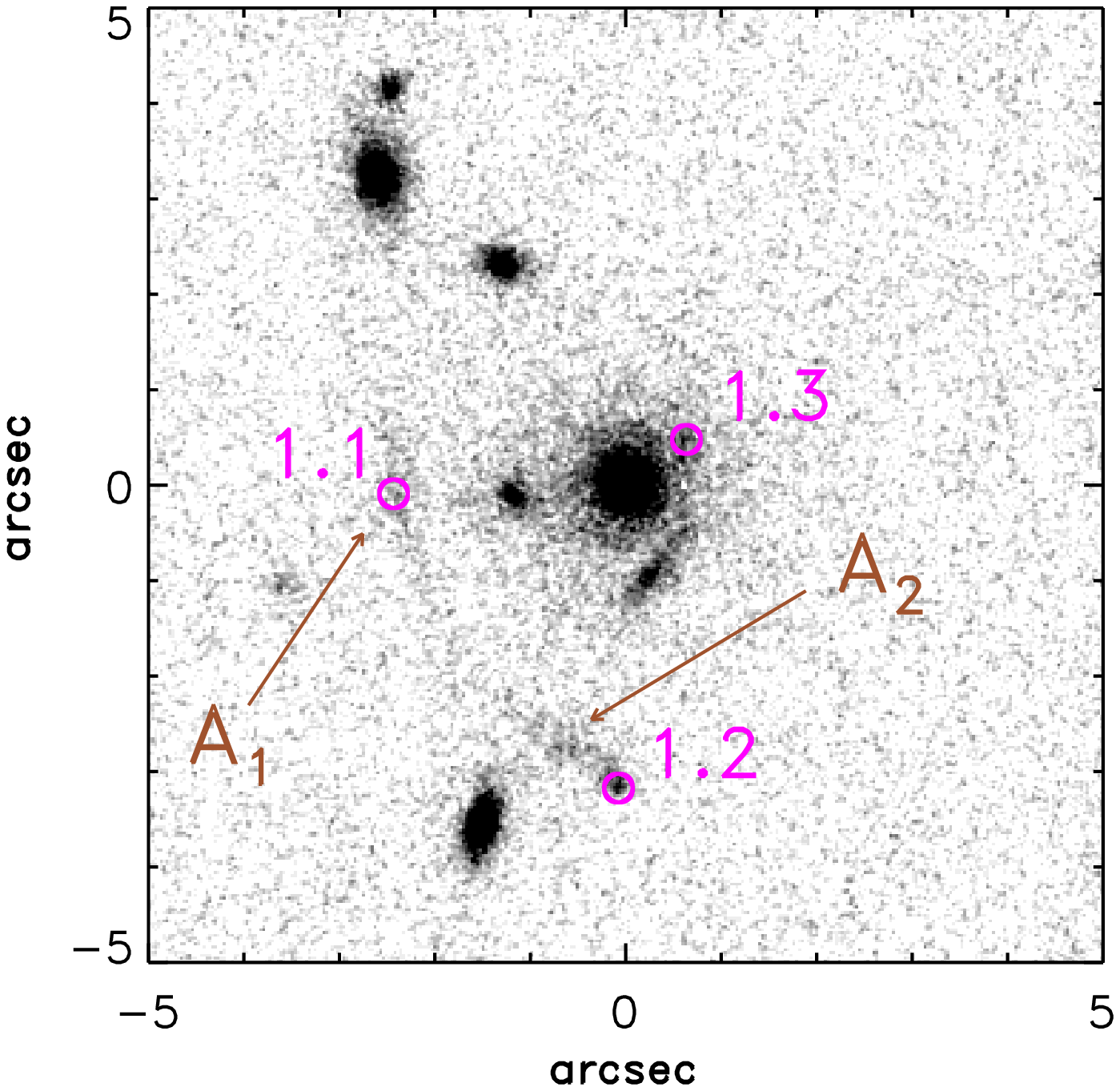}}%
  \subfloat{\includegraphics[width=5.2cm,scale=0.35]{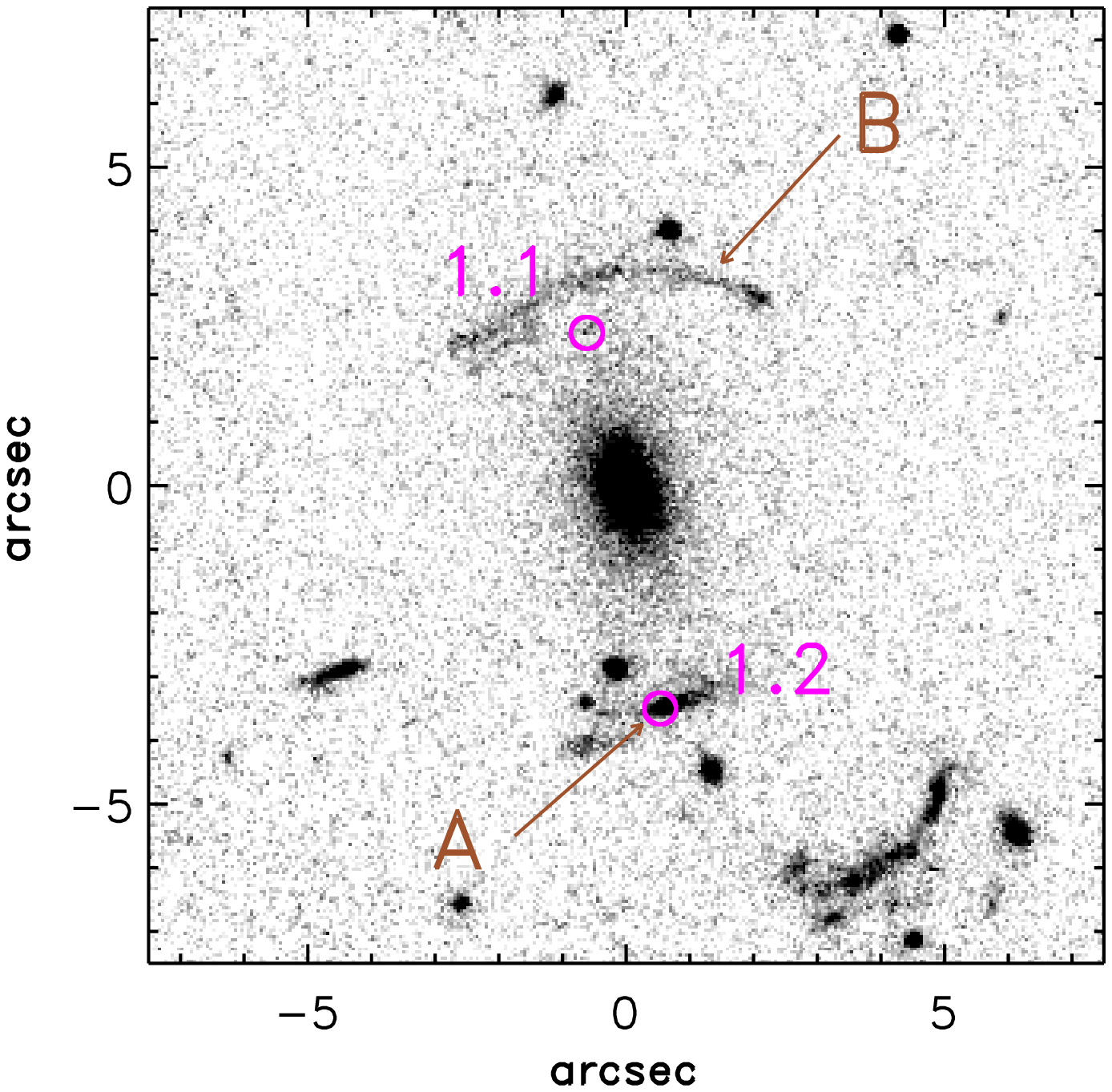}}\\ 
  \subfloat{\includegraphics[width=5.2cm,scale=0.35]{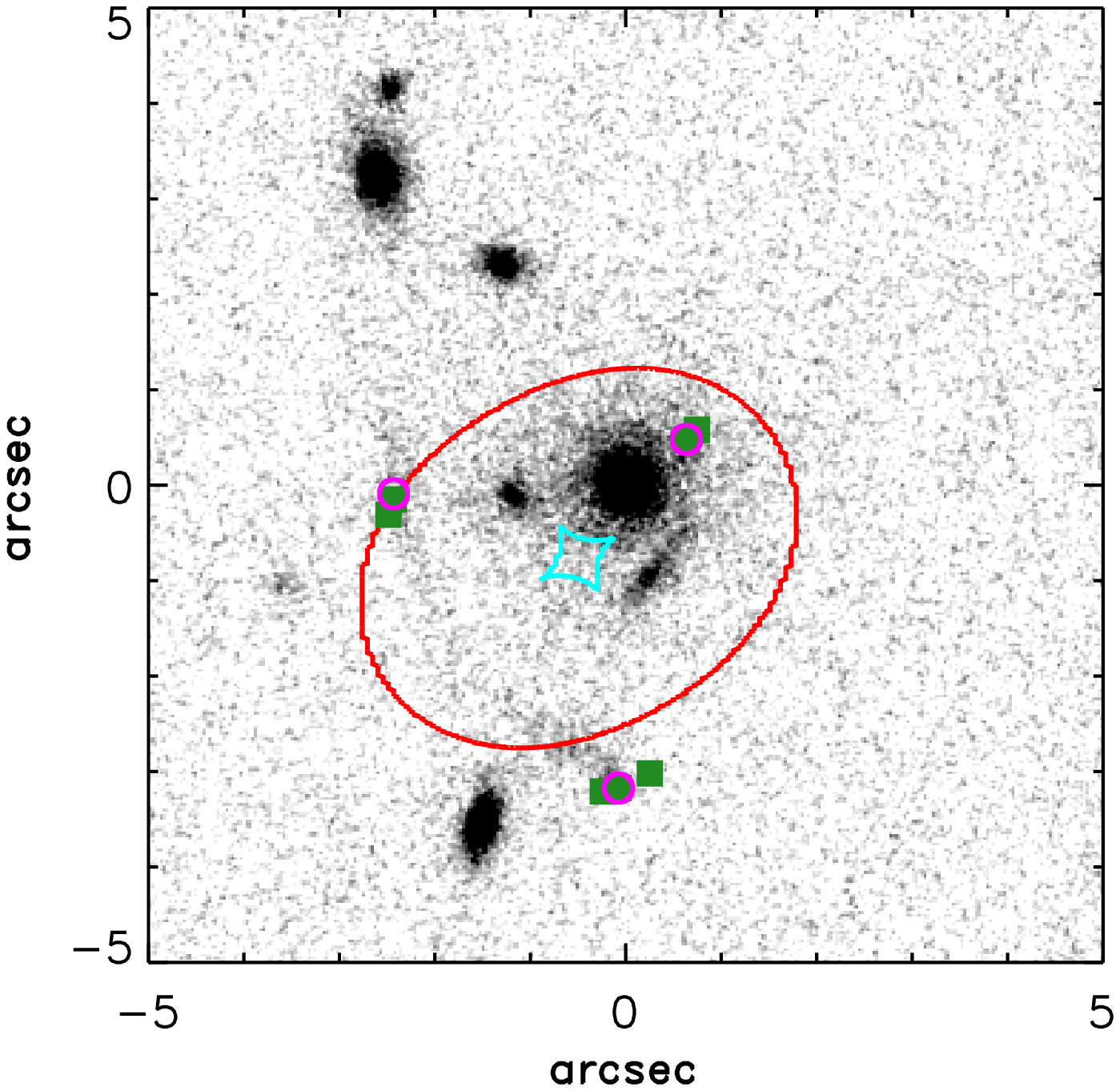}}%
  \subfloat{\includegraphics[width=5.2cm,scale=0.35]{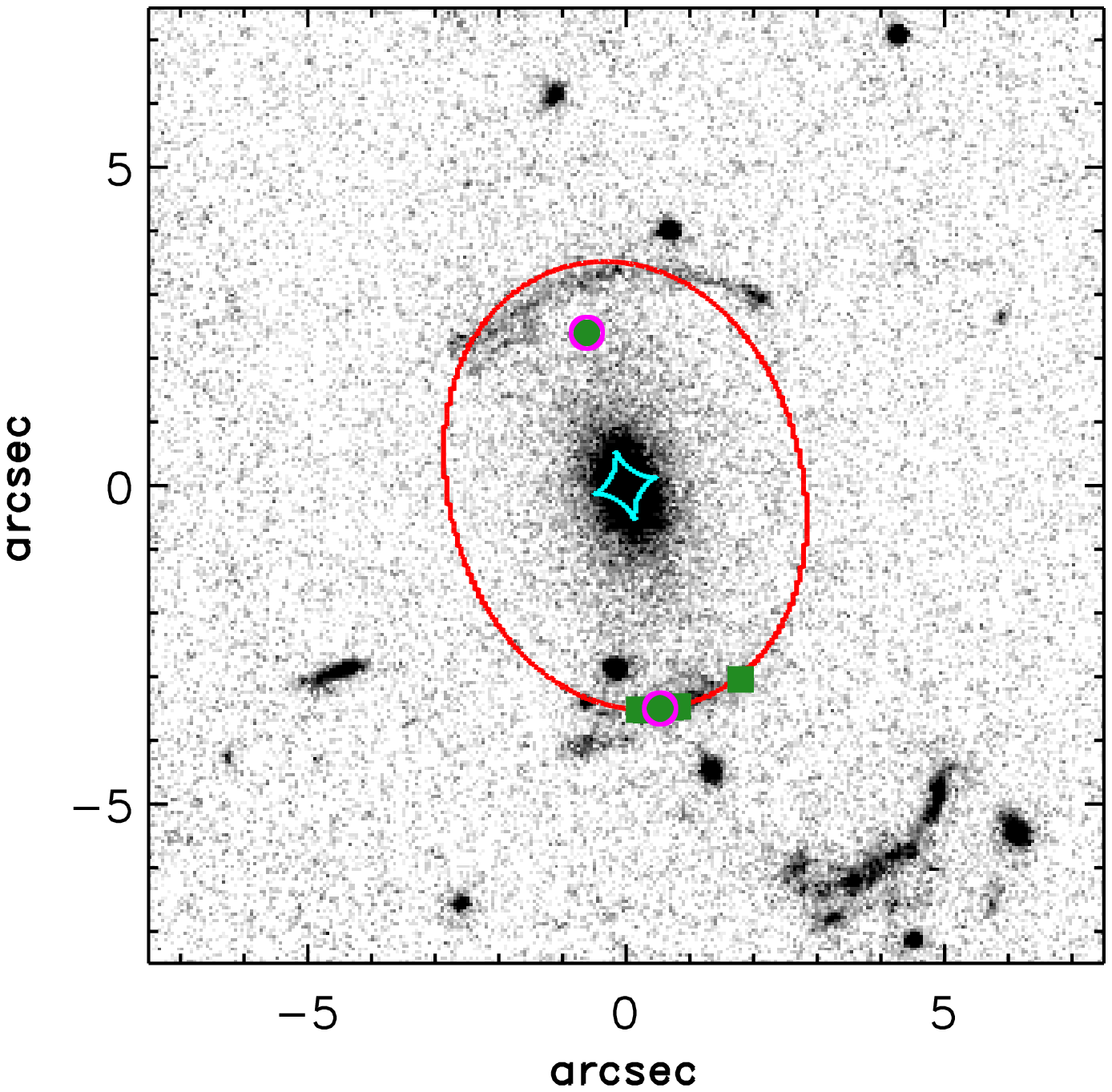}}\\ 
  \subfloat{\includegraphics[width=5.5cm,height=5.55cm]{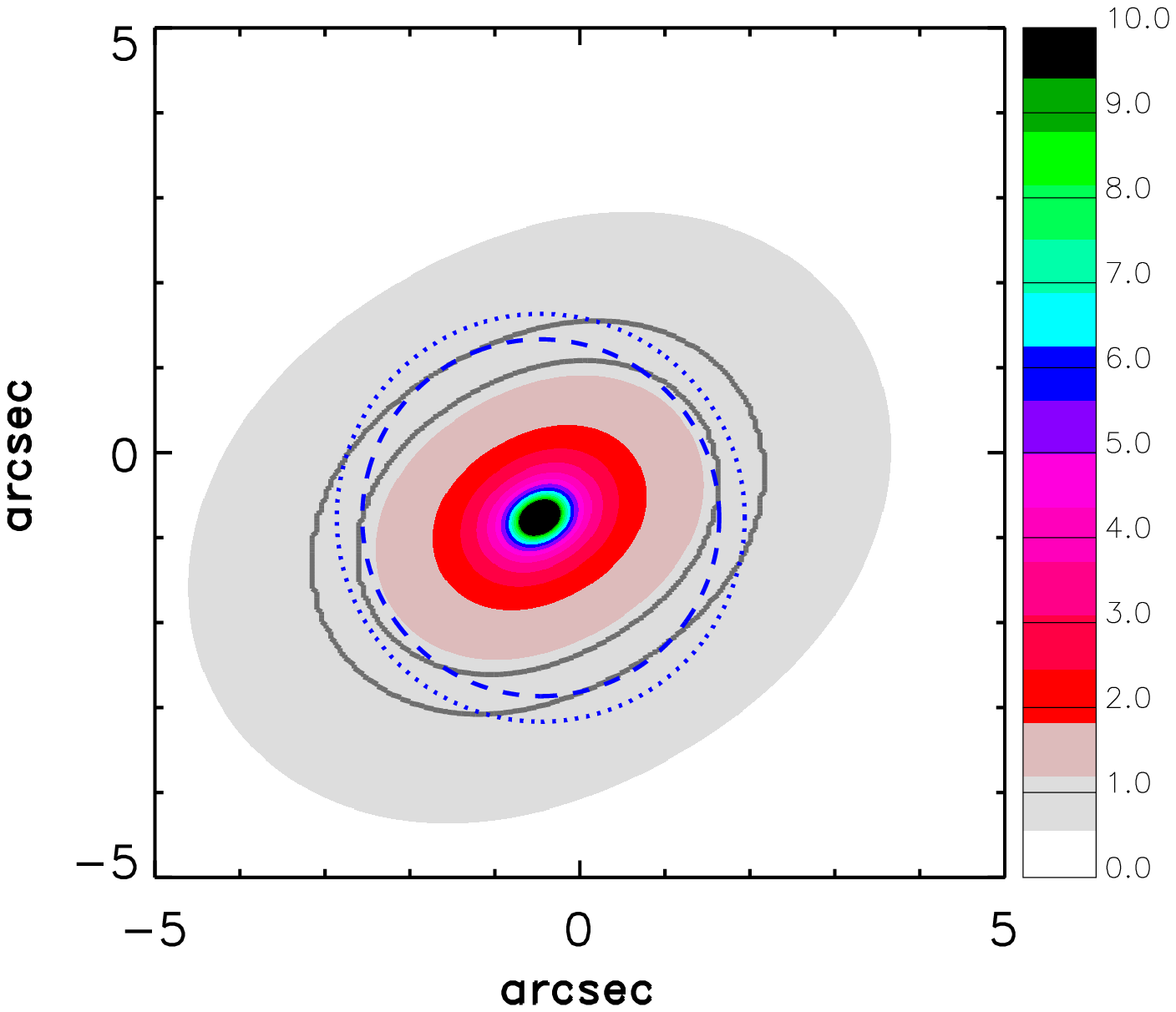}}%
   \subfloat{\includegraphics[width=5.5cm,height=5.55cm]{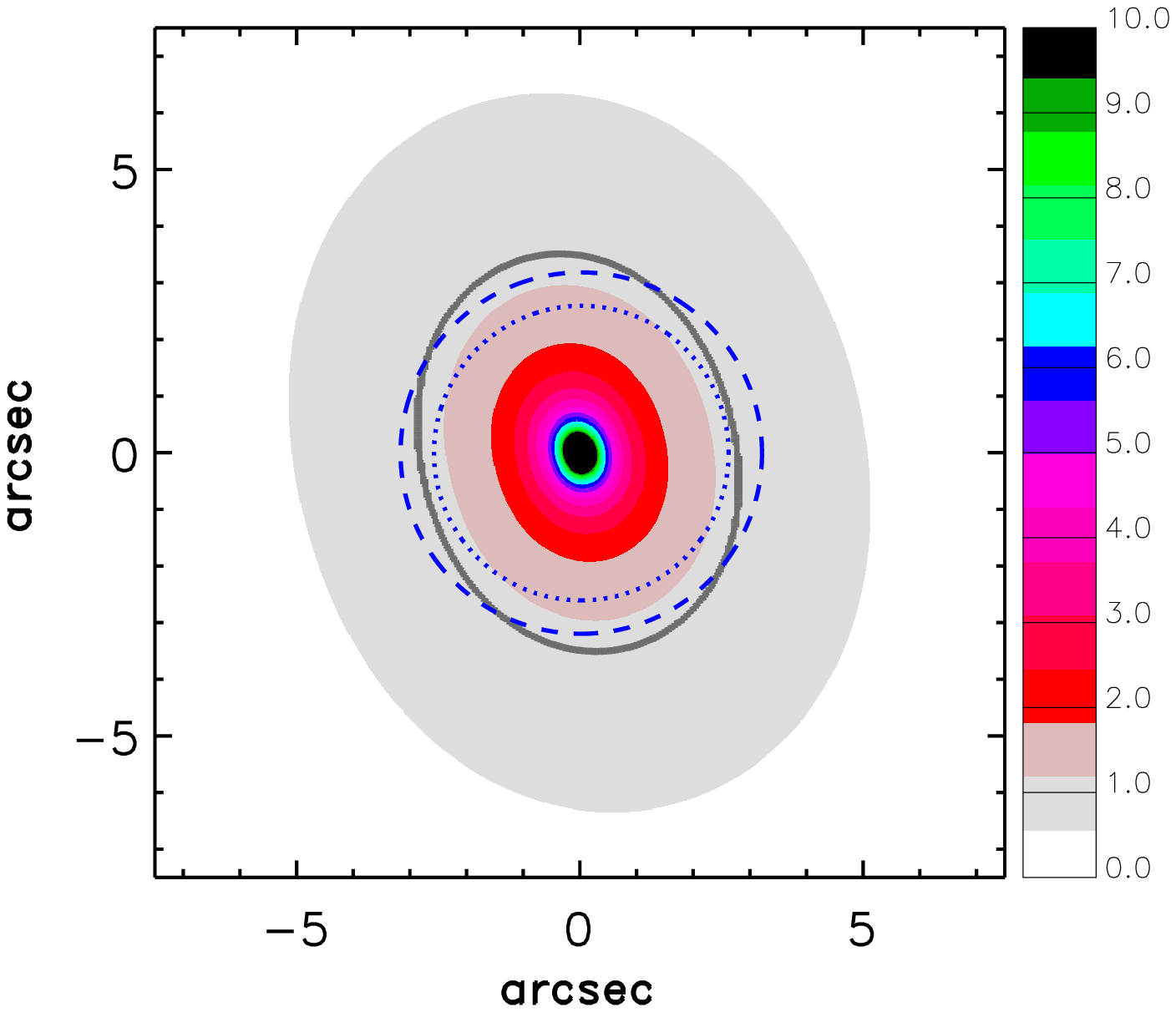}}\\  
  \qquad 
  \caption[]{Continuation. SL2S J09595+0218 (SA80) and SL2S J10021+0211 (SA83), left and right columns, respectively. \textit{First row.-} Composite CFHTLS $g$, $r$,$i$ color images (30$\arcsec$$\times$30$\arcsec$). \textit{Second row.-} Identification of the arcs and their substructure in each lens (see Section ~\ref{RT_modeling}). \textit{Third row.-} Critical (red) and caustic lines (cyan) for the strong lensing models. The magenta circles show the measured positions of the image (input data for the model) and the green filled squares the model-predicted image positions. \textit{Fourth row.-} Convergence maps. For a source located at the respective redshifts, $z_s$, given in Table ~\ref{tbl-4}. Dark gray lines shows $\kappa$ = 1 for sources located  in ($z_{phot}$ - $\delta z_{phot}$, $z_{phot}$ + $\delta z_{phot}$). Dotted blue lines and dashed blue lines depict the values for $R_A$, and $\theta_{E,III}$, respectively.}
  \label{model}
\end{center}
\end{figure*}

The middle- and bottom-right  panels of Fig.~\ref{ThetaEWL} show $\theta_{E,II}$ vs $R_A$ for the groups with 2.0" $\leq$ $R_A$ $\leq$ 8.0". As in the first method, we found a low correlation between both variables (see Table ~\ref{tbl-3} ).  As before,  if we eliminate the outliers  using a 0.5 $\leq$  $R_A$/$\theta_{E,I}$  $\leq$  2.0  cutoff, the results improve,  and we obtain  $R$ = 0.7  ($P$ = 2$\times$10$^{-7}$) and $R$ = 0.7  ($P$ = 1$\times$10$^{-6}$)  for the first and second catalog, respectively. We note again that some groups with giant arcs and strong lensing models are not close to the one-to-one correlation.

Finally, in Fig.~\ref{ThetaIIvsThetaI} we plotted $\theta_{E,II}$ vs $\theta_{E,I}$  to look for systematic differences between both estimates. It is clear that for larger values of $\theta_{E,I}$ there is a slight overestimation of $\theta_{E,II}$. This can be explained by the  large velocity dispersion calculated for the groups: the halo associated with the NFW profile needs to be more massive in order to enclose the same mass at $R_{200}$ as the one calculated from weak lensing (WL).  Similarly, the opposite is true for small values of  $\theta_{E,I}$. This trend explains the change in the slopes in the correlations obtained for $\theta_{E,I}$-$R_A$ and $\theta_{E,II}$-$R_A$, which is also clear in the three  right panels of Fig.~\ref{ThetaEWL}.

 \subsection{The ray-tracing code method}\label{RT_modeling}

In this section, the comparison between $R_A$ and $\theta_E$ is done using strong lensing models for 11 groups in the SARCS sample.  The subsample consist of: SA22 (SL2S J02140$-$0532), SA39 (SL2S J02215$-$0647), SA50 (SL2S J02254$-$0737), SA66 (SL2S J08544$-$0121), SA112 (SL2S J14300$+$5546), SA123 (SL2S J22133$+$0048), SA127 (SL2S J22214$-$0053), SA72 (SL2S J08591$-$0345), SA63 (SL2S J08520$-$0343), SA80 (SL2S J09595$+$0218), and SA83 (SL2S J10021$+$0211). The first seven groups were previously modeled and the results were presented in  \citet[][]{paperI}. The four remaining groups were selected  because they have HST images, an important asset in lensing modeling. This kind of data allows  us to resolve the features in the lensed images and to improve the constraints in the models. Other reasons for their selection are that they have different characteristics (luminosity contours, as well as number of galaxies in the center of the lens), different redshifts, and different lensing configurations (two of them, SA63 and SA80, without previously reported models). Figure~\ref{model} (first row) shows the color composite CFHT images for these four lensing groups, SL2S J08591--0345 (SA72) is the most complex, but it has the most constraints  on the lens.

For the four groups quoted above, we apply  a simple strong lensing mass modeling in order to estimate the Einstein radius using the LENSTOOL code \citep{Kneib1993,jullo07}. Since galaxy groups are generally not well suited to performing accurate lensing models because of the lack of observational constraints \citep[see the discussion in][]{paperI}, we use a singular isothermal ellipsoid (SIE), which has only five free parameters: the position ($x$ and $y$), ellipticity ($e$), position angle (PA), and Einstein radius. All the optimizations were done in the image plane. We want to stress that constructing detailed models for the lens is far from the scope of the present work, and the current data do not allow us to undertake such an analysis. To test more complex models more spectroscopic data is required (we are currently in a campaign to obtain this data for the arcs and galaxies in some of the SL2S groups).

\begin{figure}[h!]
\begin{center}
\includegraphics[scale=0.5]{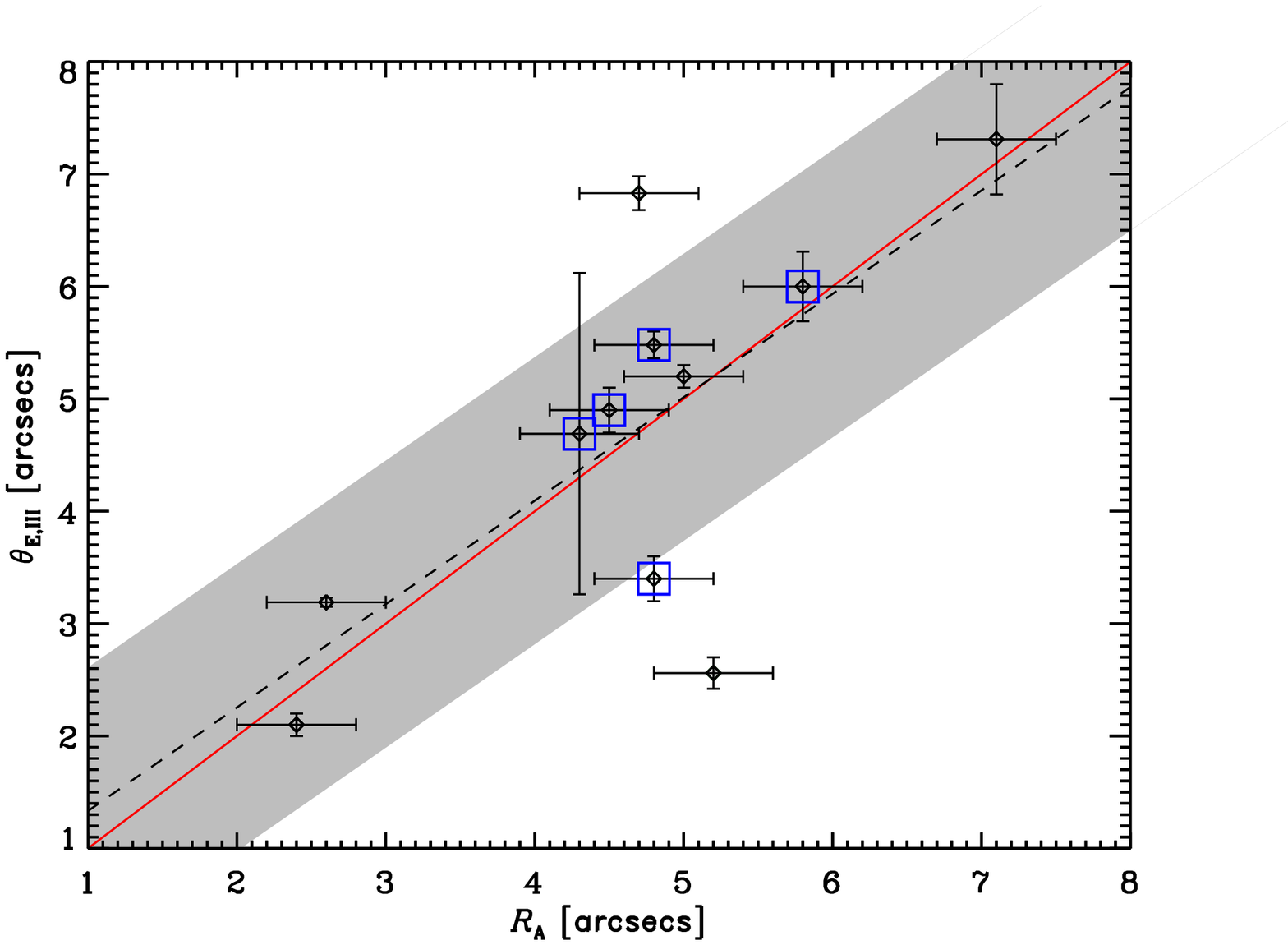}
\caption{The figure shows $\theta_{E,III}$ $vs$ $R_A$ for those groups with strong lensing models.  The black dashed line shows the fit to the data, with the 1$\sigma$-error depicted as a gray shaded region. Blue squares are groups with giant arcs (see Section ~\ref{WeakLensing}), and  the red continuous line shows the one-to-one relation.    }
\label{RA_SL}
\end{center}\end{figure}

\textit{SL2S J08591$-$0345} (SA72) at $z_l$ = 0.642$\pm$0.001 is a confirmed galaxy group \citep[see][]{Roberto2013} with three bright galaxies and two smaller ones in the center of the lens. The multiple images of this exotic lens draw an oval contour around the deflector (see top-left panel of Figure~\ref{model}). The object was modeled by \citet{Orban2009} as a four-component lens, showing the complexity of this compact group. In addition, the object was presented in the first sample of groups by  \citet{paperI}, and later cataloged in the SARCS sample \citep{More2012}.   Assuming that all the lensing features come from one single source at $z_s$ = 0.883 $\pm$0.001 ($z_{phot}$ = 1.04 $_{-0.08}^{+0.04}$), we constructed our model leaving all parameters free. Although our best model shows a large $\chi^{2}$ (see Table ~\ref{tbl-4}),  it is important to note that the complexity of a five-galaxy lens in the center of the group would require us to assume a more elaborate model than a simple SIE for its mass distribution.

\textit{SL2S J08520$-$0343} (SA63) is populated by two galaxies in the center (G1 and G2) and displays several arc features, with those labeled arc A and B being the most prominent  (see second row, right panel of Figure~\ref{model}). In the ground-based image, these arcs seems to belong to the same source, i. e., forming a system. However on the HST-ACS image this association is not so clear since arc B is  brighter and straighter than arc A. Thus, to construct our model we assume arcs A and B are different systems and we used only arc A  to perform the optimization. Given the resolution of the HST-ACS image, we can conjugate two points on arc A, increasing the number of constraints to six \citep[e.g.,][]{paperI,Verdugo2011}. Our model predicts a demagnified counter-image for arc A (see third row, right panel of Figure~\ref{model}) near galaxy G2 that can be associated with some arc-like features  close to this galaxy. The results of our best fits are summarized in Table ~\ref{tbl-4}.

\textit{SL2S J09595$+$0218 (SA80)} shows a configuration with two bright arcs on one side of the deflector and a very demagnified image on the other side of the deflector (see second row, left panel of the continuation of Figure~\ref{model}). The photometric redshift  sets the arc at $z$ $\sim$ 1.2. With only three multiple images forming the system, we have only four observational constraints. First, we set $x$ = 0 and $y$ = 0, leading to an elliptical mass distribution model with only three free parameters, namely, the ellipticity $e$, the position angle PA, and the Einstein radius. Given the impossibility of obtaining a good fit  in these conditions, we proceeded in an alternative way. We set the ellipticity and the position angle with fixed values $e$ = 0.30, and PA = 30$^{\circ}$, leaving $x$ and $y$ free. In the third row and left panel of the continuation of Figure~\ref{model} we show the predicted positions of our best model, as well as the observed positions.

\textit{SL2S J10021$+$0211 (SA83)} was discovered in the COSMOS survey \citep{Faure2008,Faure2011}. It has two arcs surrounding the central galaxy of the group, one long red arc (showing substructure) situated at the north (arc B) and another  blue and more compact one (arc A) at the south (see second row, right panel of the continuation of Figure~\ref{model}). We were able to obtain the photometric redshift only for the blue arc, z $\sim$ 1. Since arc A does not show surface brightness peaks that can be conjugated as different multiple image systems, there are not enough constraints to try even a simple model. However, using the position of a possible counter-image located below arc B (see second row, right panel of the continuation of  Figure~\ref{model}), we have two constraints, which are enough to probe the Einstein radius. Thus, following  \citet{Faure2008} we perform the optimization fixing the center of the lens as the center of the bright galaxy, $e$ = 0.25, and PA = 105$^{\circ}$ (see Table ~\ref{tbl-1}). Our model (see third row, right panel of the continuation of  Figure~\ref{model}) predicts that the counter-image of arc A will be very demagnified, which explains why it is not observed in the CFHT images. Our value of $\theta_{E,III}$ shown in Table ~\ref{tbl-4} agrees with the value $\theta_E$ = 3.14$\arcsec$ found by \citet{Faure2008}, although our $z_l$ and $z_s$ values (obtained using a different methodology) are slightly different to those reported in that work.

 The bottom panels of Figure~\ref{model}  presents the convergence map for each strong lensing model,  considering a source located at the respective redshift, $z_s$, given in Table ~\ref{tbl-4}.  Dark gray lines depict  the convergence locus, $\kappa$ = 1, considering two different sources  situated at $z_{phot}$ $\pm$ $\delta z_{phot}$. We note that such values are consistent with $R_A$ (dotted blue line) and  $\theta_{E,III}$ obtained from the strong lensing model (dashed blue line). In the case of SL2S J08591$-$0345 (SA72), the large difference between the extreme values of $z_{phot}$  arises from the source and lens proximity in redshift.

Finally, in Figure~\ref{RA_SL} we show $\theta_{E,III}$ vs $R_A$ for the eleven groups presented in this section. We found a good agreement between both values, with  $\theta_{E,III}$ = (0.4 $\pm$ 1.5) + (0.9 $\pm$ 0.3)$R_A$, a Spearman's rank correlation coefficient $R$ = 0.6,  and a significance $P$ = 6$\times$10$^{-2}$.  The two groups with slightly extreme values below the correlation are SA123 (SL2S J22133$+$0048) and  SA39 (SL2S J02215$-$0647).   This might be related to the multiple components  present in their luminosity maps (see Section ~\ref{Discussion}). On the other hand, the group with the value over the correlation, SA127 (SL2S J22214$-$0053),  is a group with  a high degree of elongation. We also note that the groups with giant arcs, except SA123 (SL2S J22133$+$0048), nearly follow the one-to-one relation.

 \section{Discussion}\label{Discussion}
 
\subsection{$R_A$ vs $\theta_E$}\label{Discussion1}

In Section~\ref{EinsteinRadius} we found a low correlation between  $\theta_{E,I}$ and $R_A$, with a large scatter. We found  $\theta_{E,I}$ = (2.2 $\pm$ 0.9) + (0.7 $\pm$ 0.2)$R_A$ with a correlation coefficient $R$ = 0.33, and a significance $P$ = 6$\times$10$^{-3}$.   It is important to note that the analysis was done in the whole sample (with 2.0" $\leq$ $R_A$ $\leq$ 8.0"),  without distinction between regular and irregular groups. We also note that our sample, as we commented in the introduction, does not only include giant arcs. This introduces scatter in the correlation, but even in a sample with only giant arcs, there are some groups far from the one-to-one correlation (see right column of Fig.~\ref{ThetaEWL} ). This is related to another  factor that we need to consider in this comparison between $\theta_{E,I}$ and $R_A$, namely that we are using information at large scale (weak lensing velocity) in a lower-scale regime (strong lensing).

\begin{figure}[h!]
\begin{center}
\includegraphics[scale=0.47]{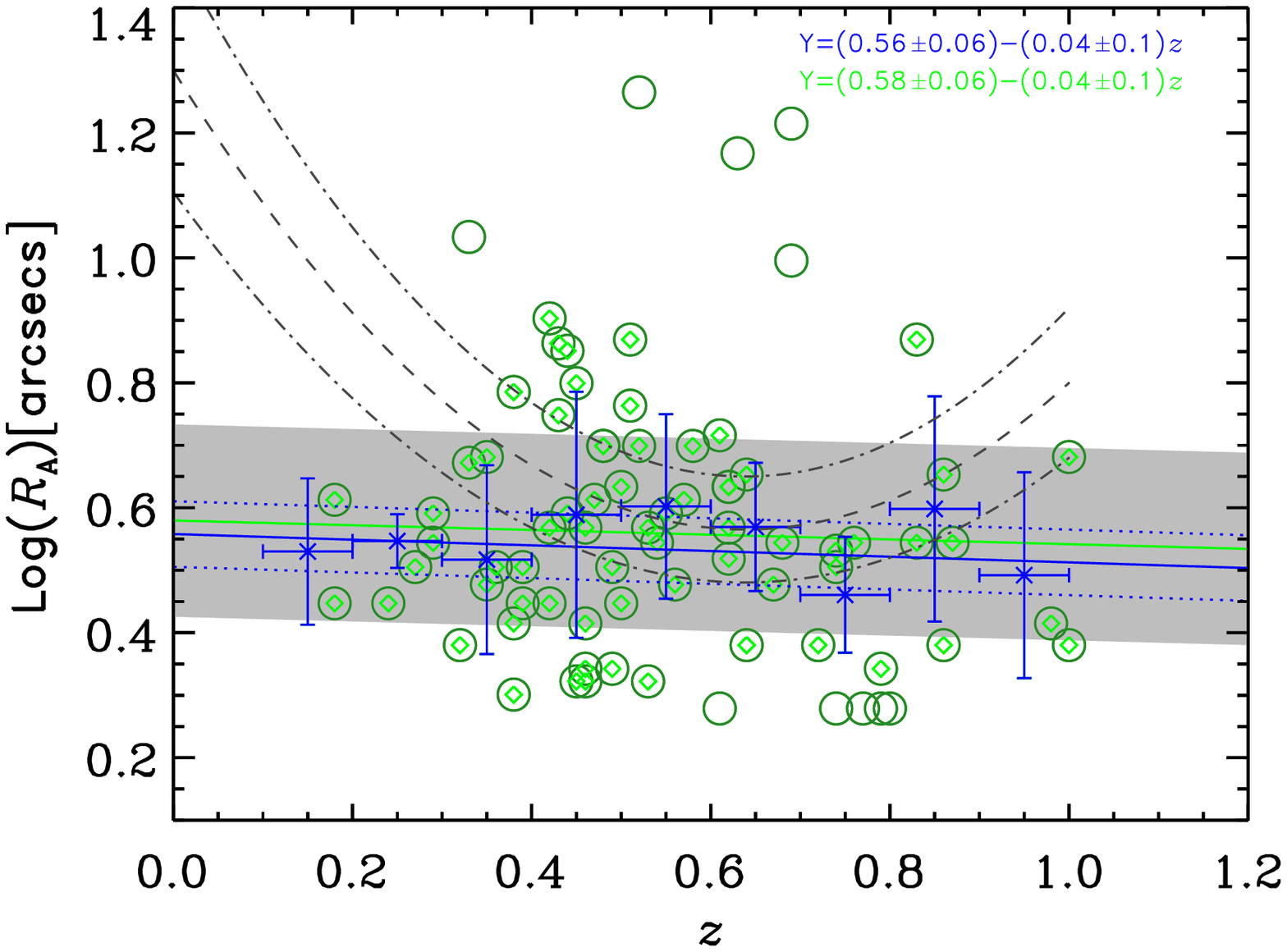}
\caption{The figure shows $R_A$ as a function of the redshift. Green circles depict the secure lens candidates \citep{Gael2013}, and green diamonds those with  2.0" $\leq$ $R_A$ $\leq$ 8.0". Blue asterisks with error bars highlight the correlation between $R_A$ and $z$ after binning the data (green diamonds). The blue continuous line shows the fit to the binned data, with 1$\sigma$-error depicted as a dotted blue line. The green continuous  line shows the fit to the green diamonds, with the 1$\sigma$-error depicted as a gray shaded region. The dashed line is a second-order polynomial function from \citet{Zitrin2012}, assuming an error of 15$\%$ (dot-dashed line).}
\label{Ra_vs_L_1}
\end{center}\end{figure}

  \Citet{Gavazzi2005} shows that departures between lensing and true mass tend to vanish at large scale. Moreover, the asphericity has a lower effect on weak lensing and X-ray-based mass estimates \Citep[see Figure 6 in][]{Gavazzi2005} than in the case of strong lensing. In this context,  as we have measured $\sigma_{WL}$ from weak lensing data  (i.e. large scale), it is possible  that, in some cases, we might not be able to recover the observed Einstein radius obtained when using the velocity  dispersion directly   in Eq.~\ref{eq:theta_eWL}.
 Thus, at radii $\sim$$\theta_E$ we are probing a different mass. As we can see in the top-left panel of  Fig.~\ref{ThetaEWL}, the scatter is present in both directions, confirming this assertion. On the other hand, even if some galaxy groups look like simple lensing objects \citep[but see][]{Orban2009,Limousin2010},  they could be dynamical complex objects with multimodal components and substructures \citep[e.g.][]{Hou2009,Hou2012,Ribeiro2013}. Some of our groups have luminous  morphologies \citep{Gael2013} that cannot be associated with regular groups (roughly circular isophotes around the strong lensing system) and have elongated (elliptical isophotes with a roughly constant position angle form inner to outer parts) or multimodal morphologies (two or more peaks in the central part of the map). Since an irregular luminous morphology it is not a robust confirmation of the  dynamical irregularity of a group (although the spectroscopic analysis of the seven objects in  \citealt{Roberto2013} is consistent with the luminous morphology presented in \citealt{Gael2013}), we carry out our analysis using groups with  both irregular and regular luminous morphology. Moreover, as mentioned in Section~\ref{WeakLensing}, if we discard objects with extreme  $R_A$/$\theta_{E,I}$  (for example with a cut off 0.5 $\leq$  $R_A$/$\theta_{E,I}$  $\leq$  2.0), the correlation coefficient and the significance improve ($R$ = 0.6, and $P$ = 1$\times$10$^{-5}$). This is consistent with the idea presented above about the comparison between strong lensing mass and weak lensing mass, and is  also consistent with the results found when we used the eleven strong lensing models (Section~\ref{RT_modeling}). For such groups with extreme ratios it is more inaccurate to associate the mass at large scale with the mass inside the Einstein radius.

The velocity dispersion $\sigma_{WL}$ is transformed  into the corresponding mass and used to compute $\theta_{E,II}$   through Eq.~(\ref{eq:ThetaESL}). As before,  this mass comes from a large-scale measurement, and is probably overestimated \citep{Roberto2013} or underestimated \Citep{Gavazzi2005}, and so it could be inappropriate to link  $R_A$ and $\theta_{E,II}$  for some groups. We note that the behavior of the points on the top-left panel of  Fig.~\ref{ThetaEWL} are inherited  by the  middle-left and  bottom-left panels of the same figure, with an increase in the scatter and a shift to lower values of the Einstein radius. The trend is almost the same, independently  of whether we use the catalog constructed using $V_{\rm rms}$ as proxy for the velocity dispersion inferred in the observed lenses, or  the one built to match the observational shape of the redshift distribution of the lenses. This is  an expected  result  because both quantities ($\theta_{E,I}$ and $\theta_{E,II}$) are correlated through Eq.~(\ref{eq:Mass_cM}). Both the shift and the scatter are related to the assumed mass, $M_{200}$, derived from the velocity dispersion of the isothermal profile; as the mass grows without  a boundary at large radii, this produces less concentrated halos, product of the $c-V_{\rm rms}$ relation. The small values of $\theta_{E,II}$ can be interpreted as a failure of $\Lambda$CDM to reproduce the observed Einstein radius \citep[e.g.,][]{Broadhurst08}, but this implication is ruled out  in our case considering that our analysis has oversimplifications, the most important being the assumption of spherical symmetry. Thus, consistent with the first result discussed above, we found a low correlation between $\theta_{E,II}$ and $R_A$.  We obtained  $\theta_{E,II}$ = (0.4 $\pm$ 1.5) + (1.1 $\pm$ 0.4)$R_A$  with a correlation coefficient $R$ = 0.4  ($P$ = 1$\times$10$^{-3}$ ), and $\theta_{E,II}$ = (0.4 $\pm$ 1.5) + (1.1 $\pm$ 0.4)$R_A$  with $R$ = 0.4  ($P$ = 1$\times$10$^{-3}$ ) for the  first and second catalog, respectively.

\begin{figure}[h!]
\begin{center}
\includegraphics[scale=0.47]{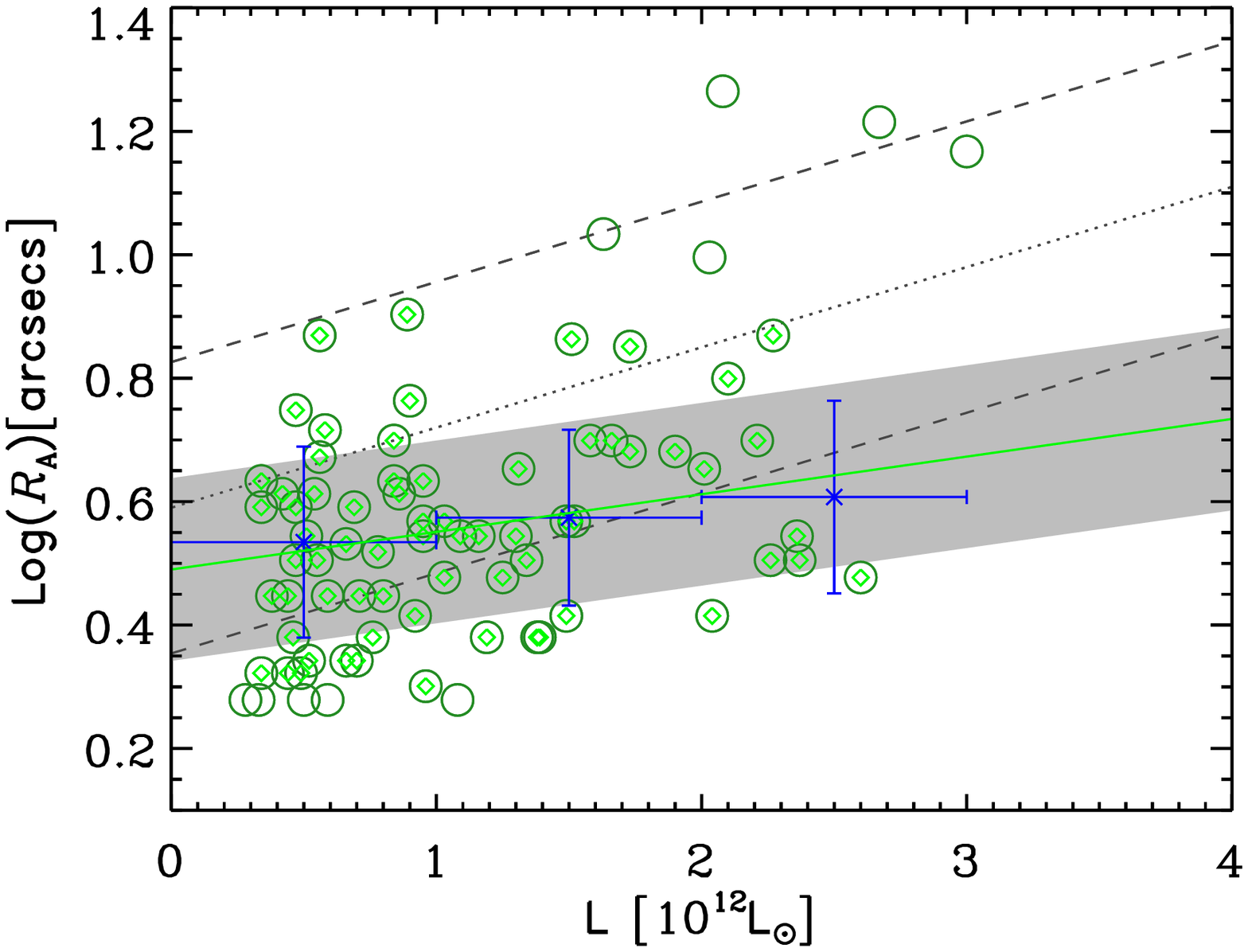}
\includegraphics[scale=0.47]{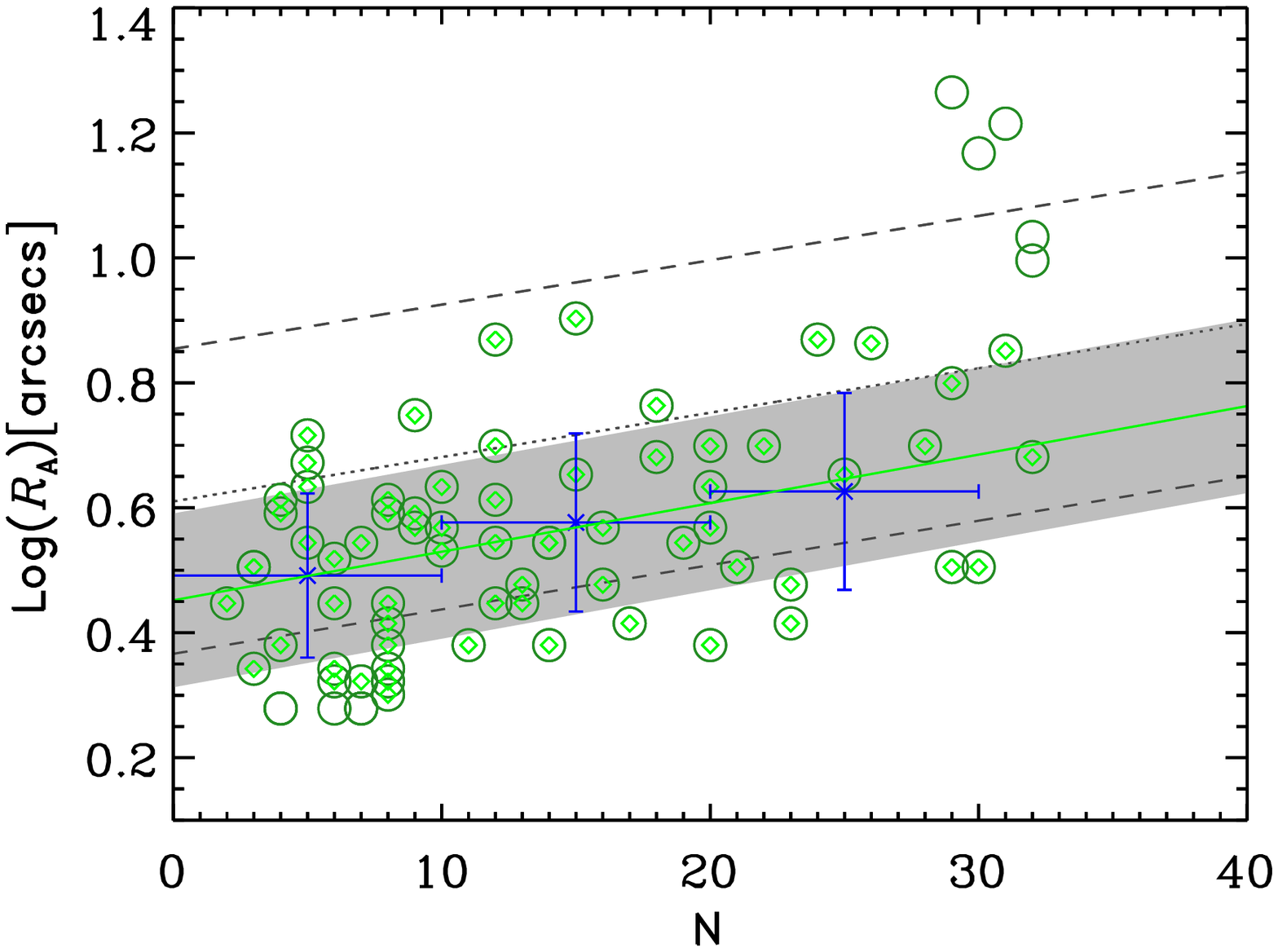}\\
\caption{\textit{Top panel}: $R_A$ as a function of the luminosity.  Green circles depict the secure group candidates in \citet{Gael2013}, green diamonds those with  2.0" $\leq$ $R_A$ $\leq$ 8.0". \textit{Bottom panel}: $R_A$ as a function of optical richness. The quantities were measured within an aperture of 0.5 Mpc. The green continuous  line shows the fit to the green diamonds, with the 1$\sigma$-error depicted as a gray shaded region. Blue asterisks with error bars are plotted to highlight the correlation between $R_A$ and luminosity or optical richness after binning the data. The dotted line shows the relations reported by \citet{Zitrin2012} with a dispersion depicted by the two dashed lines (see Sect.\,\ref{Discussion2}). }
\label{Ra_vs_L}
\end{center}\end{figure}

Furthermore, the two bottom plots in the right panel of Fig.~\ref{ThetaEWL} show the same behavior in the correlation: statistically $\theta_{E,II}$ is slightly larger than  $R_A$.  The same  behavior occurs for small values of  $R_A$ for $\theta_{E,I}$ (see top-left panel of Fig.~\ref{ThetaEWL}). This trend is probably related  to the fact that our sample does not only contain giant arcs. For instance, it is possible that some images  do not form at the tips of the critical lines, which biases the comparison. However, this factor probably has less influence than the one discussed before, namely, the fact that we are using the information at large scale to infer the Einstein radius. This  assertion will be clear below, when we discuss the third method.

\begin{table*}
\caption{Least-squares fitting results for the scaling relations}
\label{tbl-5} 
\centering 
\begin{tabular}{ccccc}
\hline\hline 
\\

\multicolumn{1}{c}{Scaling law}  & \multicolumn{1}{c}{a $\pm$ $\delta$ a}  & \multicolumn{1}{c}{b $\pm$ $\delta$ b}    
& \multicolumn{1}{c}{$\chi^{2}$/$\nu$}      & \multicolumn{1}{c}{$Q$($\frac{N-2}{2}$, $\chi^2/2$)}    \\

\multicolumn{1}{c}{} & \multicolumn{1}{c}{ } & \multicolumn{1}{c}{ } & \multicolumn{1}{c}{ } & \multicolumn{1}{c}{ }
       
\\
\hline 
\\
Log[$R_A$] - N & 0.45 $\pm$ 0.04 & 0.007 $\pm$ 0.002  &  0.85  &  0.82 \\

Log[$R_A$] - L & 0.49 $\pm$ 0.04 & \phantom{0}0.06 $\pm$ 0.03$^{\dagger}$  &  0.96  &  0.60 \\

Log[$R_A$] - (N/L)$_{N}$ & 0.44 $\pm$ 0.06 & 0.2 $\pm$ 0.1  &  0.94  &  0.62 \\
\\
\hline 
\end{tabular}
\tablefoot{
($\dagger$): $\times$10$^{12}$\\
Column (1) lists the scaling law. Columns (2) and (3) list the coefficient values in the relation $Y$ = $a$ + $b$$X$. Columns (4) and (5) list the reduced $\chi^{2}$ and the statistical significance, respectively.}
\end{table*}

The analysis of our strong lensing models for SL2S J08591--0345 (SA63),  SL2S J08520--0343 (SA72),  SL2S J09595+0218 (SA80), and  SL2S J10021+0211 (SA83), and the models previously reported by \citet[][]{paperI}  show that $R_A$ can be used as a proxy of $\theta_E$, which is quantitatively supported by the fit: $\theta_{E,III}$ = (0.4 $\pm$ 1.5) + (0.9 $\pm$ 0.3)$R_A$ with a correlation coefficient $R$ = 0.6. 
In Figure~\ref{RA_SL} we can see that the one-to-one relation is not only followed by those groups strictly defined as having giant arcs.
 If we compare this figure with the right column of  Fig.~\ref{ThetaEWL}, we see that groups with giant arcs are only near the one-to-one relation if we take into account the big error bars in $\theta_{E,I}$ and $\theta_{E,II}$. The fact that in our third method we found a better correlation (with less scatter) between $\theta_{E,III}$ and $R_A$  is explained because we are using the arc positions in the models, i.e. even using simple SIE models we are directly constraining the mass inside the Einstein radius. Unfortunately, we cannot improve our statistics because we do not have  enough data to perform such strong lensing models for more groups in our sample.  Thus, we can regard the first and second method as a complementary study despite their lower robustness.

In Appendix~\ref{sec:ap1} we present the luminosity density contours for the four  new  groups modeled in the present work. These contours were obtained using luminosity maps constructed following \citet[][]{paperI} and \citet{Gael2013}. From Figures ~\ref{fig:A1}, ~\ref{fig:A2},  and ~\ref{fig:A3}  it is clear that SL2S J08591--0345 (SA63),  SL2S J08520--0343 (SA72),  and SL2S J09595+0218 (SA80) are regular groups with circular isophotes around the strong lensing deflector.  The fourth group, SL2S J10021+0211 (SA83),  has a weak-lensing detection less than 3$\sigma$, thus  it is not part of the final sample of \citet{Gael2013} and  it does not appear in Fig.~\ref{ThetaEWL}. However, this object is a good candidate to be a strong lensing group in \citet{More2012}.  This group also shows very elongated and elliptical isophotes, with multimodal peaks (see Fig.~\ref{fig:A4}). It is probably   part of a large-scale structure, since a group or cluster is clearly present at a distance of less than 1 Mpc. There are some  other examples of potentially large-scale structures (Fo{\"e}x et al. in preparation) in the sample of galaxy groups, which is not surprising since such structures have been reported in local large galaxy surveys \citep{Colless2001,Pimblet2004} or detected  around massive clusters at higher redshifts \citep[e.g.,][]{Limousin2012}. Despite the complex morphology, $R_A$ agrees with the value of $\theta_E$ computed from the strong lensing model, but as we noted in Fig.~\ref{ThetaEWL}, and discussed before, irregular groups can have $R_A$$\approx$$\theta_E$.

\citet{Puchwein2009}   studied the cross-sections for giant arcs, using the Millennium simulation, and showed that the radial distribution of tangential arcs is broad and could extend out to several Einstein radii.  Their results (see their Figure 10) are in agreement with the work presented here in the sense that the ratio $R_A$/$\theta_{E}$ could be in some cases greater than 4 (see Fig.~\ref{ThetaEWL}). On the other hand, if we analyze our sample of eleven groups discussed in Section~\ref{RT_modeling}, we found $R_A$/$\theta_{E,III}$ = 1 with a standard deviation of  $\approx$35$\%$. It is interesting that an analogous result is found  for $R_A$/$\theta_{E,I}$ and $R_A$/$\theta_{E,II}$  when the outliers  are eliminated (i.e.,  with a 0.5 $\leq$  $R_A$/$\theta_{E,I}$  $\leq$  2.0 cutoff). It is beyond the scope of the present paper to analyze  the impact on the Einstein radii of other effects such as the orientation \citep[e.g.,][]{Hennawi2007}, the substructures, or structures along the line of sight \citep[][]{Meneghetti2007,Puchwein2009}, the bias in concentration \citep{Oguri2009} and the effect of galaxies in the lensing properties of the groups \citep[][]{Puchwein2009}.  That investigation could be done in the future if more space-based images become available (optical and X-ray) to construct accurate lensing models, and more spectroscopic data is obtained to confirm the group candidates and add additional constraints to the models \citep[see][and the discussion about multi-wavelength approaches]{Limousin2013}. However, we want to point out that the inclusion of galaxy-scale halos in our models can boost the lensing efficiency, between 40$\%$ and 10$\%$ given the redshift of our groups and sources  \citep[see][]{Puchwein2009}. Thus we will use the 35$\%$ of scatter calculated in our analysis as an estimate of all  the aforementioned possible  sources of error  in the use  of $R_A$ as a proxy of $\theta_E$  and its relationship with the luminosity and richness.

\subsection{$R_A$ as a proxy}\label{Discussion2}

In the light of the above discussion, we assume that statistically $R_A$ can be used as a proxy of the Einstein radius, especially if the Einstein radius is estimated through lens modeling and not indirectly, as in the first two methods discussed in this paper. Thus we can employ it to study some properties in galaxy groups.

\subsubsection{$R_A$ and $z$}

As we mentioned in Section~\ref{Intro},  \citet{Zitrin2012} found a possible Einstein radius evolution with redshift. In Fig.~\ref{Ra_vs_L_1} we depicted Log($R_A$) as a function of the redshift. It is evident that, if we restrict the analysis to lenses with  2.0" $\leq$ $R_A$ $\leq$ 8.0",  it is difficult to conclude the existence of an evolution of $R_A$ with redshift, since our sample has few groups below $z$ = 0.2. However,  there is a trend showing a weak anti-correlation between $R_A$ and $z$, which partially supports the preliminary results of \citet{Zitrin2012},  i.e. galaxy groups (as clusters) are expected to be more concentrated at lower redshifts (see Sect.\,\ref{Simulations}). 
We note that we cannot  extend our conclusions to the cluster regime because of the low number of massive groups in our sample. However, if we consider as massive groups those with $R_A$ $>$ 8", a slight  increment  is present in $R_A$ for  $z$ $>$ 0.4, agreeing again with \citet{Zitrin2012}.  For the binned data, we obtained a  weak anti-correlation between $R_A$ and $z$; with Log$R_A$ = (0.56$\pm$0.06) - (0.04$\pm$0.1)$z$. Similarly, for the un-binned data we found Log$R_A$ = (0.58$\pm$0.06) - (0.04$\pm$0.1)$z$. In both cases, the Spearman's rank correlation coefficient was $\sim$ 0.1, but such weak statistical dependence between the two variables is  probably a result of assuming $R_A$ as a linear function of $z$. As our sample is poor below $z$ = 0.3, we  do not attempt to perform a second-order polynomial fitting like the one in  \citet{Zitrin2012}.   In spite of this, it is  important to point out the good agreement between their second-order polynomial fitting and the present work around redshift 0.5.

Considering that the background sources have a wide range of redshifts (within each bin on redshift of the groups),  we can quantify the contribution of this geometrical effect in the scatter in Fig.~\ref{Ra_vs_L_1}.   Lets consider a fixed mass for the lens in  Eq.~\ref{eq:theta_eWL},  $R_A$ $\propto$ $D_{LS}$/$D_{OS}$. Taking a bin centered at  $z$ = 0.45 (i.e.,  setting $z_l$ = 0.45) we found $\Delta$Log$R_A$ = 0.07, using the maximum and minimum  $z_s$ values for the arcs included in that bin. Therefore, the geometrical effect, estimated through  $\Delta$Log$R_A$, is considerably less than the dispersion observed in  Log$R_A$ (less than the 1$\sigma$-error on the correlation). Similar or even smaller  values for  $\Delta$Log$R_A$ are found if we use the other bins, indicating that this effect is negligible.

\begin{figure}[h!]
\begin{center}
\includegraphics[scale=0.5]{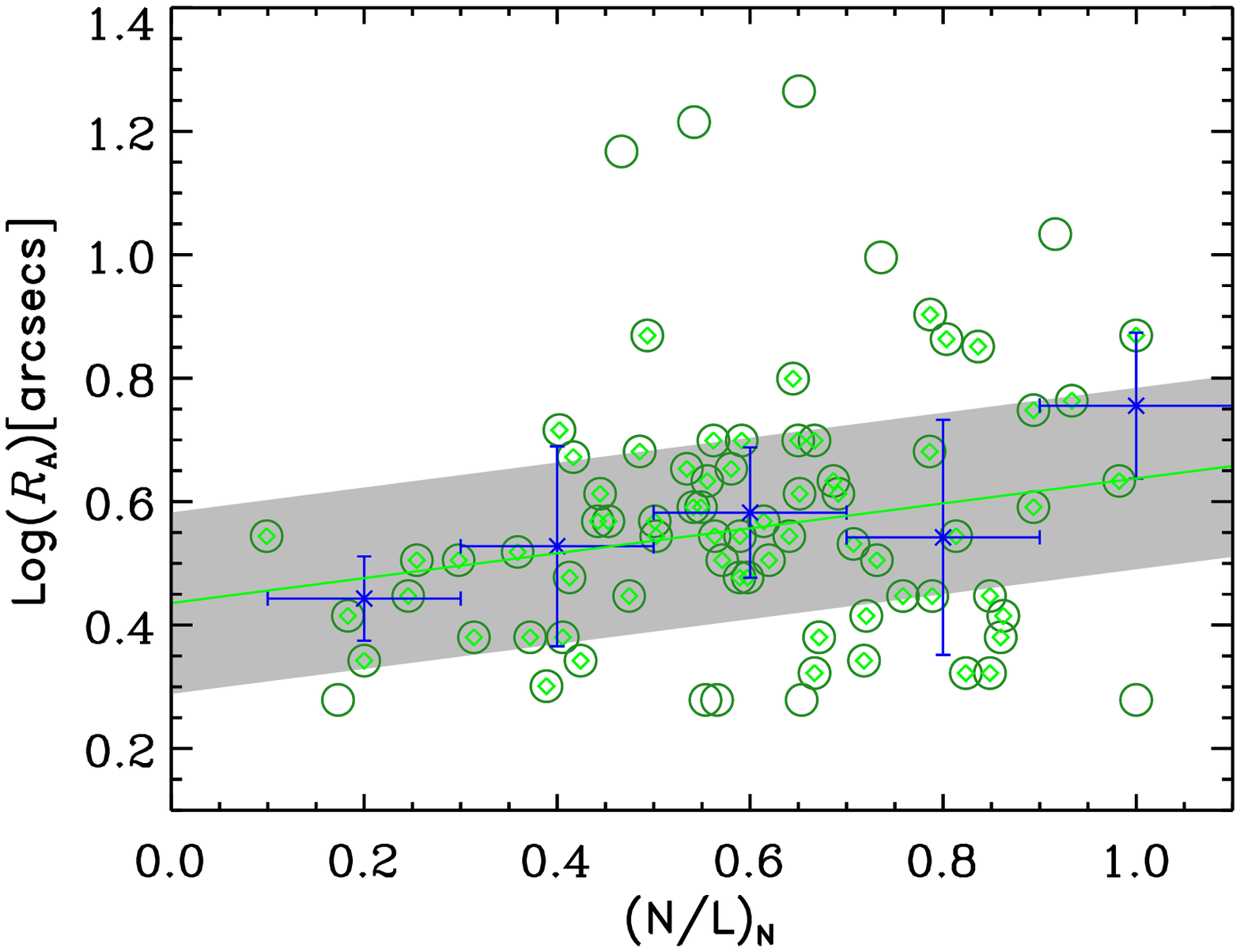}\\
\caption{$R_A$ as a function of the (N/L)$_{N}$ ratio (see Sect.\,\ref{Discussion2}).  Green circles depict the secure group candidates in \citet{Gael2013}, green diamonds those with  2.0" $\leq$ $R_A$ $\leq$ 8.0". The green continuous  line shows the fit to the green diamonds, with the 1$\sigma$-error depicted as a gray shaded region. Blue asterisks with error bars represent the binned data.}
\label{Ra_vs_NL}
\end{center}\end{figure}

\subsubsection{$R_A$, luminosity and richness.}

The  top and bottom panels in Fig.~\ref{Ra_vs_L} show $R_A$ as a function of luminosity and optical richness,  respectively. Both parameters were measured at 0.5 Mpc from the center \citep[see the discussion in][]{Gael2013}. The correlations depicted in Fig.~\ref{Ra_vs_L} are consistent with the results presented in \citet{Gael2013}.  The same trends in richness and luminosity were discussed in  \citet{Zitrin2012}. Nevertheless, we want to  stress the main differences between both works. First, we are presenting the analysis in a sample of galaxy groups. Although the mass regime is included in  \citet{Zitrin2012}, we have a smaller limit than their lower limit of eight members  \citep[this limit  set by ][since their catalog includes clusters with at least this number of members within 0.5 Mpc]{Hao2010}. Second, our groups are a  bona fide sample of strong lensing groups \citep[in some cases confirmed through spectroscopy,][]{Thanjavur2010,Roberto2013}. In Fig.~\ref{Ra_vs_L}  we  also depicted the correlations reported by \citet[][see their Figure 13]{Zitrin2012}, taking into account the 1-$\sigma$ width in the distribution. Our results are in good agreement with their work in particular for Log[$R_A$] - N. In the case of Log[$R_A$] - L our correlation is slightly shallower,  possibly reflecting the different mass regime  being analyzed;  i.e. our sample has a  cutoff at $R_A$ = 8.0".   This cutoff is more noticeable in luminosity because  for massive groups we probably miss  some luminous members when using a 0.5 Mpc radius. 

Estimating the error in using $R_A$ as a proxy of $\theta_E$ is complicated because $R_A$ is measured directly from the images \citep[see][]{More2012}, in contrast to $\theta_E$ which is estimated through  modeling. Additionally, we need to consider the wide distance between the positions of the tangential arcs, i.e  $R_A$, and the Einstein radius,  produced by substructures or structures along the line of sight, 
the bias in concentration, or orientation effects \citep[][]{Hennawi2007,Meneghetti2007,Puchwein2009,Oguri2009}. Thus, we choose a 35$\%$  error (see section ~\ref{Discussion1}) in order to encompass the influence of such factors.
This uncertainty on the Einstein radii is less conservative than the 25$\%$ imposed by  \citet[][]{Auger2013} in their analysis of  strong gravitational lens candidates.  The errors are omitted in Fig.~\ref{Ra_vs_L} for clarity. In Table ~\ref{tbl-5} we present the least-squares fitting results for both scaling relations, Log[$R_A$] - N and Log[$R_A$] - L. We also calculate the statistical significance of the fit through the  $Q$[($N-2$)$/2$, $\chi^2/2$] value. Although the probability is high,  indicating a strong correlation, it is important to  stress the potential overestimation of errors.

As $R_A$ correlates  with $N$ and $L$,  it is natural to infer some dependence between $R_A$ and the ratio N/L. 
Figure~\ref{Ra_vs_NL} shows Log[$R_A$] as a function of (N/L)$_{N}$, i.e. the  optical richness-to-light ratio, normalized to the largest value. Table ~\ref{tbl-5} show the results  for our fit. Because  a mass-richness  relationship is found in galaxy clusters \citep[e.g.,][and references therein]{Gladders2007,Rozo2009,Giodini2013}, we expect a correlation between N/L  and  the mass-to-light ratio (M/L), and thus a dependence between $R_A$ and N/L, since the Einstein radius correlates with its enclosed mass.  In particular, our group sample show $R_A$ scale with the mass-to-light ratio (Fo{\"e}x et al. in preparation). However, going beyond such connections, we want to understand the  physical meaning of this correlation. Big groups, i.e. those with larger $R_A$, have a higher N/L ratio probably because they are less dynamically evolved (without enough time to build big galaxies),  they are not relaxed,  and they exhibit substructure. Thus,  these groups are strong lenses, which means a large $R_A$. At the other end of the correlation,  in small groups (small $R_A$) the luminosity is dominated by the brightest galaxy  in the group, thus the luminosity of the group is well represented by only a few galaxy members.  Given the scatter, we cannot go any further in the analysis of this correlation, and we only establish our results as tentative. Nevertheless, we are able to discuss two selection effects that would bias the correlation. First, we could have groups nearer in redshift or larger with a large $R_A$, thus,  going down to much lower luminosities in the luminosity function (LF) would increase N and decrease L. This scenario is ruled out because both N and L are estimated within a given range in absolute magnitude (the bright part of the LF) for every redshift/velocity  \citep[see][]{Gael2013}. Second, both N and L could have significant uncertainties due to contamination by interlopers. However, L suffers more contamination than N because, for instance, a couple of bright-foreground interlopers have a large effect on L  but only change N by a comparatively small amount. We cannot  completely rule out this second bias effect, and it could be associated with the discrepancies between \citet{Zitrin2012} and the correlation obtained for top panel  in Fig.~\ref{Ra_vs_L}.

\citet{Gael2013} showed that it was possible to use the optical-scaling relations as reasonable mass proxies to analyze large samples of lensing galaxy groups, and to obtain cosmological constraints. They explored the relation between $\sigma_{v}$ and the main optical properties of the SARCS sample and found good correlations,  i.e. the more massive systems are richer and more luminous (despite a 35$\%$ scatter). The analysis presented here is one more step in the analysis of strong lensing galaxy groups. The correlations discussed in the preceding paragraphs demonstrate the potential of using $R_A$ as a complementary proxy to study a sample of lensing groups. Being galaxy groups less complex than galaxy clusters from a lensing viewpoint,  the $R_A$  and the $\sigma_{v}$ obtained from weak lensing will be important tools to characterize galaxy groups. These are valuable assets in the forthcoming era of big surveys (LSST, DES, EUCLID) since spectroscopic data for all the objects would be not always available.

 \section{Conclusions}\label{Conclusions}
 
We analyzed for the first time, using different approaches, the Einstein radii in a sample of objects that belongs to the SL2S \citep[][]{Cabanac2007,More2012} and where selected as secure group candidates in \citet{Gael2013}. The task was done using observational data, image and spectroscopy (CFHT, HST, IMACS), and numerical simulations (\verb"Multidark"). Our main results can be divided in two parts and summarized as follows:

\begin{itemize}
\item Despite the scatter, we found a correlation between $R_A$ and the Einstein radius in galaxy groups.

\begin{enumerate}

   \item Using weak lensing data \citep{Gael2013} we show $\theta_{E,I}$ correlates with $R_A$ with a large scatter; $\theta_{E,I}$ = (2.2 $\pm$ 0.9) + (0.7 $\pm$ 0.2)$R_A$ with  $R$ = 0.33. However, when we eliminate extreme values (outliers) in the ratio $R_A$/$\theta_{E,I}$, both the correlation coefficient and the significance improve ($R$ = 0.6, and $P$ = 1$\times$10$^{-5}$). Since  the distribution of tangential arcs extends beyond the Einstein radius \citep[e.g.][]{Puchwein2009},  some scatter is expected.  In our sample the scatter comes mainly because we are using the information at large scale (weak lensing velocity) in a  lower-scale regime (strong lensing).

   \item Using numerical simulations, we constructed two different catalogs to mimic our sample of galaxy groups. One with $V_{\rm rms}$ as proxy for the velocity dispersion measured in the observed lenses and the second one built to match the shape of the observational redshift distribution of the lenses. We found $\theta_{E,II}$ = (0.4 $\pm$ 1.5) + (1.1 $\pm$ 0.4)$R_A$  with a correlation coefficient $R$ = 0.4  ($P$ = 1$\times$10$^{-3}$ ), and $\theta_{E,II}$ = (0.4 $\pm$ 1.5) + (1.1 $\pm$ 0.4)$R_A$  with $R$ = 0.4  ($P$ = 1$\times$10$^{-3}$ ) for the  first and second catalog, respectively. Thus, for the second method we also found a correlation between $\theta_{E,II}$  and $R_A$.

    \item We presented strong lensing models obtained using the LENSTOOL code  for SL2S J08591--0345 (SA63), SL2S J08520-0343 (SA72), SL2S J09595+0218 (SA80), and SL2S J10021+0211 (SA83),  showing that for these groups $R_A$ $\sim$ $\theta_{E,III}$.   The first three groups have regular  morphologies, while the last one  exhibits a complex morphology, probably because it is part of a large-scale structure. With the information obtained from these new models, as well as the one obtained from \citet[][]{paperI}, we found $\theta_{E,III}$ = (0.4 $\pm$ 1.5) + (0.9 $\pm$ 0.3)$R_A$ with a correlation coefficient $R$ = 0.6.  This method shows more clearly that there is a correlation between $\theta_{E}$  and $R_A$. This better agreement can be explained because these models make use of arc positions,  which directly constrain the mass inside the Einstein radius.

\end{enumerate}

\item We found that the proxy $R_A$ is useful to characterize some properties such as luminosity  and richness in galaxy groups.

\begin{enumerate}

   \item Analyzing Log$R_A$ as a function of $z$  we found Log$R_A$ = (0.56$\pm$0.06) - (0.04$\pm$0.1)$z$, and  Log$R_A$ = (0.58$\pm$0.06) - (0.04$\pm$0.1)$z$ for the binned and un-binned data, respectively.  Since  our sample has few groups below $z$ = 0.2, we  are unable to confirm the existence of an anti-correlation between $R_A$ and $z$. However,  using groups with $R_A$ $>$ 8", we found a slight  increment  in $R_A$  for $z$ $>$ 0.4, suggesting  a possible evolution of the Einstein radius with redshift  in agreement with \citet{Zitrin2012}.

 \item  It is shown that $R_A$ is correlated with luminosity, and richness.  The more luminous and richer the group is, the larger the $R_A$. We found, Log$R_A$ = (0.45$\pm$0.04) + (0.007$\pm$0.002)N and Log$R_A$ = (0.49$\pm$0.04) + (0.06$\pm$0.03)L. This is consistent with the weak lensing analysis of our sample presented in  \citet{Gael2013}, and  it is  an expected  result given that the Einstein radius is related to the mass of  those systems.

\item We also found a possible correlation between $R_A$ and the N/L ratio. Groups with higher N/L ratio have greater $R_A$. However,  considering our sample might suffer from contamination effects, we emphasize these results are only tentative and require further analysis  (Fo{\"e}x et al. in preparation).

\end{enumerate}

\end{itemize}

SL2S offers a unique sample for the study of  a strong lensing effect in the galaxy-group range \citep{Cabanac2007,Tu09,paperI,Limousin2010,Thanjavur2010,Verdugo2011,More2012,Roberto2013,Gael2013}. The present paper is an additional contribution to these efforts. Currently, our research  team is working on different lines, such as photometric analysis of the groups members, dynamical analysis through spectroscopy, X-ray gas distribution \citep{Gastaldello2014}, large scale structure, and numerical simulations \citep{Fernandez2014}. New results will come, incrementing our understanding of lensing galaxy groups.

 \begin{acknowledgements}
 We thank the anonymous referee for thoughtful suggestions. The authors also acknowledge  A. Jordan for giving  us part of his time in Magellan to observe our targets. We thank R. Gavazzi for allowing us to use his HST reduced images. We also thank K. Vieira for helping with proofreading. T. Verdugo acknowledges support from CONACYT through grant 165365 and 203489 through the program Estancias posdoctorales y sab\'aticas al extranjero para la consolidaci\'on de grupos de investigaci\'on. V. Motta gratefully acknowledges support from FONDECYT through grant 1120741. G. Fo\"ex acknowledges support from FONDECYT through grant 3120160. R.P. Mu\~noz acknowledges support from CONICYT CATA-BASAL and FONDECYT through grant 3130750.  M. Limousin acknowledges the Centre National de la Recherche Scientifique for its support, and the Dark Cosmology Centre, funded by the Danish National Research Foundation. G.F., M.L., and V.M. also acknowledge support from ECOS-CONICYT C12U02.
\end{acknowledgements}

\bibliographystyle{aa} 
\bibliography{references}

\clearpage

\onecolumn

\begin{appendix}
\section{Luminosity density contours}
\label{sec:ap1}
 
\begin{figure*}[!htp]
\centering
\includegraphics[width=0.6\textwidth]{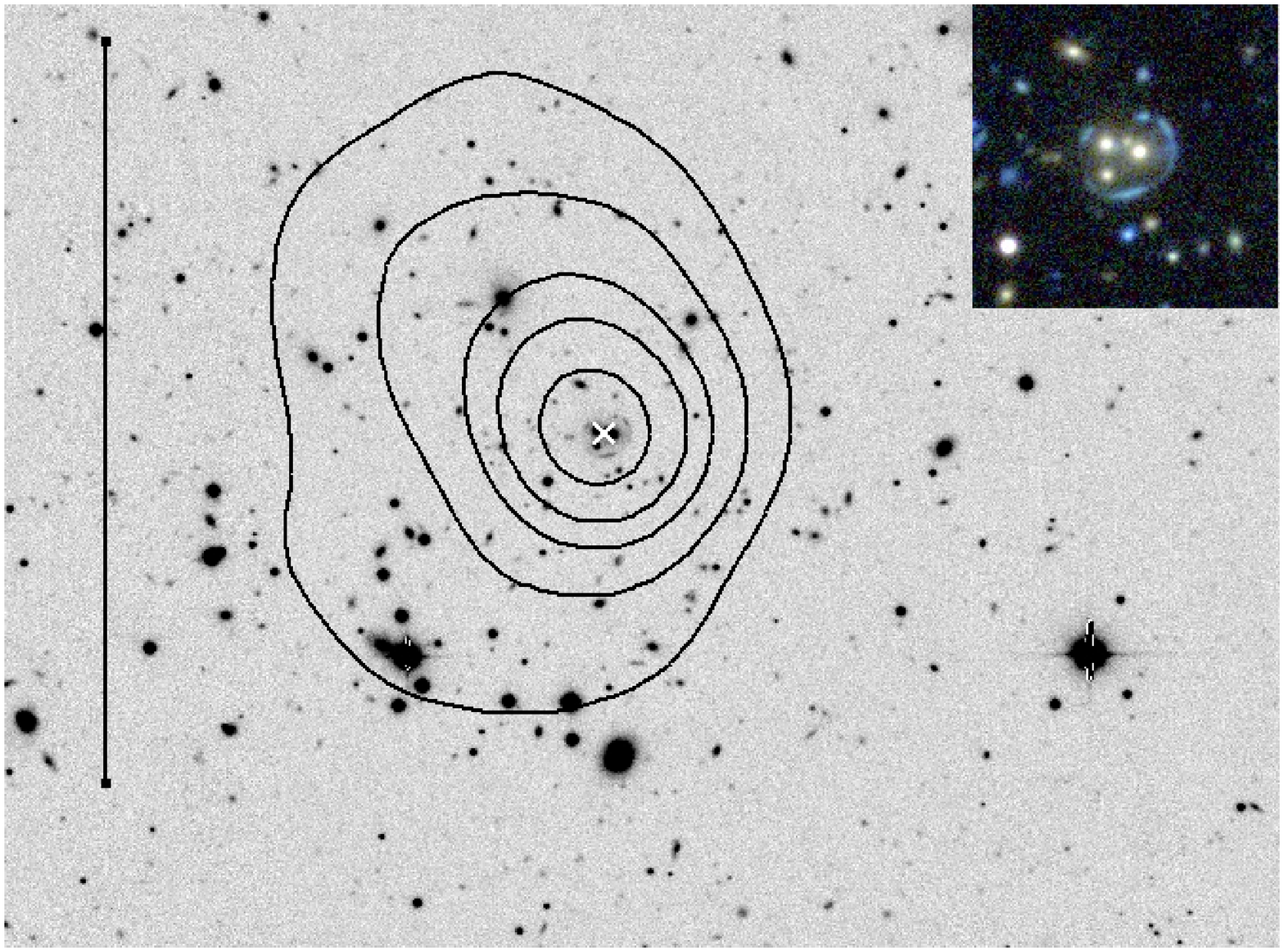}
\caption{Luminosity density contours for SL2S J08591--0345 (SA63). They represent $2\times10^6$, $4\times10^6$, $7\times10^6$ and $10^7$, and $1.5\times10^7\;$L$_\odot\,$kpc$^{-2}$ from outermost to innermost. The white cross marks the galaxy at the center of the strong lensing system. The black vertical line on the left is 1 Mpc long. The stamp in the top-right corner shows a 30$\arcsec$$\times$30$\arcsec$  CFHTLS false color image of the system.
}
\label{fig:A1}
\end{figure*}

\begin{figure*}[!htp]
\centering
\includegraphics[width=0.6\textwidth]{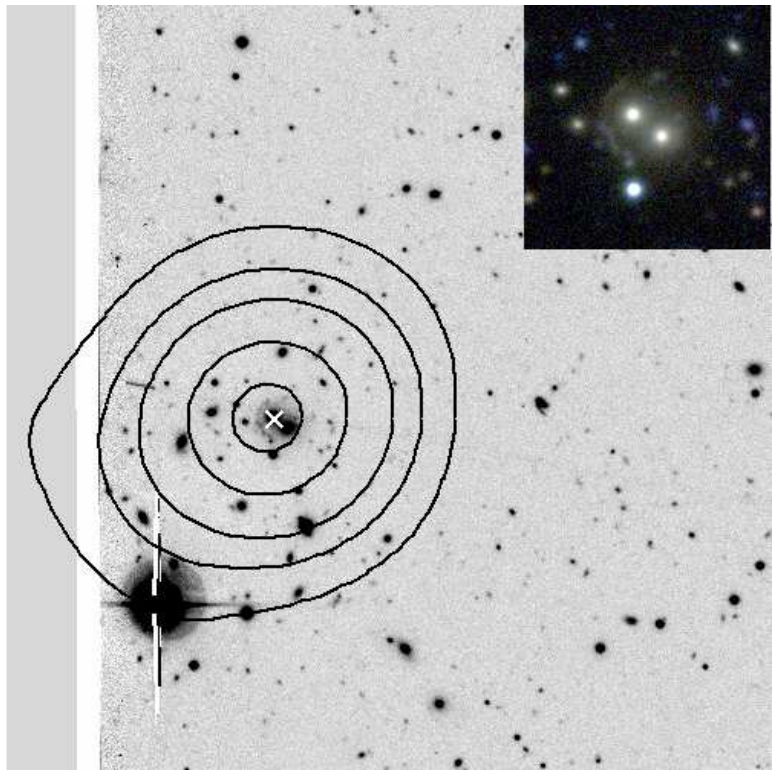}
\caption{Same as Fig. A.1 for SL2S J08520--0343 (SA72)}
\label{fig:A2}
\end{figure*}

\begin{figure*}[!htp]
\centering
\includegraphics[width=0.6\textwidth]{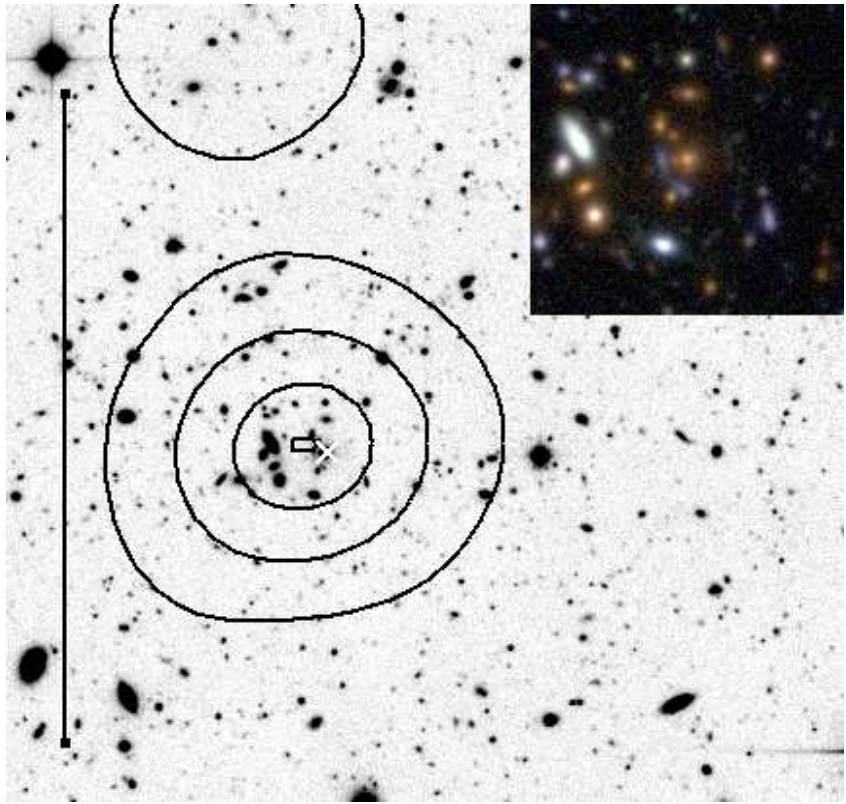}
\caption{Same as Fig. A.1 for SL2S J09595+0218 (SA80).
}
\label{fig:A3}
\end{figure*}

\begin{figure*}[!htp]
\centering
\includegraphics[width=0.6\textwidth]{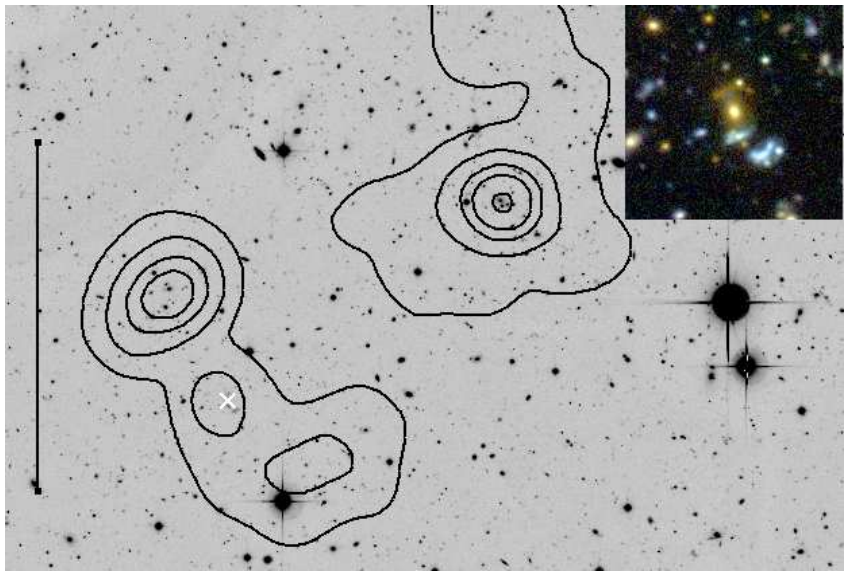}
\caption{Same as Fig. A.1 for SL2S J10021+0211 (SA83).
}
\label{fig:A4}
\end{figure*}

\end{appendix}

\end{document}